\documentclass[apj]{emulateapj}

\usepackage{apjfonts}
\usepackage{lscape}
\usepackage{color}
\usepackage{amsmath}
\usepackage{textcomp}
\usepackage{url}
\usepackage{graphicx, subfigure}

\newcommand{\kms}{\hbox{km~s$^{-1}$}}
\newcommand{\Mjup}{$M_{\mathrm{Jup}}$}
\newcommand{\Msol}{$M_{\odot}$}
\newcommand{\masyr}{$\mathrm{mas}\ \mathrm{yr}^{-1}$}

\slugcomment{To appear in ApJ.}

\shorttitle{Very Low Mass Candidate Members to Young Kinematic Groups}
\shortauthors{Gagn\'e et al.}

\begin{document}

\title{BANYAN. II. VERY LOW MASS AND SUBSTELLAR CANDIDATE MEMBERS TO NEARBY, YOUNG KINEMATIC GROUPS WITH PREVIOUSLY KNOWN SIGNS OF YOUTH}

\author{Jonathan Gagn\'e,\, David Lafreni\`ere,\, Ren\'e Doyon,\, Lison Malo, {\scriptsize AND} \'Etienne Artigau}
\affil{D\'epartement de Physique and Observatoire du Mont-M\'egantic, Universit\'e de Montr\'eal, C.P. 6128 Succ. Centre-ville, Montr\'eal, Qc H3C 3J7, Canada}

\begin{abstract}

We present Bayesian Analysis for Nearby Young AssociatioNs~II (BANYAN~II), a modified Bayesian analysis for assessing the membership of later-than-M5 objects to any of several Nearby Young Associations (NYAs). In addition to using kinematic information (from sky position and proper motion), this analysis exploits 2MASS-\emph{WISE} color-magnitude diagrams in which old and young objects follow distinct sequences. As an improvement over our earlier work, the spatial and kinematic distributions for each association are now modeled as ellipsoids whose axes need not be aligned with the Galactic coordinate axes, and we use prior probabilities matching the expected populations of the NYAs considered versus field stars. We present an extensive contamination analysis to characterize the performance of our new method. We find that Bayesian probabilities are generally representative of contamination rates, except when a parallax measurement is considered. In this case contamination rates become significantly smaller and hence Bayesian probabilities for NYA memberships are pessimistic. We apply this new algorithm to a sample of 158 objects from the literature that are either known to display spectroscopic signs of youth or have unusually red near-infrared colors for their spectral type. Based on our analysis, we identify 25 objects as new highly probable candidates to NYAs, including a new M7.5 bona fide member to Tucana-Horologium, making it the latest-type member. In addition, we reveal that a known L2$\gamma$ dwarf is co-moving with a bright M5 dwarf, and we show for the first time that two of the currently known ultra red L dwarfs are strong candidates to the AB Doradus moving group. Several objects identified here as highly probable members to NYAs could be free-floating planetary-mass objects if their membership is confirmed.

\end{abstract}


\section{INTRODUCTION}

Nearby Young Associations (NYAs) provide a unique means of studying the formation processes and physical properties of stars and brown dwarfs (BDs) at ages ranging from 8~Myr to 120~Myr. Since these associations are close-by and believed to have formed coevally, each of them consists of an easily accessible sample of objects at the same age. Furthermore, their relative youth means that they have not dispersed significantly yet, and hence that their members still share similar space velocities, within a few \kms. The advent of the \emph{Hipparcos} catalog has revealed several NYAs within 100 pc. The main ones that are well-defined and younger than 120~Myr include TW Hydrae (TWA; 8 - 12~Myr; \citealp{2004ARA&A..42..685Z}), $\beta$ Pictoris ($\beta$PMG; 12 - 22~Myr; \citealp{2001ApJ...562L..87Z}), Tucana-Horologium (THA; 20 - 40~Myr; \citealp{2000AJ....120.1410T}, \citealp{2001ASPC..244..122Z}), Carina (CAR; 20 - 40~Myr; \citealp{2008hsf2.book..757T}), Columba (COL; 20 - 40~Myr; \citealp{2011ApJ...732...61Z}),  Argus (ARG; 30 - 50~Myr; \citealp{2011ApJ...732...61Z}) and AB Doradus (ABDMG; 70 - 120~Myr ; \citealp{2004ApJ...613L..65Z}). However, since \emph{Hipparcos} is limited to bright stars, it uncovered only the most massive (F, G and K) members to NYAs. Since the initial mass function (IMF) peaks around 0.3 \Msol ($\gtrsim$~M3), most of the members to NYAs remain to be identified, a challenge that has only recently been tackled (\citealp{2004ARA&A..42..685Z}, \citealp{2008hsf2.book..757T}, \citealp{2009AJ....137.3345C}, \citealp{2013ApJ...762...88M}, \citealp{2013ApJ...774..101R}, \citealp{2013prpl.conf2G024F}, \citealp{2013ApJ...777L..20L} and references therein). Finding these low-mass members would be of great interest for several reasons. It would allow us to study the low-mass end of the IMF in different environments while providing a unique test bench for evolutionary models at young ages, in addition to providing a sample of age-calibrated young systems in the solar neighborhood. The latter is particularly interesting for the dynamic field of exoplanet imaging: low-mass stars (LMSs) or BDs are intrinsically fainter than their more massive equivalents, and young planets are hotter (thus brighter) than older ones because of the thermal energy stored during their initial contraction. Those two effects both reduce the contrast ratio between a planet and its host star, thus facilitating their detection. Yet the identification of such low-mass objects is a difficult task because (1) members of NYAs are spread over very large portions of the sky, and (2) their colors can be confused with those of the overwhelmingly more numerous field stars and BDs. In the case of the youngest NYAs, objects later than $\sim$ L1 could have masses down into the planetary regime, which would provide an easy way of studying the atmosphere of such objects. NYAs represent interesting test benches for planetary formation theories, since 10 and 30~Myr respectively correspond to the formation timescales of giant and terrestrial planets \citep{2003ApJ...599..342S}.\\

Recently, \cite{2013ApJ...762...88M} proposed a new quantitative method, Bayesian Analysis for Nearby Young AssociatioNs (BANYAN),  to assess the probability that a given object belongs to such NYAs through Bayesian inference. With the use of this method, they identified an M5 + M6 binary bona fide member to the $\beta$PMG, 16 very strong K5~\textendash~M5 candidates to NYAs with radial velocity and parallax measurements, as well as 167 strong candidates without available radial velocity or parallax measurements. We define bona fide members in a way similar as Malo et al. (2013, Section 4.3; see also Section \ref{sec:bonafide} of this paper)~:  we thus consider that bona fide members are objects with a good measurement of proper motion, radial velocity and parallax which show Galactic position, space motion and youth indicators consistent with the properties of a NYA. \\

Later-type candidates could not be efficiently uncovered with the method of \cite{2013ApJ...762...88M}, because they made use of the $I_C - J$ colors to calibrate the probabilities over the distances considered, where $I_C$ magnitude is generally not available for very low-mass stars and BDs. Adapting the tool of \cite{2013ApJ...762...88M} to enable the identification of very low-mass stars and BDs in NYAs is the main focus of this work. Since the spectral energy distribution (SED) shifts to the near-infrared (NIR) at later spectral types, it is thus necessary to use yet redder colors to identify the latest members of NYAs. For this purpose, we use here two colors based on filters from the 2MASS and \emph{WISE} surveys. We also implement several other modifications to the approach of \cite{2013ApJ...762...88M} to bring the Bayesian probabilities closer to physically meaningful values. The new method presented here has already identified a candidate free-floating planetary-mass object (\emph{planemo}) member to  ABDMG \citep{2012A&A...548A..26D} and a binary M5 candidate to THA around which a 12\textendash 14 \Mjup\ object was directly imaged (\citealp{2013A&A...553L...5D}; J. Gagn\'e et al., in preparation). \\

This paper starts by describing the current known population of late type ($>$ M5) dwarfs showing signs of youth or NIR colors redder than normal. Then, we describe the Bayesian statistical method used for finding new candidate members to NYAs. Since this statistical tool needs an input model for every hypothesis under test, namely the membership to a given NYA or to the field, we describe how to build photometric, spatial and kinematic models that can be compared against observables. This is followed by a Monte Carlo analysis to assess the reliability of the probabilities yielded by this Bayesian method. Finally, we apply this analysis to our sample to identify several new very low-mass, highly probable candidate members to NYAs, one new bona fide member, as well as a bright co-moving M5 dwarf to a known, young L2$\gamma$ dwarf.

\section{YOUNG LATE-TYPE OBJECTS IN THE LITERATURE}\label{sec:youngl}

Several LMSs and BDs have been previously identified as young objects either because (1) their optical or NIR spectra display  lower-than normal \ion{Na}{1} (8183 and 8195 \AA; 1.13 and 1.14 $\mu$m), \ion{K}{1} (7665 and 7699 \AA; 1.17 and 1.24 $\mu$m), FeH (8692 \AA; 0.98 and 1.19 $\mu$m), TiO (8432 \AA) or CrH (8611 \AA) equivalent widths due to a lower pressure in their photosphere (due to low surface gravity ; \citealp{2009AJ....137.3345C}), (2) their spectra show stronger-than-normal VO bands, indicative of lower surface gravity \citep{2007ApJ...657..511A}, (3) their NIR spectra display a triangular-shaped $H$-band continuum due to decreased H$_2$ collision-induced absorption which is also a consequence of low gravity, (4) they display signs of accretion, (5) they display lithium at a temperature where old objects would have completely destroyed it, (6) they are over-luminous because of their inflated radius, (7) they display unusually red NIR colors for their spectral type because of a greater amount of dust in their photosphere, (8) they are fast rotators, and/or (9) they display a high level of chromospheric activity, either through high levels of X-ray, radio, UV or H$\alpha$ emission. Based on our review of the literature, we have compiled a list of 158 currently known later-than-M5 young objects ; the observational properties of these candidates are given in Table~\ref{tab:input}, along with the NYA association to which they were previously identified, when applicable. Since the 2MASS and \emph{WISE} catalogs provide a sufficiently good baseline (typically $\approx 11$ yr) to achieve proper motion measurements with errors typically lower than $10$ \masyr, we have used them to measure the proper motion for all objects in our sample and combined them with already existing NIR proper motion measurements when available. For some cases where a parallax solution had been measured for a given object, a very precise proper motion measurement was available and was preferred over the less accurate proper motion provided by 2MASS and \emph{WISE}. There are two exceptions where a proper motion could not be measured this way: \emph{G~196--3B} because the \emph{WISE} source is masked by its bright primary, and \emph{2MASS~J00250365+4759191} because it is absent from the \emph{WISE} catalog. For both of them, other measurements were available in the literature so we have used those. We have included in Table~\ref{tab:input} a subsample of "Possibly Young Objects" with marginal indicators of youth, yet with NIR colors unusually red for their spectral type. This subsample includes the 11 URLs that have been identified by \cite{2008ApJ...686..528L}, \cite{2008ApJ...689.1295K}, \cite{2010ApJS..190..100K}, \cite{2013ApJS..205....6M} and \cite{2013PASP..125..809T}. These URL objects display very red colors but no other signs of low-gravity, which brings the question whether they are unusual young objects, or just old objects with very dusty atmospheres. It has also been hypothesized that these objects could have an anomalously high metallicity. In Section~\ref{sec:results}, we will assess whether those objects could plausibly be members of NYAs using a modified Bayesian analysis.

\section{A MODIFIED BAYESIAN INFERENCE}\label{sec:bayes}

The new method presented here is a modified version of the Bayesian analysis described in \cite{2013ApJ...762...88M}, based on a naive Bayesian classifier. This statistical tool has already shown its high potential in other branches of astrophysics (see \citealp{2001ApJ...548..219B}, \citealp{2004ApJ...607..721N}, \citealp{Zhang:2004tf}, \citealp{2005ESASP.576..467P}, \citealp{2007ASPC..371..429P}, \citealp{2008AN....329..288M}, \citealp{2010MNRAS.407..339B} and \citealp{2011ApJS..194....4B}). We use the position and proper motion of a given object, along with its spectral type and 2MASS $J$, $H$, $K_s$ and \emph{WISE} $W1$ and $W2$ magnitudes, altogether defining a set of observables $\{O_i\}$, to assess the probability that it is a member of any of several NYAs, or to the field (old or young; see Section~\ref{sec:field}); these possibilities define the set of hypotheses $H_k$. When such a measurement is available, radial velocity and/or parallax can be added to the observables to get an updated membership probability that is subject to less false positives. However, since these measurements are generally not available, the general case is developed whereby both radial velocity and distance are treated as marginal parameters. \\

By following the principles of a naive Bayesian classifier i.e. by treating every observable as an independent variable, one can write a generalization of Bayes' theorem including a set of $N$ hypotheses $\{H_k\}$ and $M$ observables $\{O_i\}$ associated with a single astrophysical object $\mathcal{O}$, where its unknown radial velocity $\nu$ and trigonometric distance $\varpi$ are treated as two additional marginal parameters. Following Bayes' theorem, the probability that $\mathcal{O}$ satisfies $H_k$ given its observables $\{O_i\}$ (the set $\{O_i\}$ does not include $\nu$ and $\varpi$) is :

\begin{equation}\label{eq:bayes}
	P(H_k|\{O_i\}) = \frac{P(H_k)}{P(\{O_i\})}\int_0^\infty \int_{-\infty}^\infty P(\{O_i\},\nu,\varpi|H_k)\ d\nu\ d\varpi.
\end{equation}

The \emph{i} and \emph{j} indices always refer to an observable whereas \emph{k} and \emph{l} always refer to an hypothesis. The list of hypotheses $H_k$ considered here are given in Table~\ref{tab:groups}. The \emph{prior probability} $P(H_k)$ is the \emph{a priori} probability that $\mathcal{O}$ respects hypothesis $H_k$ before having performed the Bayesian analysis, and is discussed in Section~\ref{sec:prior}. $P(\{O_i\})$ is called the \emph{evidence}, and acts as a normalization factor. It represents the probability that an object displays the set of observables $\{O_i\}$ irrespective of the hypothesis $H_k$ it verifies. It is simply given by the sum of those probabilities over each hypothesis considered :

\begin{equation}\label{eq:separation}
	P(\{O_i\}) = \sum_{l=1}^{N} P(H_l) \int_0^\infty \int_{-\infty}^\infty \ P(\{O_i\},\nu,\varpi|H_l)\ d\nu\ d\varpi.
\end{equation}

\begin{figure}
	\centerline{\includegraphics[width=0.5\textwidth]{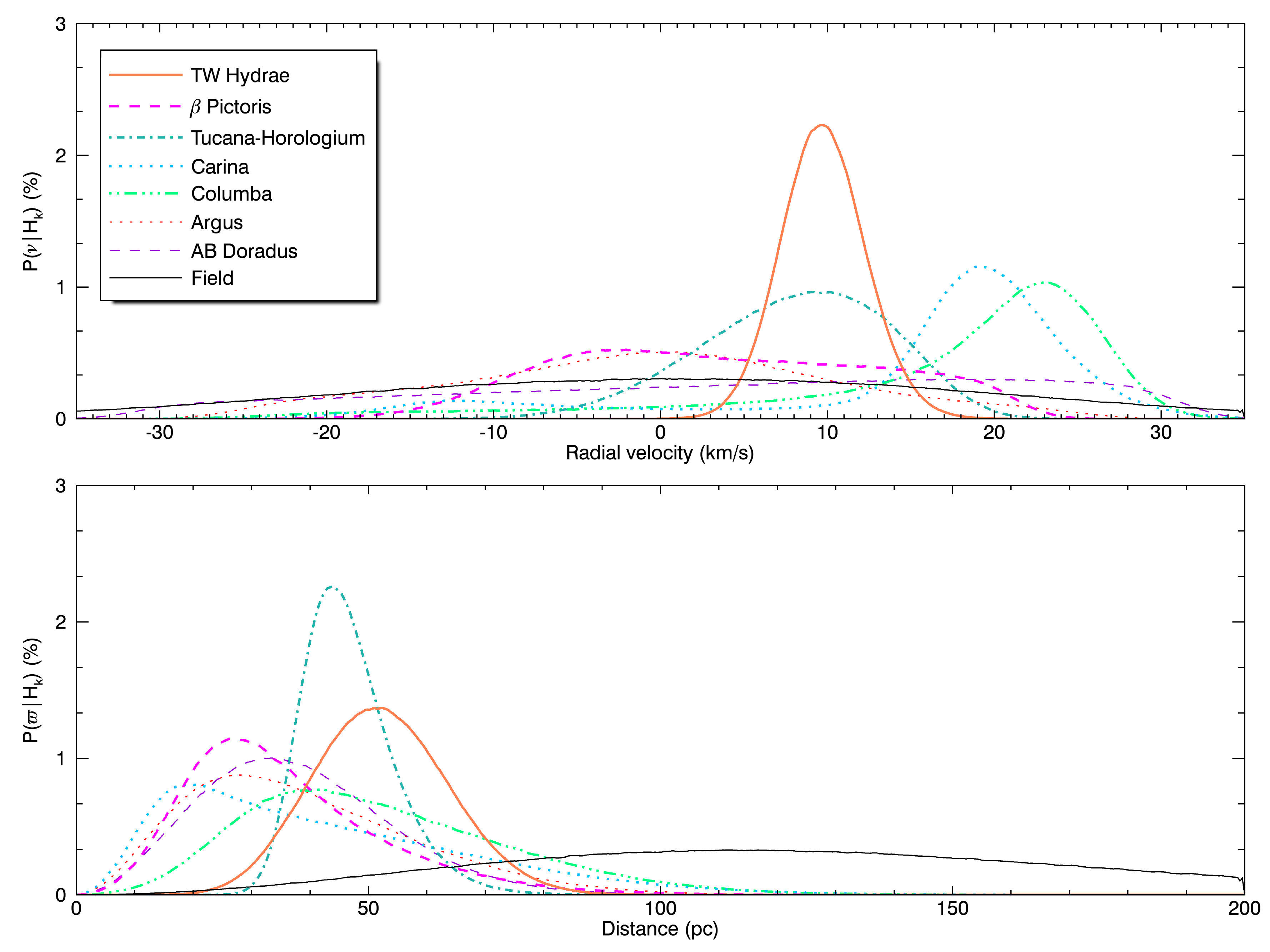}}
	\caption{Prior distributions $P(\nu|H_k)\ d\nu$ and $P(\varpi|H_k)\ d \varpi$ for the two marginalized parameters in our analysis : distance and radial velocity. It can be clearly seen that most of these distributions would be poorly represented by a normal PDF. Each distribution is normalized such that the total area under its curve is equal to unity. We did not show separately young and old field populations, since their prior distributions are similar.}
	\label{fig:priors}
\end{figure}

In practice, a numerical integration of Equation~(\ref{eq:bayes}) is done on a regular $500 \times 500$ grid of distances and radial velocities varying from 0.1 to 200 pc and --35 to 35 kms$^{-1}$ , respectively. These intervals ensure that no object in our sample has a prior or likelihood probability density function (PDF) that peaks near or outside the limits of the grid. At each position of this grid, we evaluate the PDF of the \emph{likelihood} that an hypothesis $H_k$ generates the set of observables $\{O_i\}$ by making the assumption that $\{O_i\}$, $\nu$ and $\varpi$ are independent :

\begin{equation}
	P(\{O_i\},\nu,\varpi|H_k) = P(\nu|H_k)\ P(\varpi|H_k)\prod_{j=1}^{M'}P(Q_j|H_k,\nu,\varpi),
\end{equation}

where $\{Q_j\} = \{Q_j(\{O_i\},\nu,\varpi)\}$ is a set of $M^\prime$ quantities obtained through a transformation of the $M$ observables $\{O_i\}$ and/or $\nu$ and $\varpi$. The purpose of transforming observables is to obtain quantities $Q_j$ which can be represented by a normal distribution for each hypothesis $H_k$ :

\begin{equation}\label{eq:gauss}
	P(Q_j|H_k,\nu,\varpi) = \frac{1}{\sqrt{2\pi}\sigma_j}e^{-(Q_j-\bar{Q_j})^2/2\sigma_j^2},
\end{equation}
	
where $\bar{Q}_j$ and $\sigma_j$ are the mean value and standard deviation describing the normal PDF of $Q_j$ if $\mathcal{O}$ respects the hypothesis $H_k$. The transformed quantities $Q_j$ considered in this work are described in sections \ref{sec:skm} and \ref{sec:photometry}. The quantities $P(\nu|H_k)\ d\nu$ and $P(\varpi|H_k)\ d\varpi$ are generally not well represented by normal distributions, but rather by complex PDFs arising from the transformation of several normal PDFs. These distributions are determined through a numerical Monte Carlo analysis. Each time, we draw a million synthetic objects from the spatial and kinematic models (SKMs) of each $H_k$ (see Section~\ref{sec:skm}) and compute the radial velocity and distance of each one of them. We then build a normalized PDF for $\nu$ and $\varpi$ on the same grid as previously described (see Figure~\ref{fig:priors}). The $P(\{O_i\},\nu,\varpi|H_k)$ represent 2D PDFs for the radial velocity and distance of an object verifying hypothesis $H_k$ (see Figure~\ref{fig:density} for an example). The position of the peak and its characteristic width give the most probable radial velocity and parallax of the object if the hypothesis is true, along with their respective 1$\sigma$ errors. When the radial velocity and/or the distance are known, we remove them from the set of marginal parameters and insert them back into the set of observables $\{O_i\}$. We take measurement errors $\{\Delta O_i\}$ into account by propagating them to the modified observables $\{Q_j\}$, and then by widening their PDFs (see Equation~(\ref{eq:gauss})) by replacing $\sigma_j$ with $\sigma_j^\prime = \sqrt{\sigma_j^2 + \Delta Q_j^2}$. For simplicity, we will refer to the Bayesian probabilities with the $P_{H_k}$ notation instead of $P(H_k|\{O_i\})$ in the remainder of this work. \\

\subsection{The Definition of Prior Probabilities}\label{sec:prior}

The prior probability $P(H_k)$ represents the probability that an object $\mathcal{O}$ verifies the hypothesis $H_k$ before having performed Bayesian inference. Hence, this quantity should depend on the population of objects from hypothesis $H_k$ that could mimic the properties of $\mathcal{O}$. For simplicity reasons, we only consider observables that significantly affect this population estimate, namely the magnitude of proper motion, the Galactic latitude, radial velocity and distance. We define the population fraction $\xi_{O_i;\ k}$ of objects from hypothesis $H_k$ that have the observable $O_i$ comparable to $\mathcal{O}$'s measurement $O_{i;\ m}$ that has a measurement error $\sigma_{i;\ m}$ as :

\begin{equation}
	\xi_{O_i;\ k} = \frac{1}{\sqrt{2\pi}\sigma_{i;\ m}} \int e^{-\ (x\ -\ O_{i;\ m})^2/2\sigma_{i;\ m}^2} \ P(O_i = x|H_k)\ dx,
\end{equation}

where the integral is performed over the range where $O_i$ is defined, and $P(O_i = x|H_k)$ represents the value for the likelihood PDF $P(O_i|H_k)$ at $O_i = x$. For example, the population fraction $\xi_{\varpi;\ k}$ corresponding to an object $\mathcal{O}$ with a distance measurement $\varpi \pm \sigma_\varpi$ would be :

\begin{equation}
	\xi_{\varpi;\ k} = \frac{1}{\sqrt{2\pi}\sigma_\varpi} \int_0^\infty e^{-\ (x\ -\ \varpi)^2/2\sigma_\varpi^2} \ P(\varpi = x|H_k)\ dx.
\end{equation}

In an ideal case where the measurement error would be strictly zero, one would find :

\begin{equation}
	\xi_{O_i;\ k} = P(O_i = O_{i;\ m}|H_k).
\end{equation}

We thus define the prior probability that an object $\mathcal{O}$ verifies $H_k$ by :

\begin{equation}
	P(H_k) = \frac{N_k \prod_i \xi_{O_i;\ k}}{\sum_l N_l \prod_i \xi_{O_i;\ l}},
\end{equation}

where $N_k$ is the expected total population of objects that verify $H_k$ and indice $i$ runs over all available observables from the magnitude of proper motion, Galactic latitude, radial velocity and distance. The denominator serves as a normalization factor so that all prior probabilities sum up to unity. In order to estimate $N_k$, we define our sample as dwarfs later than M5, younger than 1~Gyr and lying within 200~pc of the Sun. We choose 1~Gyr as a conservative limit to ensure that any field object that could imitate the properties of young NYA members is included in the \emph{young field} hypothesis. The reason for choosing such an old limit compared to the oldest NYA considered (ABDMG at 70~\textendash~130~Myr) is that BDs (especially objects with masses around $\sim 80$~\Mjup) significantly younger than 1~Gyr might not have reached their equilibrium radius yet \citep{2001RvMP...73..719B}, which means that they could display signs of low-gravity. A conservative limit is preferred since spectral properties of low-mass objects do not allow to make a precise statement on their age. The 200~pc limit was chosen to match with the grid over which we marginalize distance (see Equation~(\ref{eq:separation}) as well as explanations following it). \\

We cannot estimate the number of NYA members in this sample in a precise manner since their population is still largely incomplete for such late type objects. For this reason, we estimate $N_k$ by supposing that NYAs are complete in the A0~\textendash~M0 spectral type range, then using a log-normal IMF with $m_c = 0.25$ \Msol and $\sigma = 0.5$ dex (\citealp{2012EAS....57...45J}; \citealp{2005ASSL..327...41C}) to estimate the expected number of objects later than M5 in each NYA. To avoid small number statistics, we have combined together bona fide members of all NYAs considered here, and estimated that the total expected late-type population should be approximately 616 objects. Since we do not want to make any predictive statement on the relative population of each NYA, we have thus used an averaged population $N_k = 88$ for every of the seven NYAs considered here. We do not state that this necessarily represents the real low-mass end of the IMF, since it is not well known yet. We rather use this as the best \emph{a priori} estimate that one can make at this time. \\

We define $N_{\mathrm{field}}$ as the total number of objects in our field model (see Section~\ref{sec:skm}). It is probable that some A0~\textendash~M0 stars are still missing in the census of NYAs, the effect here would be that we may underestimate Bayesian probabilities $P(H_k)$ for the NYA hypotheses, and hence that our membership probabilities, as well as our contamination rates (see Section \ref{sec:contam}) would be too conservative. It should be stressed that including such priors in our analysis does not significantly affect the relative classification ranking of different objects, but changes the absolute values of the Bayesian probabilities that each objects are members of a specific NYA. In particular, Bayesian probabilities calculated this way will be significantly lower than those reported in \cite{2013ApJ...762...88M}, who set all priors to unity. In the present work, we use Bayes' theorem to try and assess the probability that objects belong to several NYAs, consisting of our different hypotheses $H_k$. However, since we use a naive Bayesian classifier in the sense that we treat input parameters as independent variables, we expect that the Bayesian probabilities $P(H_k|\{O_i\})$ we derive this way will be biased \citep{Hand:2001tr}. Because of this, we will perform a Monte Carlo analysis (see Section~\ref{sec:contam}) to estimate un-biased membership probabilities, as well as the recovery rate of our method. We strongly advise that the Bayesian probabilities should always be interpreted together with the prior assumptions that were made, and the reader should keep in mind that even if the relative ranking of each hypothesis is preserved for a given object, the absolute Bayesian probabilities remain inevitably biased.

\subsection{The Equal-luminosity Binary Hypothesis}\label{sec:binaries}

In the case of objects for which youth is uncertain, we expect that part of the false-positive candidate members identified with our method will be unresolved field binaries, since such objects would fall higher than the old sequence in a color-magnitude diagram (CMD), and could thus be misinterpreted as earlier, brighter and/or redder (young) objects. For this reason, for each group in our analysis, including the field, we add an \emph{equal luminosity binary} hypothesis, which has the exact same SKM, but with a CMD shifted up by $0.7$ magnitudes. This ensures that objects falling above the old CMD sequence but with position or kinematics not coherent with any NYA would not be interpreted as candidate members. Hence, our membership probabilities will be more conservative by including those binary hypotheses. Higher probabilities for the binary hypotheses (compared to the single-object hypotheses) will also flag the potentially unresolved binaries in our sample. However, since the photometric properties of young systems are not very well defined yet, we do expect a fraction of false-positives amongst the systems we flag as possible binaries. Objects for which the binary hypothesis of the most probable NYA has a higher probability than the single-object hypothesis are indicated as \emph{possible binaries} in the following. For simplicity, we did not use different priors for single and binary hypotheses. This is equivalent to the prior supposition that the binary fraction of young or old, late-type objects is 50\%, regardless of their membership.

\subsection{Modeling Field Stars}\label{sec:field}

We have used a Besan\c{c}on Galactic model (A. C. Robin et al. 2013, in preparation; \citealp{2012A&A...538A.106R}) to compute the values in Tables~\ref{tab:groups} and \ref{tab:NYA_coord}, for both the \emph{field} and \emph{young field} hypotheses, consisting of objects with ages more or less than 1~Gyr, respectively. The main differences between those two populations are (1) that the old one is larger in number and has a larger kinematic scatter, and (2) that younger objects have different photometric properties (early-type objects are intrinsically brighter, whereas late-type objects are redder; see Section \ref{sec:photometry}). When one computes the Bayesian probability that an object is a member of NYAs, both field hypotheses should be included in the Bayesian algorithm, unless the object displays evidence for low-gravity, hence youth. In the latter case, the \emph{old field} hypothesis should not be included. As explained earlier, we have included only objects within 200 pc having spectral types M5 or later and luminosity class V (see Section~\ref{sec:prior}). Since these models do not include objects at spectral types later than M9, we have used the same IMF as described in Section \ref{sec:prior} to estimate the population of objects later than M9, which are included in the numbers reported in Table~\ref{tab:groups}. We thus find that the expected number of objects for the young and old field populations are $390\ 007$ and $1\ 601\ 130$, respectively. Since the estimated field population is much higher than that of NYAs, the Bayesian probability that any object belong to NYAs will be significantly decreased in comparison with \cite{2013ApJ...762...88M} where they set prior probabilities to unity. This reflects the fact that an object randomly chosen in an all-sky sample with the aforementioned properties has a much larger probability to be a field object than being a member of a NYA.

\section{MODELING NEARBY, YOUNG ASSOCIATIONS}\label{sec:model_nyas}

In the current model, we have included only NYAs younger than 130~Myr that lie within 100 pc of the Sun and have at least 6 bona fide members. Those associations, along with some of their properties, are listed in Tables~\ref{tab:groups} and \ref{tab:NYA_coord}. In the following sections, we will refer only to associations in this list when we use the term NYAs.

\renewcommand{\tabcolsep}{1.3mm}
\begin{deluxetable}{lcccc}
\tablewidth{0pt}
\tablecolumns{3}
\tablecaption{Properties of Young Local Associations \label{tab:groups} }
\tablehead{
\colhead{Name} & \colhead{Age\tablenotemark{a}} & \colhead{Distance ($\varpi$)}  & \colhead{RV ($\nu$)} & \colhead{Bona Fide} \\
\colhead{of Group} & \colhead{Range (Myr)} & \colhead{(pc)} & \colhead{(\kms)} & \colhead{Members\tablenotemark{b}}
}
\startdata
TW Hydrae (TWA) & $8 \-- 12$ & $40 \-- 62$ & $7 \-- 12$ & $18$ \\
$\beta$ Pictoris ($\beta$PMG) & $12 \-- 22$ & $18 \-- 40$ & $-9 \-- 16$ & $33$ \\
\scriptsize{Tucana-Horologium} (THA) & $20 \-- 40$ & $38 \-- 51$ & $3 \-- 14$ & $52$ \\
Columba (COL) & $20 \-- 40$ & $26 \-- 63$ & $19 \-- 26$ & $21$ \\
Carina (CAR) & $20 \-- 40$ & $11 \-- 42$ & $16 \-- 23$ & $8$ \\
Argus (ARG) & $30 \-- 50$ & $15 \-- 48$ & $-10 \-- 9$ & $11$ \\
AB Doradus (ABDMG) & $70 \-- 120$ & $19 \-- 50$ & $-11 \-- 29$ & $54$ \\
Young Field & $0 \-- 1000$ & $66 \-- 169$ & $-19 \-- 19$ & $\--$ \\
Old Field & $1000 \-- 8000$ & $70 \-- 177$ & $-34 \-- 32$ & \ \ \  $\--$
\enddata
\tablenotetext{a}{We do not suggest those as robust age estimates for NYAs, which is out of the scope of this work. These age ranges result instead from a collection of the different ages proposed in the literature for each NYA. The relative age of the different associations should be correct, however.}
\tablenotetext{b}{See Section~\ref{sec:bonafide}}
\end{deluxetable}

\subsection{A New Spatial and Kinematic Model for Young Moving Groups}\label{sec:skm}

\begin{figure*}
\begin{center}
		\subfigure{
		 \includegraphics[width=0.46\textwidth]{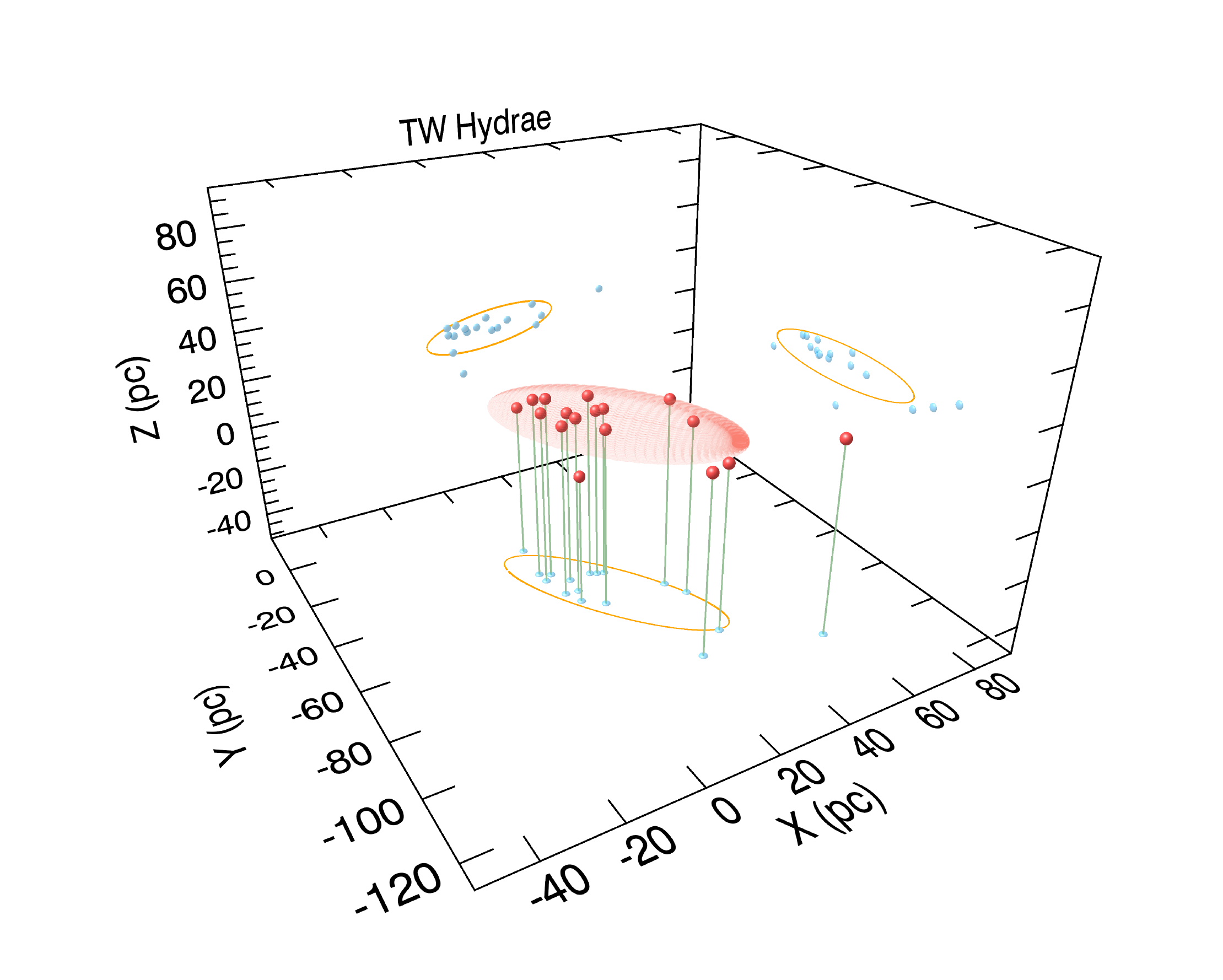}
		}
		\subfigure{
		 \includegraphics[width=0.46\textwidth]{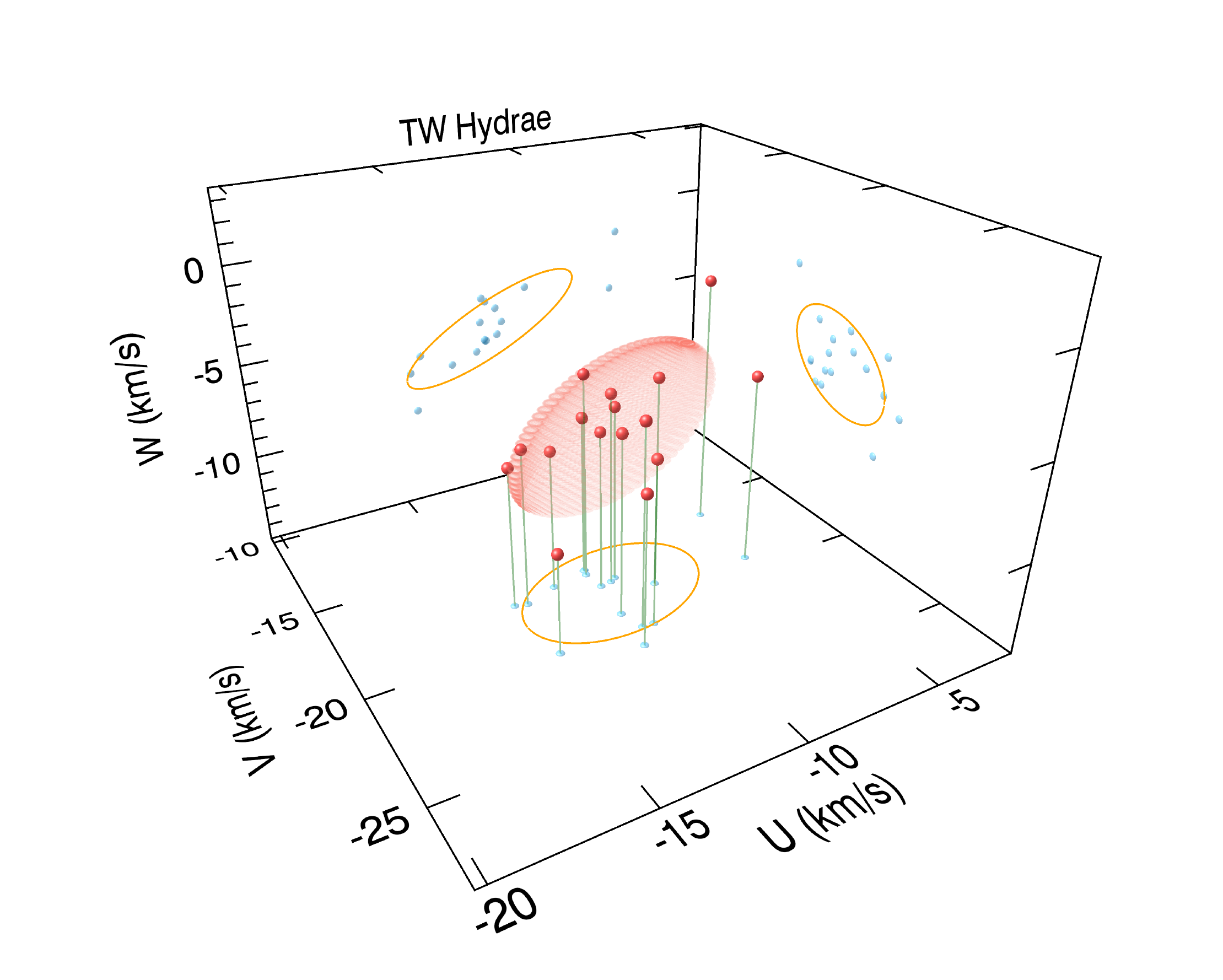}
		}
\end{center}
 \caption{Spatial and kinematic models for TWA (red ellipsoids) derived from its bona fide members (red dots). We show their respective projections as orange lines and blue dots, such that the misalignment of TWA with the local galactic coordinate axes is obvious. Similar figures for all NYAs considered here are available at our group's website \emph{www.astro.umontreal.ca/\textasciitilde gagne}.}
 \label{fig:XYZ_TWA} \label{fig:UVW_TWA}
\end{figure*}

\renewcommand{\tabcolsep}{0.6mm}
\begin{deluxetable*}{lcccccc}
\tabletypesize{\scriptsize}
\tablewidth{0pt}
\tablecolumns{7}
\tablecaption{Mean Galactic Motion and Position in Rotated Reference Frames\label{tab:NYA_coord}}
\tablehead{
\colhead{Name} & \colhead{$U V W$} & \colhead{$\phi_{\sc D}\theta_{\sc D}\psi_{\sc D}$} & \colhead{$\sigma_{\sc U^\prime V^\prime W^\prime}$} & \colhead{$X Y Z$} & \colhead{$\phi_{\sc S}\theta_{\sc S}\psi_{\sc S}$} & \colhead{$\sigma_{\sc X^\prime Y^\prime Z^\prime}$}\\
\colhead{of group} & \colhead{(\kms)} & \colhead{(\textdegree)} & \colhead{(\kms)} & \colhead{(pc)} & \colhead{(\textdegree)} & \colhead{(pc)}
}
\startdata
TW Hydrae (TWA) & $-11.12,-18.88,-5.63$ & $-158.7,-55.3,-5.4$ & $0.90,1.56,2.78$ & $19.10,-54.16,21.54$ & $25.3,60.8,80.4$ & $4.98,7.16,22.57$\\
$\beta$ Pictoris ($\beta$PMG) & $-11.03,-15.61,-9.24$ & $-113.0,-70.3,76.6$ & $1.38,1.72,2.50$ & $7.58,-3.52,-14.53$ & $-90.2,65.1,-77.9$ & $8.22,13.52,30.67$\\
Tucana-Horologium (THA) & $-9.70,-20.47,-0.78$ & $-52.0,-30.2,1.6$ & $1.05,1.68,2.38$ & $6.74,-21.79,-36.05$ & $-28.2,263.1,21.1$ & $3.90,10.62,20.10$\\
Columba (COL) & $-12.14,-21.29,-5.61$ & $143.4,22.7,-68.8$ & $0.51,1.27,1.69$ & $-28.11,-25.78,-28.56$ & $-25.7,-35.5,-62.2$ & $10.55,17.63,28.33$\\
Carina (CAR) & $-10.72,-22.23,-5.67$ & $-68.0,-61.6,-86.4$ & $0.31,0.65,1.08$ & $10.09,-51.63,-14.85$ & $18.4,-16.5,-64.9$ & $5.78,11.34,29.79$\\
Argus (ARG) & $-21.54,-12.24,-4.63$ & $76.1,55.9,29.4$ & $0.87,1.67,2.74$ & $15.04,-21.69,-8.09$ & $-12.4,-73.0,-51.9$ & $12.07,15.51,27.43$\\
AB Doradus (ABDMG) & $-6.96,-27.23,-13.90$ & $-54.4,185.8,10.7$ & $1.18,1.68,1.94$ & $-2.53,1.28,-16.34$ & $57.3,51.9,88.7$ & $16.33,19.95,23.47$\\
Young Field & $-11.21,-18.57,-6.94$ & $69.0,-89.3,-69.8$ & $7.74,12.46,19.58$ & $2.82,0.07,-13.14$ & $52.6,-30.0,27.5$ & $79.63,80.37,80.81$\\
Old Field & $-11.00,-37.25,-6.93$ & $68.8,-89.8,-69.0$ & $18.59,28.73,40.15$ & $2.55,0.01,-2.71$ & $-113.1,0.3,0.0$ & $79.43,79.56,96.06$
\enddata
\end{deluxetable*}
\renewcommand{\tabcolsep}{0.75mm}

In the previous Bayesian inference method described in \cite{2013ApJ...762...88M}, the SKM was defined by  fitting an error function to the cumulative density function (CDF) of the Galactic position $XYZ$ and spatial velocities $UVW$ distributions of the bona fide members in each association. Then, it was assumed that the SKM could be described as a normal distribution having the corresponding mean and standard deviation, for each of the aforementioned parameters. In other words, it was assumed that both the \emph{3D} $XYZ$ and $UVW$ ellipsoids fitting the bona fide members' positions and velocities \emph{necessarily had their principal axes aligned with the local Galactic coordinate axes}. As can be seen in Figure~\ref{fig:XYZ_TWA}, this is generally not the case. To address this issue, we have modified the SKM used here in the following way. (1) For each association, we use the \emph{krEllipsoidFit} IDL procedure\footnote[1]{\emph{krEllipsoidFit} uses a special algorithm for 3D ellipsoids fitting from Ronn Kling and Jerry Lefever, described at \url{http://www.rlkling.com}} to find the "centers of mass", respectively $C_D$ (dynamic) and $C_S$ (spatial), and principal axes of the UVW and XYZ distributions of the bona fide members, as well as the standard deviation of the distribution in the direction of the principal axes. (2) We calculate the sets of three $\phi_{\sc D}\theta_{\sc D}\psi_{\sc D}$ (dynamic) and $\phi_{\sc S}\theta_{\sc S}\psi_{\sc S}$ (spatial) Euler angles needed to make the rotations that bring each ellipsoid's principal axes along the local Galactic reference frame's axes\footnote[2]{Two different rotated reference frames are defined: one for the XYZ and another one for the UVW coordinates.}. The correct procedure to transform $UVW$ coordinates to the $U^\prime V^\prime W^\prime$s is to (1) subtract the $C_D$ center of mass to the $UVW$s, (2) build a rotation matrix from the $\phi_{\sc D}\theta_{\sc D}\psi_{\sc D}$ Euler angles\footnote[3]{A sample IDL routine to achieve this is provided in the electronic version of this paper.}, (3) apply it to the $UVW$s and (4) add back $C_D$ to the result of this rotation. The $XYZ$ coordinates are transformed in the same way. For each association, the principal axes of the $X^\prime Y^\prime Z^\prime$ (or $U^\prime V^\prime W^\prime$) distribution of bona fide members should then fall along the axes of those new frames of reference. In the Bayesian inference method described in the previous section, the $X^\prime Y^\prime Z^\prime$ and $U^\prime V^\prime W^\prime$ coordinates belong to the set $\left\{Q_j\right\}$ of transformed observables, whose PDFs can be represented by normal distributions. The parameters of these reference frames and the associated coordinates of NYAs are listed in Table~\ref{tab:NYA_coord}. The parameters determined for the Carina SKM deserve close examination as they are based on only 7 bona fide members, compared to more than 15 for all other associations. By fitting ellipsoids using only subsets of the other associations, we determined that having only 7 objects yields an uncertainty of up to a factor of 2 in the velocity dispersions, while the effect on the spatial distribution is much smaller. In Figure~\ref{fig:UVW_TWA}, we show the adopted ellipsoids for TWA as an example.

\subsection{Photometric Properties as a Function of Age}\label{sec:photometry}

\begin{figure*}
	\begin{center}
		\subfigure{
		 \includegraphics[width=0.45\textwidth]{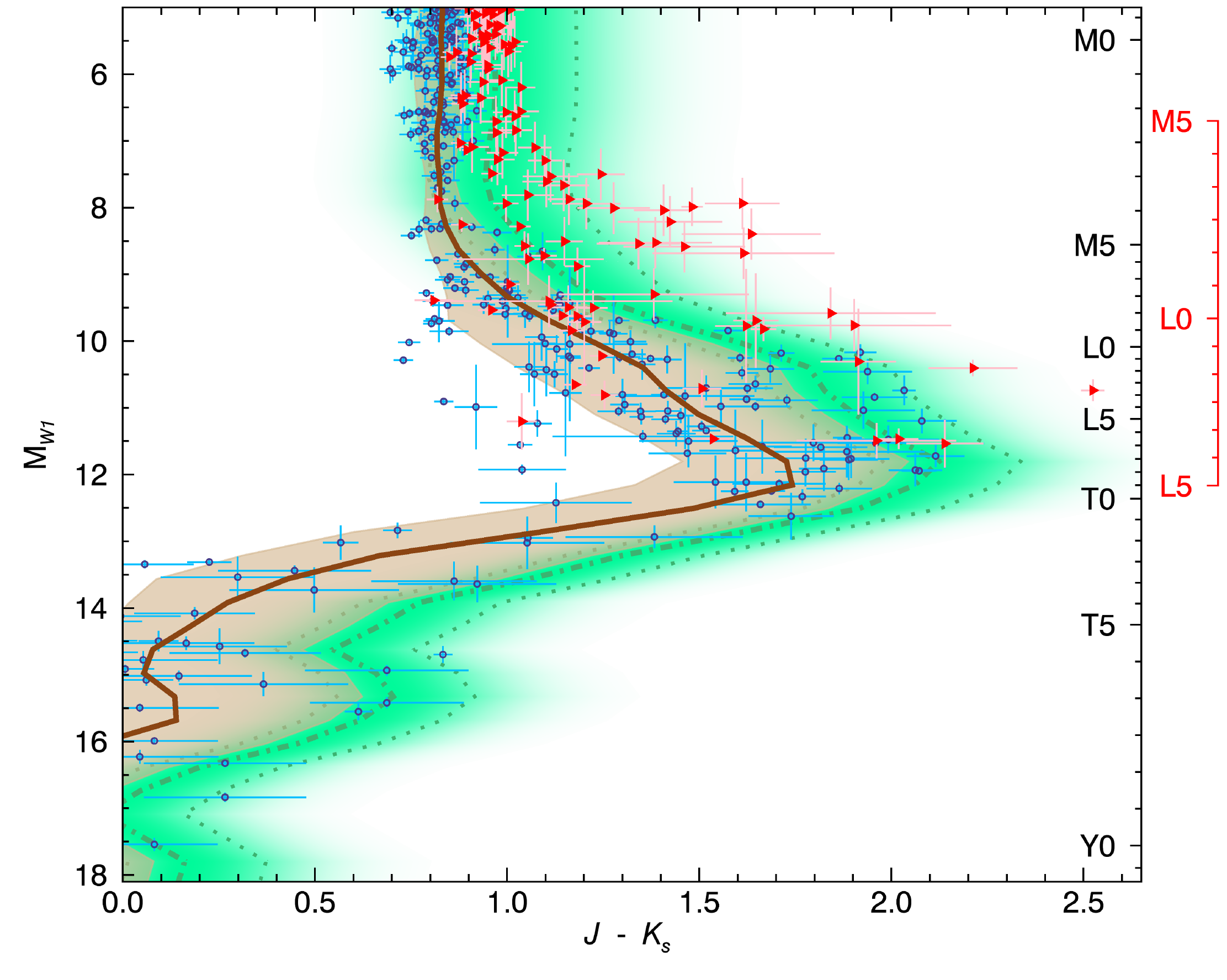}
		}
		\subfigure{
		 \includegraphics[width=0.45\textwidth]{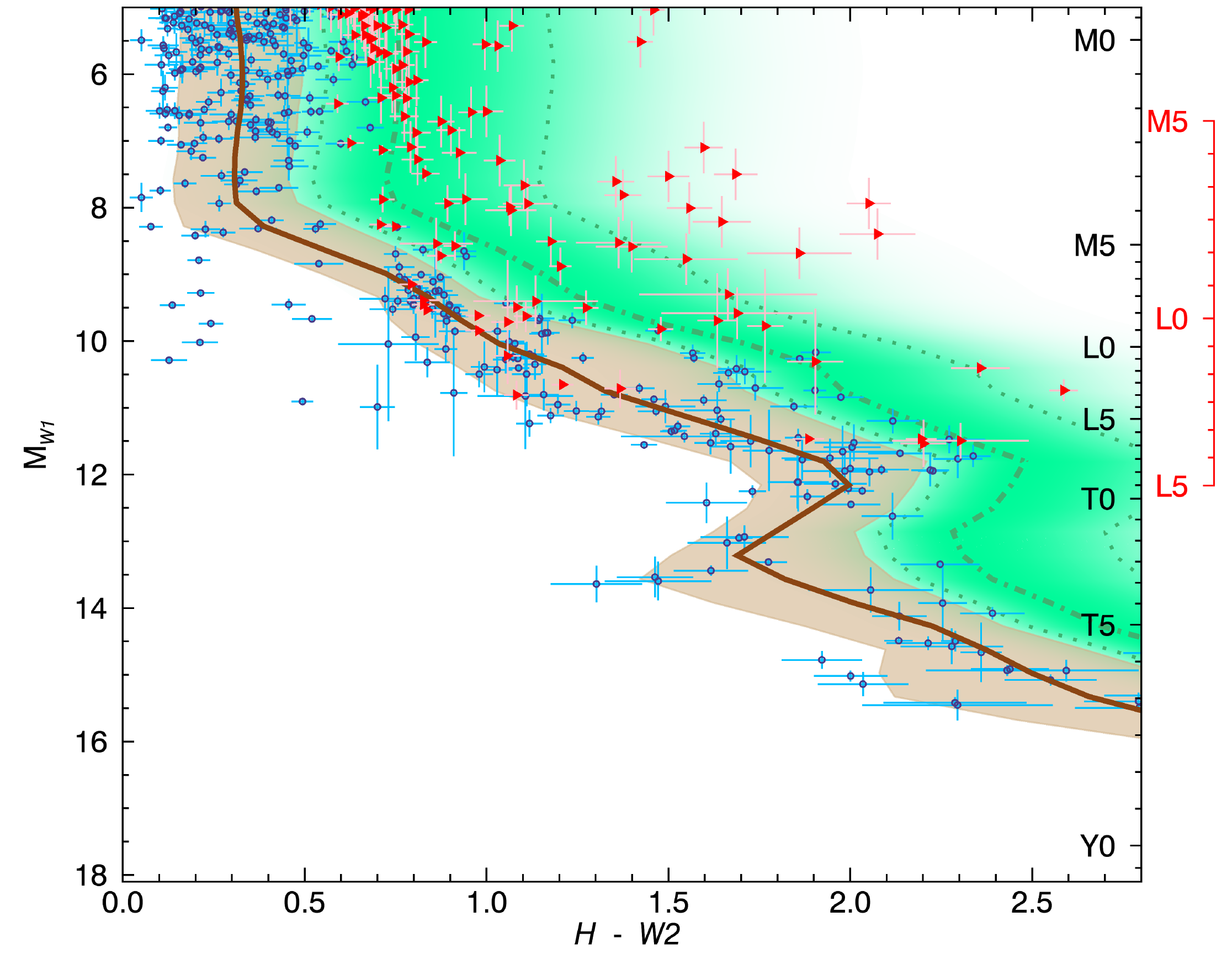}
		}
	\end{center}
	\caption{Color-magnitude diagrams for young (red dots) and old (blue dots) objects with parallax measurements. The thick, brown line and its shaded region respectively represent the old, field sequence and its 1$\sigma$ scatter. See the text for a description of the way the young sequence PDF (green region) was constructed. The thick dash-dotted green line is the field sequence and both dotted green lines delimit its $\pm$1$\sigma$ scatter regions. The rightmost black (red) axis indicates the spectral type of an old (young) dwarf at this absolute \emph{W1} magnitude.}
	\label{fig:photom_sequence}
\end{figure*}

Using a set of known old field LMS later than M5 and BDs with parallax measurements from the Dwarfarchives\footnote[4]{\url{http://ldwarf.ipac.caltech.edu}} (\citealp{2012ApJS..201...19D}; \citealp{2012ApJ...752...56F}), along with similar young Upper Scorpius objects from \cite{2011A&A...527A..24L} and \cite{2011MNRAS.418.1231D}, we have defined two CMDs based on 2MASS and \emph{WISE} photometry that best separate the old and young subsets. These two CMDs are (1) $M_{W1}$ versus $J - K_s$ and (2) $M_{W1}$ versus $H - W2$ (see Figure~\ref{fig:photom_sequence}). In both cases, the average color of the old sequence was defined by minimizing the reduced $\chi^2$ of data points in bins of 0.7 mag in the vertical ($W1$) direction. The scatter associated with this value has been computed by finding the values at which the reduced $\chi^2$ has a $p$-value of 68\%. Since there are only a few young objects, especially at the red end of both CMDs, we have proceeded in a different way to build the young sequence PDF. The shape of the young sequence is taken to be the shape of the +1$\sigma$ old sequence, but shifted to the right.  The reason why we used the shape of the rightmost 1$\sigma$ limit of the field sequence to build the young PDF is that it becomes redder at later spectral types, which is more representative of the general distribution of young objects in the CMDs, especially in the case of $J$ - $K_s$. The shift was determined in the following manner~: first, we built a 2D PDF distribution composed of a sum of 2D normal distributions, located at the positions of each young data point (the red dots in Figure~\ref{fig:photom_sequence}). The vertical and horizontal characteristic widths of each normal distribution were set respectively to the vertical and horizontal measurement errors of the corresponding data points. Then, we determined what horizontal shift to the +1$\sigma$ field sequence was needed so that half of the total area of the previously described 2D PDF was to its left. The width of the young PDF was then taken as the width for which 68\% of the total area of the 2D distribution was encompassed. The resulting young PDF is shown in Figure~\ref{fig:photom_sequence} for each of the two CMDs. We do not pretend that young objects should necessarily fall along these defined sequences, but rather use them only to represent the fact that younger objects are redder (and/or brighter) than the old sequence. \\

\begin{figure}
	\begin{center}
		\includegraphics[width=0.5\textwidth]{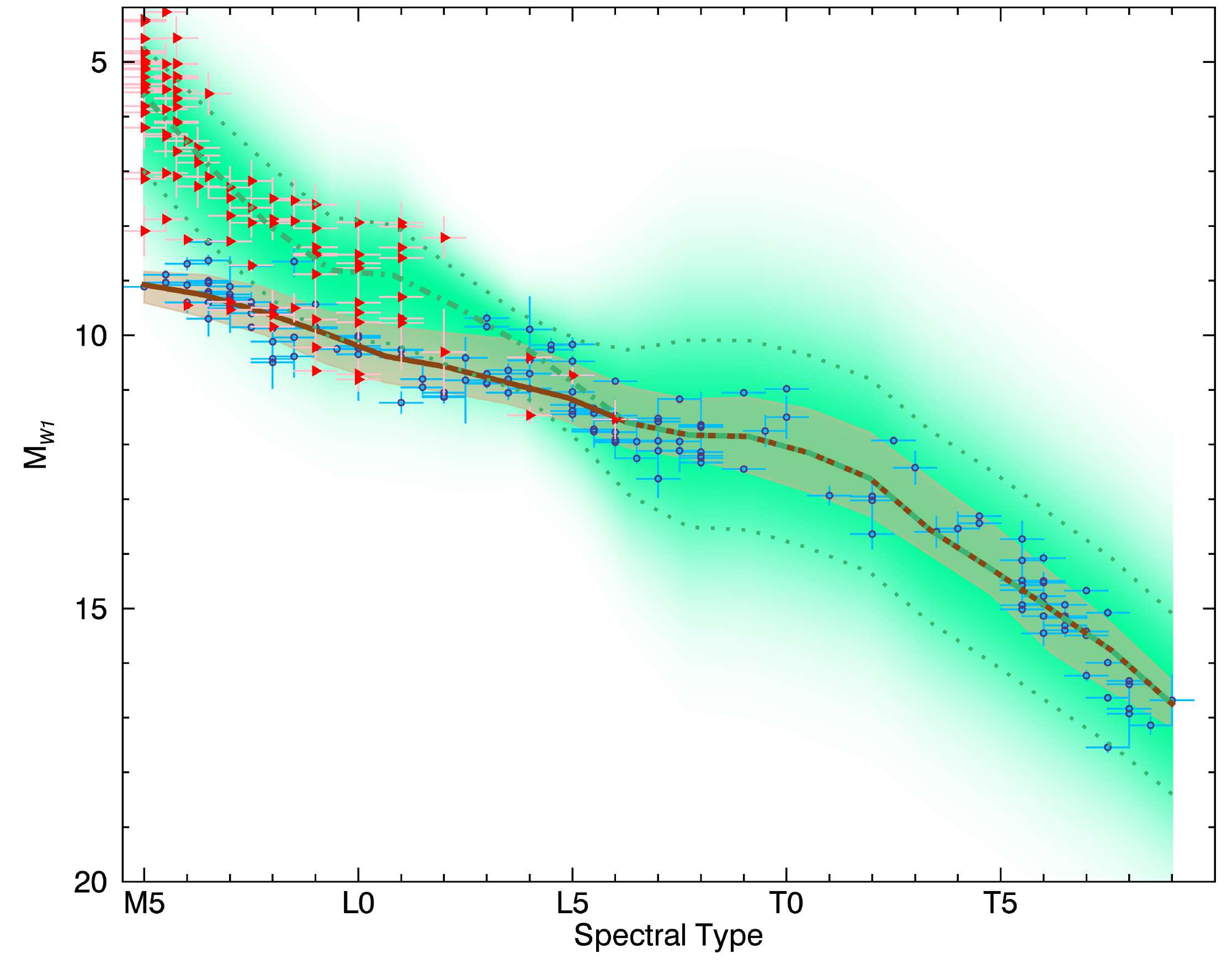}
	\end{center}
	\caption{Absolute Wise $W1$ magnitude as a function of spectral type for young (red dots) and old (blue dots) objects with parallax measurements. The old sequence is defined by the thick brown line and its 1$\sigma$ scatter represented by the shaded region. The young sequence (green dash-dotted line) was built from young objects only for spectral types $<$ L6. We have set it equal to the old sequence for later objects, but with a larger scatter (1.5 mag was added in quadrature to the field scatter), since the over- or under-luminosity of very late, young objects is not well known yet. The dotted green lines delimit the young sequence $\pm$1$\sigma$ scatter limits. The green region represents the young sequence PDF. Both sequences serve as spectroscopic distance calibrators in our Bayesian analysis.}
	\label{fig:spt_sequence}
\end{figure}

We have built an absolute magnitude \textendash\ spectral type sequence in a similar way (see Figure~\ref{fig:spt_sequence}). For young objects later than L6, no data with a parallax measurement is currently available. Hence, in this domain we have set the young sequence equal to the old one with a larger scatter to account for the fact we do not know well how those objects behave. Thus, any young candidate with spectral type later than $\sim$ L6 unveiled from our analysis should be taken with caution. These three sequences serve as photometric models in the Bayesian inference method described in the last section. More precisely, the absolute $W1$ magnitude is computed at each distance on the grid (which is described in Section~\ref{sec:bayes}) and then, for this value of $W1$, we draw expected $J$ - $K_s$ and $H$ - $W2$ colors from the magnitude \textendash spectral type sequence, and compare them to the actual measurements. Thus, $J$ - $K_s$, $H$ - $W2$ and the spectral type are included in the set of observables $\left\{Q_j\right\}$. Including such photometric models has the effect of providing a spectrophotometric distance calibration, as well as increasing the probability that very red objects belong to moving groups or the young field hypothesis (in cases where youth is not well established prior to the Bayesian inference).

\subsection{Definition of NYAs' Bona Fide Members}\label{sec:bonafide}

In order to define a robust subset of bona fide members to NYAs from which we will build their SKMs, we have started with a sample containing only objects with (1) signs of youth that are consistent with the age of the NYA they belong to, (2) a radial velocity measurement with an error $<$ 5 \kms, (3) a parallax measurement with an error $<$ 7 pc and (4) a proper motion measurement with a significance higher than 5$\sigma$. This first set of filters has removed 7 members that are considered as bona fide members to NYAs in \cite{2013ApJ...762...88M}, namely : \emph{HIP~22738} and \emph{WX~Col~A} from the ABDMG~;~\emph{2MASS~J06085283--2753583} from the $\beta$PMG~; \emph{HIP~46063} from CAR~; \emph{TWA~19~A} from the TWA~; \emph{HIP~1910~AB}, \emph{HIP~3556} and \emph{HIP~104308} from the THA. Here we consider multiple objects as only one system, so that we do not artificially double the weight for their position or kinematics. We then build a SKM model from the resulting list and compute the \emph{XYZUVW} standard deviation of each object with respect to its SKM model, and reject those with a standard deviation greater than 4. We repeat these steps independently for each NYA until no further objects are removed. This has removed 9 additional objects from our subset~: \emph{HD~178085} from ABDMG~; \emph{HIP~50156} and \emph{HIP~95261~A} from $\beta$PMG~; and \emph{HIP~17782}, \emph{HIP~24947}, \emph{GJ~490}, \emph{HIP~83494}, \emph{HIP~84642} and \emph{HIP~105404} from THA. We do not want to state that those rejected objects are not members. Instead, we consider that either we need more precise measurements or that they are possibly kinematic outliers, even if they were members. By rejecting such objects, we will get SKM models that have smaller dispersions and we will reduce the number of false-positives, with the price of possibly missing some new outlier members. We have also removed \emph{$\kappa$ And} from the COL bona fide members, since new estimates for this system's age are inconsistent with that of COL \citep{2013ApJ...779..153H}. We have added 16 new bona fide members not present in the list of \cite{2013ApJ...762...88M} either from the objects that they propose as new bona fide members, or from new members identified in \cite{2013ApJ...762..118W} and \cite{2012ApJ...758...56S} : \emph{G~269--153~A}, \emph{HIP~107948}, \emph{CD-35~2722} and \emph{BD+20~1920} in ABDMG~: \emph{2MASS~J03350208+2342356},  \emph{2MASSJ01112542+1526214}, \emph{HIP~23418~ABCD} and \emph{GJ~3331} in $\beta$PMG~: \emph{TWA~28}, \emph{TWA~2~A}, \emph{TWA~12}, \emph{TWA~13~A}, \emph{TWA~5~A}, \emph{TWA~23}, \emph{TWA~25} and \emph{TWA~20} in TWA. We have verified that all of these objects fall within $4\sigma$ of the SKM of their corresponding NYA. The membership of \emph{TWA~9} system has recently been subject to discussion~: \cite{2013ApJ...762..118W} indicated that its space motion does not agree with other TWA members in a traceback analysis. Another problem concerning this system is its discrepant age (63~Myr for TWA~9~A, 150~Myr for TWA~9~B) from BCAH98 models fitting, reported by \cite{1999ApJ...512L..63W}.  More recently, \cite{2013ApJS..208....9P} proposed that the \emph{Hipparcos} distance of this object might be off by at least $3\sigma$, which would explain both its kinematic and photometric (and thus age estimate) discrepancies. They also suggested that \emph{TWA~9} should still be considered as a bona fide member to TWA. Because of these uncertainties, we chose not to include this object in our construction of the SKM model of TWA to be more conservative. The final SKM obtained through this procedure are the ones used for all further analyses in this paper; their properties are given in Table \ref{tab:NYA_coord}.

\subsection{A Summary of Differences in This Modified Analysis}\label{sec:differences}

We briefly summarize here the differences between the analysis presented here and that of \cite{2013ApJ...762...88M}.

\begin{itemize}
\item We use $W1$ versus $H - W2$ and $W1$ versus $J - K_s$ CMDs instead of $I_C$ versus $I_C - J$, which allows to apply the method to objects later than M5.
\item When available, we use the spectral types in the input observables.
\item We consider the fact that the positions and kinematics of NYAs might be spread as ellipsoids whose major axes are not aligned with axes of the Galactic position reference frame (see Section~\ref{sec:skm}).
\item We include the error on measurements that feed the Bayesian analysis.
\item We have slightly modified the list of bona fide members to define a more robust and conservative list of core members (see Section~\ref{sec:bonafide}).
\item We estimate prior probabilities with the ratio of expected number of objects in each hypotheses, instead of setting them all to unity. Because of this, Bayesian probabilities associated to NYA hypotheses in this work are dramatically lower compared with those reported in \cite{2013ApJ...762...88M}.
\item We use a Besan\c{c}on Galactic model \citep{2012A&A...538A.106R} to build the young and old field hypotheses.
\item We consider a "young field" hypothesis consisting of $<$~1~Gyr field objects from the Besan\c{c}on Galactic model.
\item The Bayesian analysis directly compares $X^\prime Y^\prime Z^\prime U^{\prime} V^{\prime} W^{\prime}$ instead of the proper motions, the former being better represented by normal distributions. A consequence is the need to marginalize radial velocity and distance, which in turn necessitates the use of prior distributions displayed in Figure~\ref{fig:priors}.
\end{itemize}

\section{CONTAMINATION RATES}\label{sec:contam}

\begin{figure*}
	\begin{center}
		\subfigure[No radial velocity ($\nu$) or distance ($\varpi$)]{\label{fig:recova}
		 \includegraphics[width=0.45\textwidth]{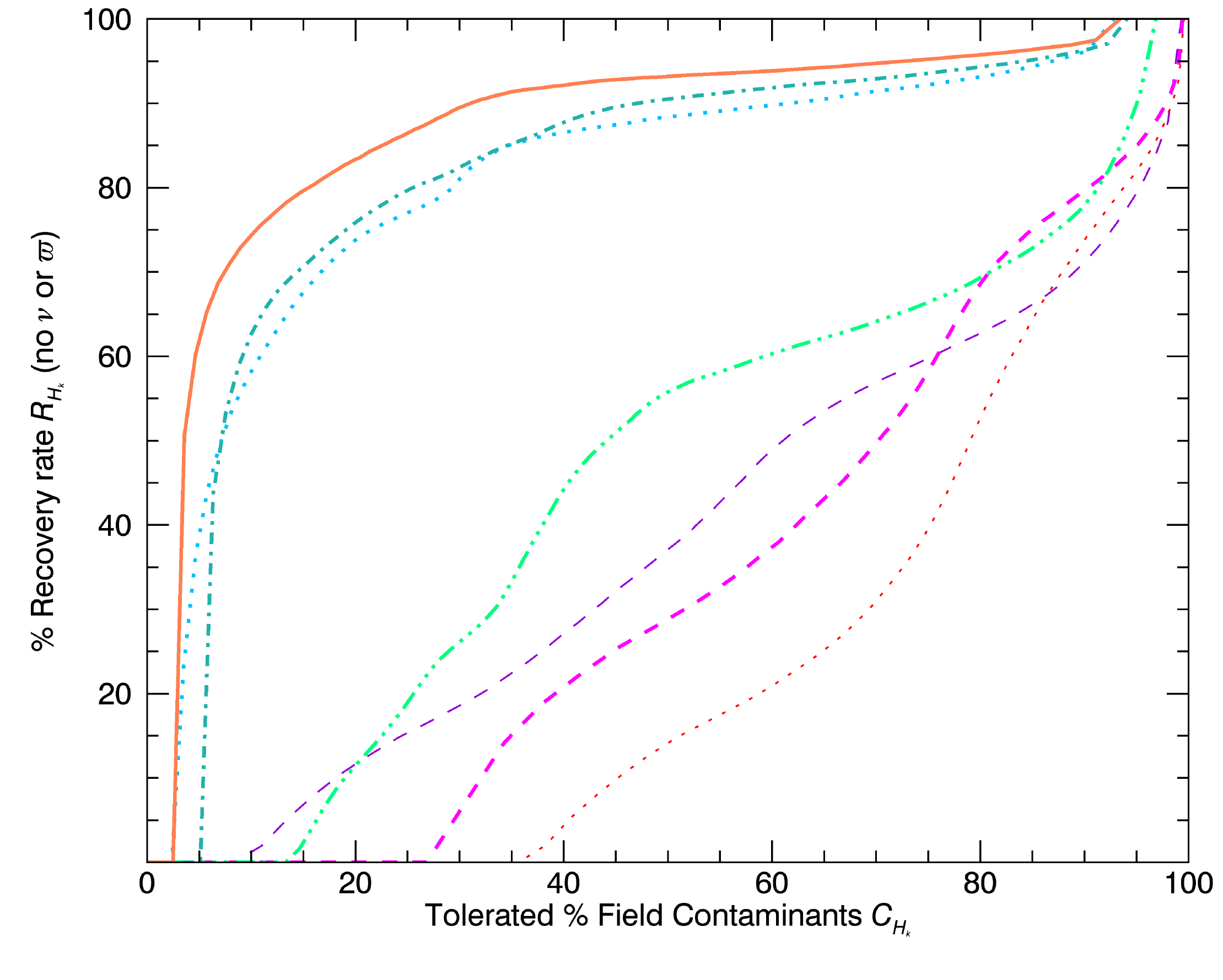}
		}
		\subfigure[Radial velocity ($\nu$) only]{\label{fig:recovb}
		 \includegraphics[width=0.45\textwidth]{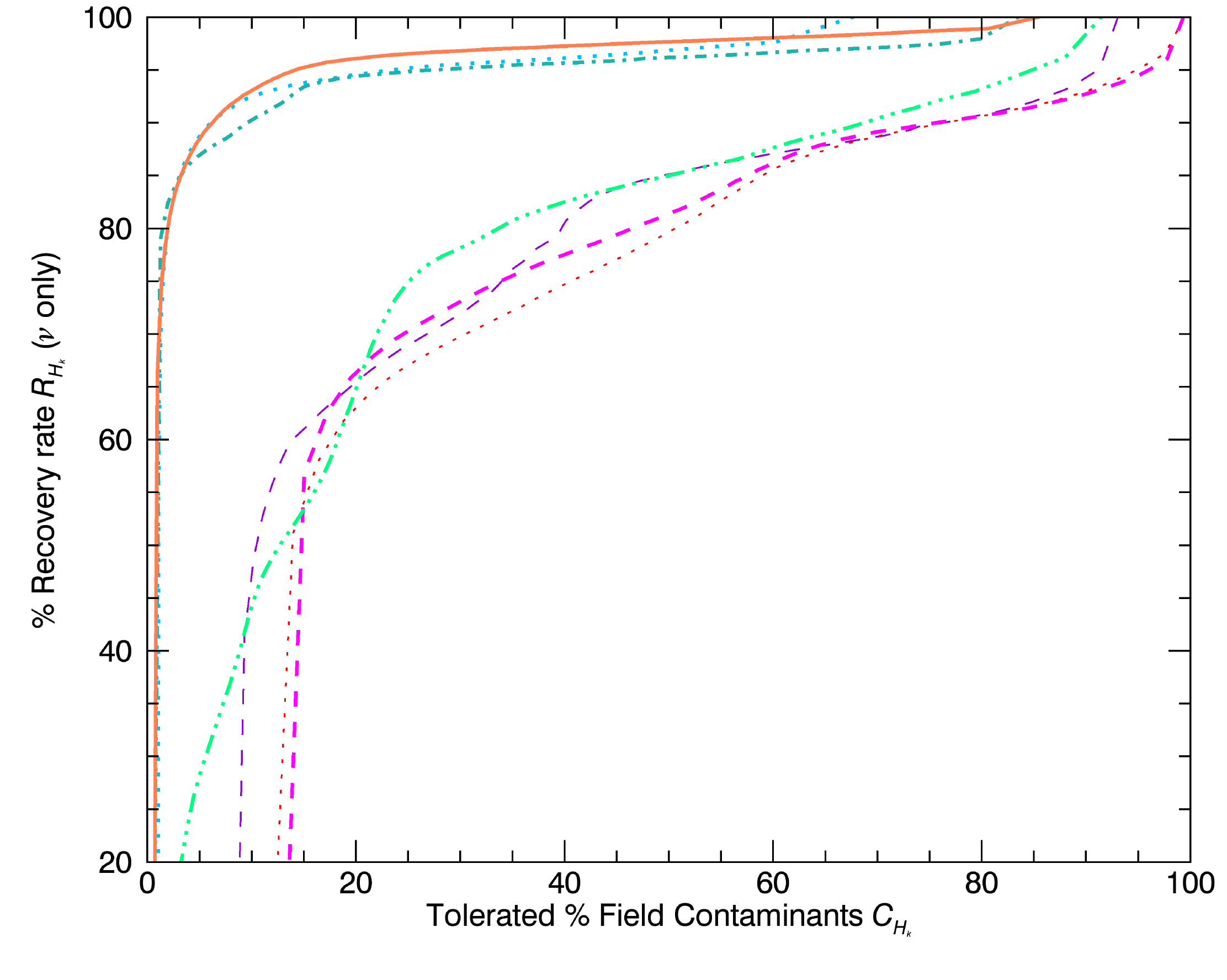}
		}
		\subfigure[Distance ($\varpi$) only]{\label{fig:recovc}
		 \includegraphics[width=0.45\textwidth]{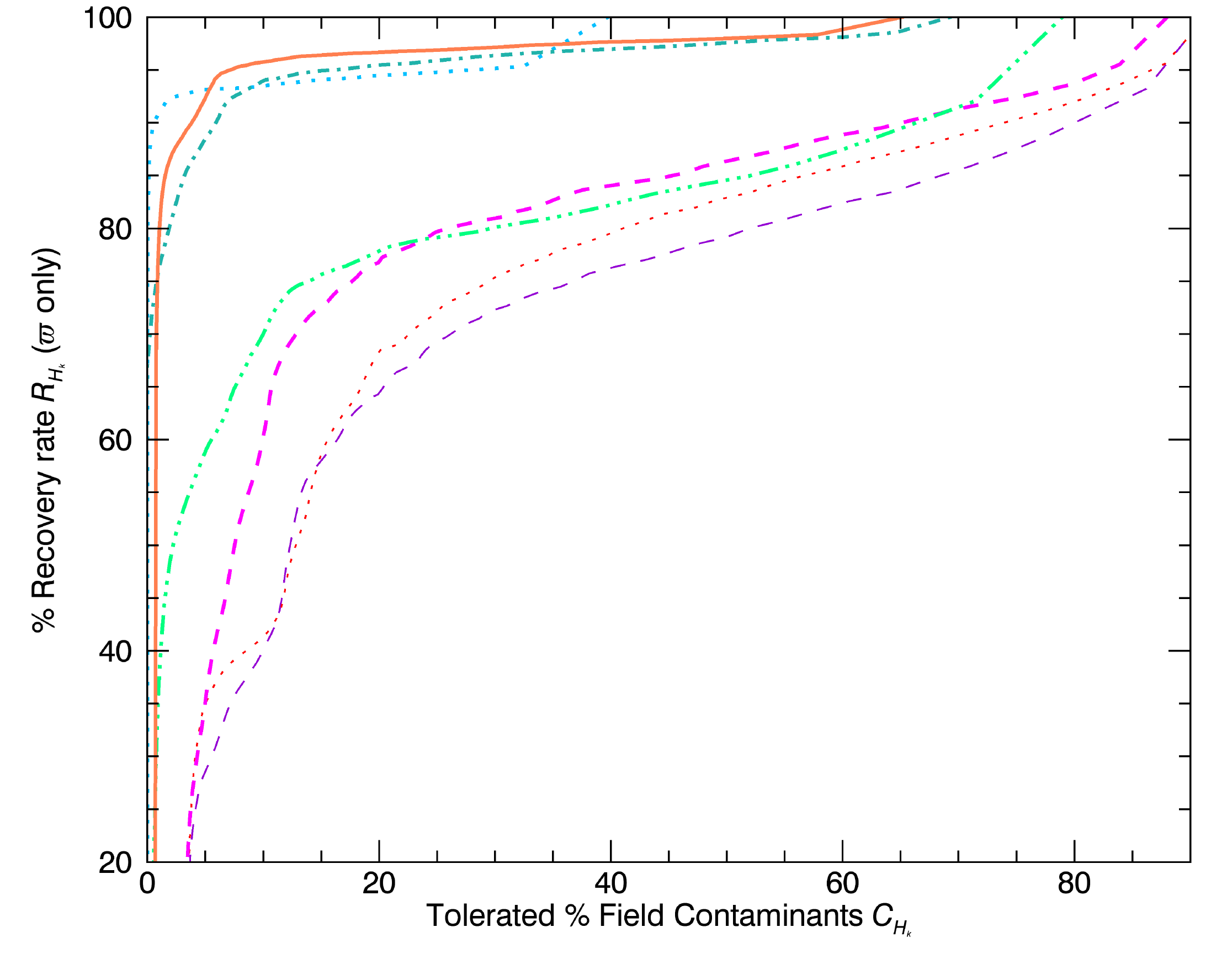}
		}
		\subfigure[Radial velocity ($\nu$) and distance ($\varpi$)]{\label{fig:recovd}
		 \includegraphics[width=0.45\textwidth]{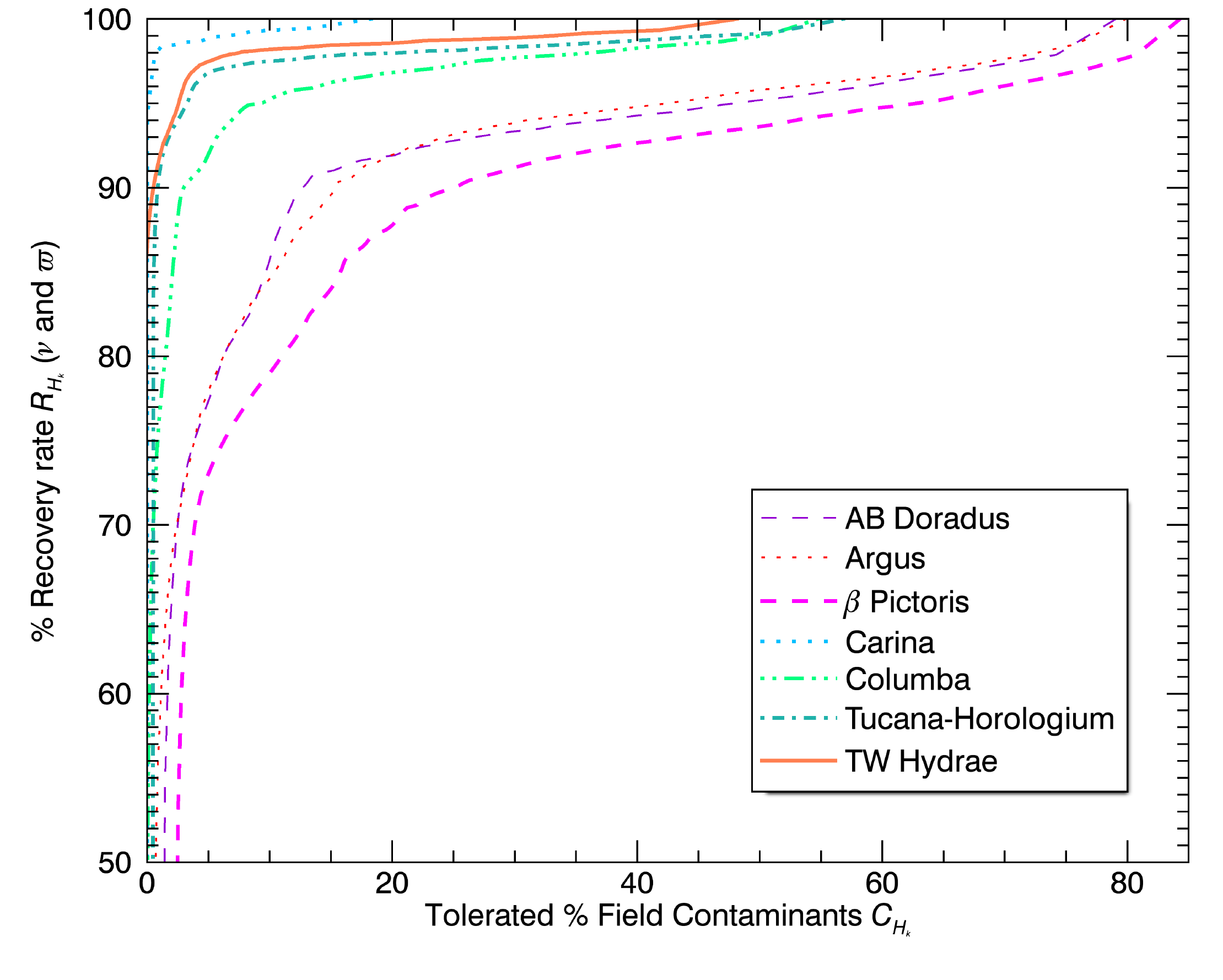}
		}
	\end{center}
	\caption{Recovery rates in different NYAs, as a function of the tolerated field contamination $C_{H_k}$. A fraction of objects $R_H(P_{\mathrm{low}})$ originating from hypothesis $H$ will be recovered by our method with a Bayesian probability $P_{H_k} > P_{\mathrm{low}}$ allowing in a fraction $C_{H_k}$ of field contaminants. The members of the closest NYAs such as $\beta$PMG, ARG and ABDMG are harder to recover without prior knowledge of radial velocity or distance, because their prior PDFs for radial velocity resemble that of the field (see Figure~\ref{fig:priors}).}
	\label{fig:recov}
\end{figure*}

\begin{figure*}
	\begin{center}
		\subfigure[No radial velocity ($\nu$) or distance ($\varpi$), part 1]{\label{fig:cont_betaa}
		 \includegraphics[width=0.45\textwidth]{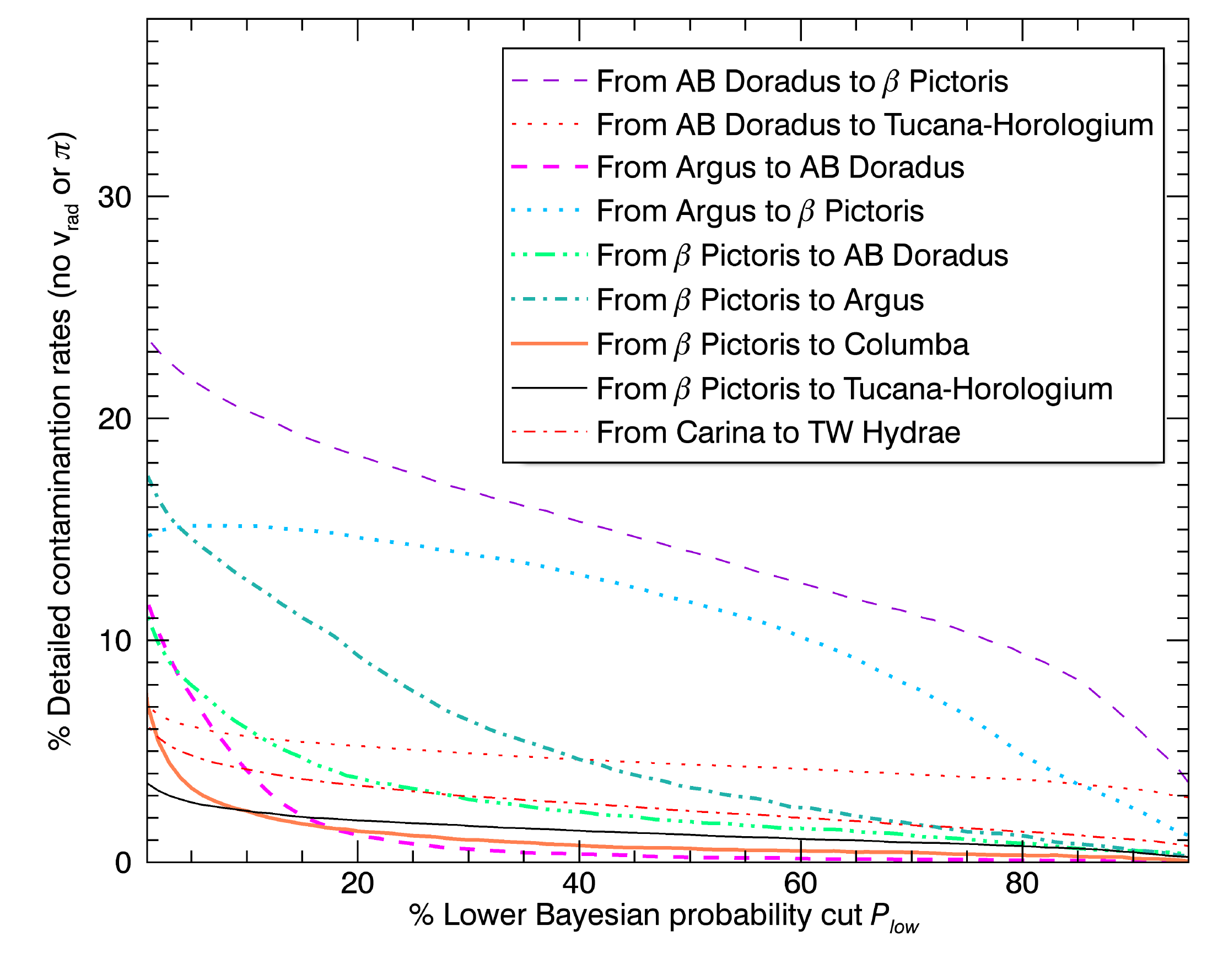}
		}
		\subfigure[No radial velocity ($\nu$) or distance ($\varpi$), part 2]{\label{fig:cont_betab}
		 \includegraphics[width=0.45\textwidth]{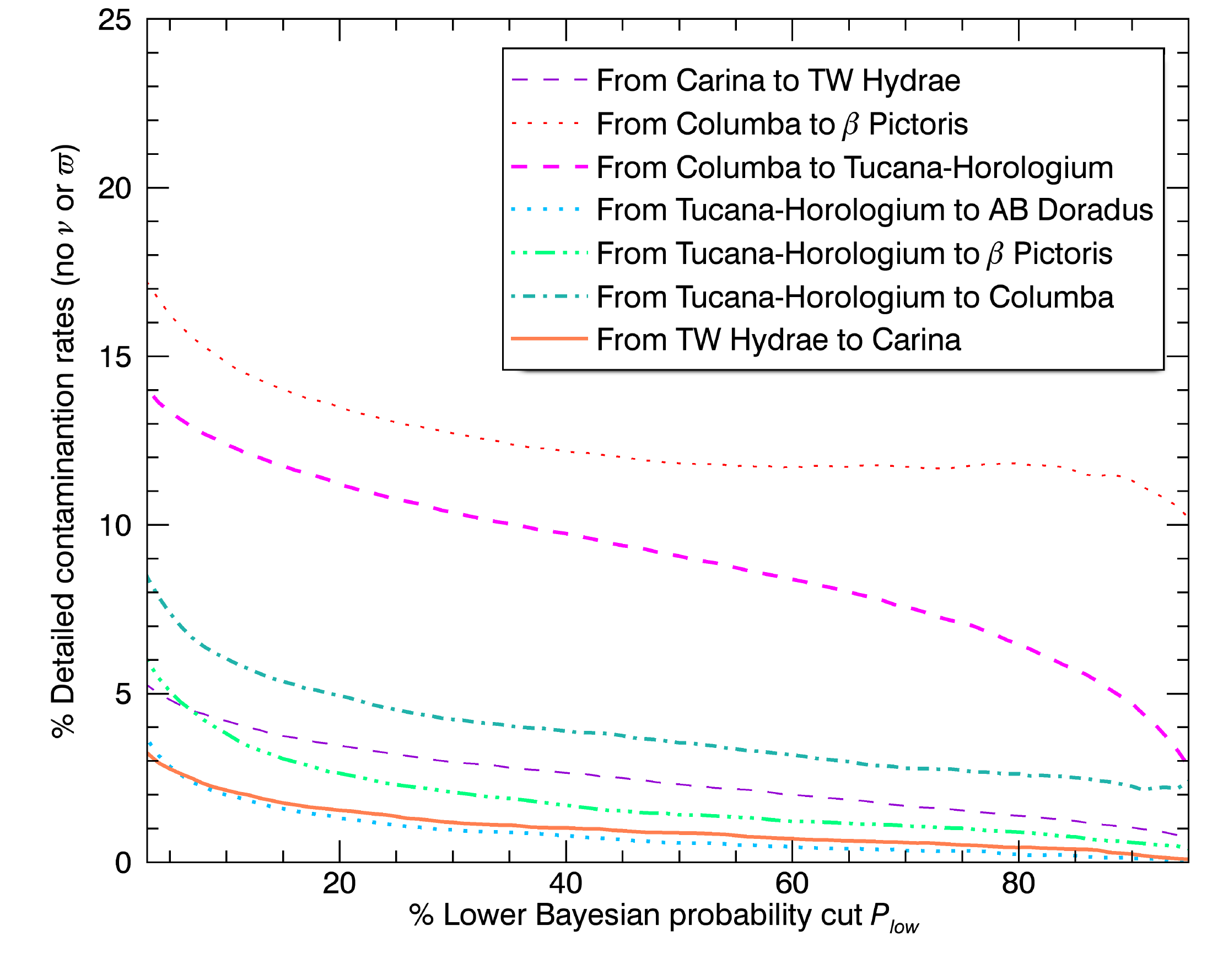}
		}
		\subfigure[Radial velocity ($\nu$) only]{\label{fig:cont_betac}
		 \includegraphics[width=0.45\textwidth]{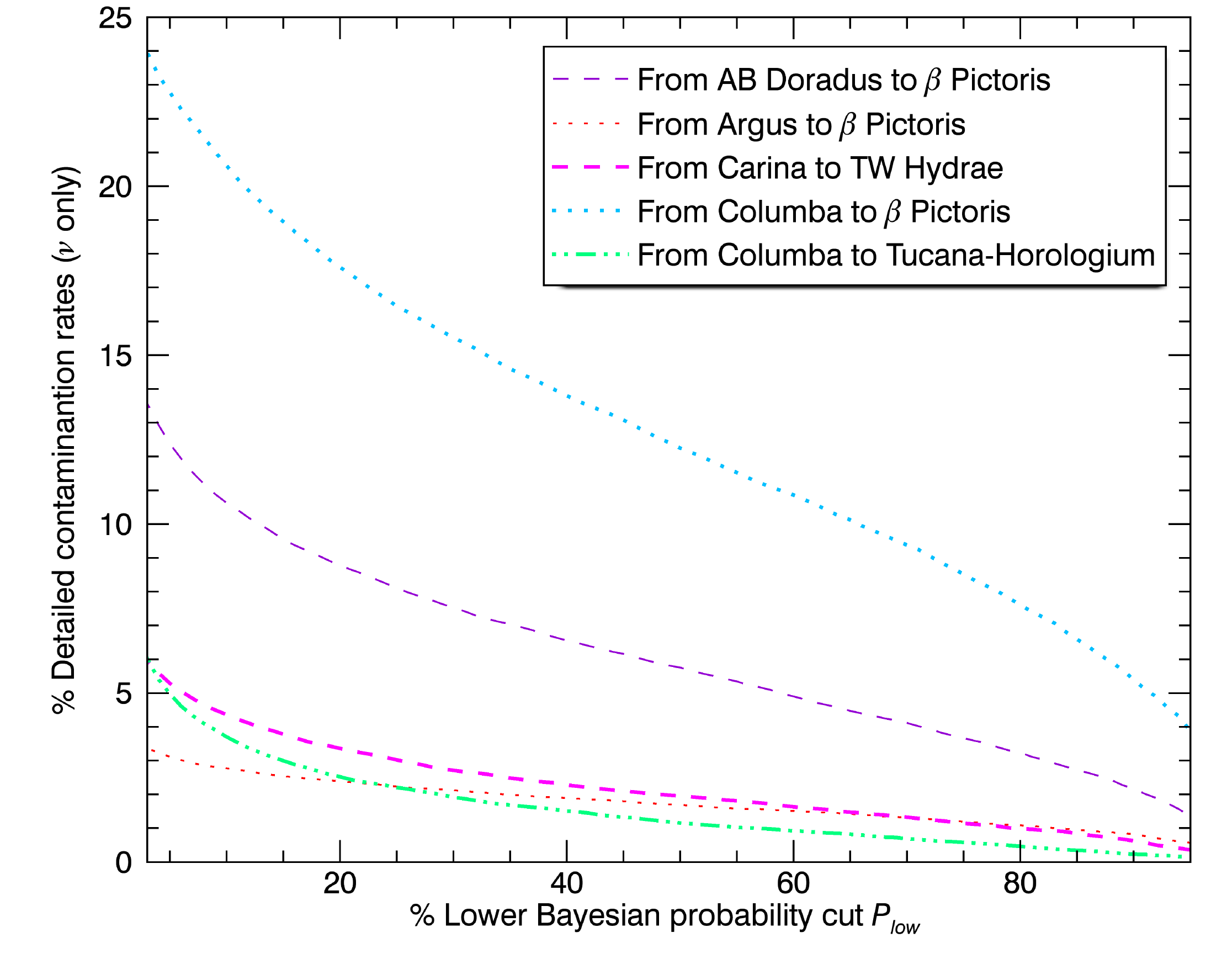}
		}
		\subfigure[Distance ($\varpi$) only]{\label{fig:cont_betad}
		 \includegraphics[width=0.45\textwidth]{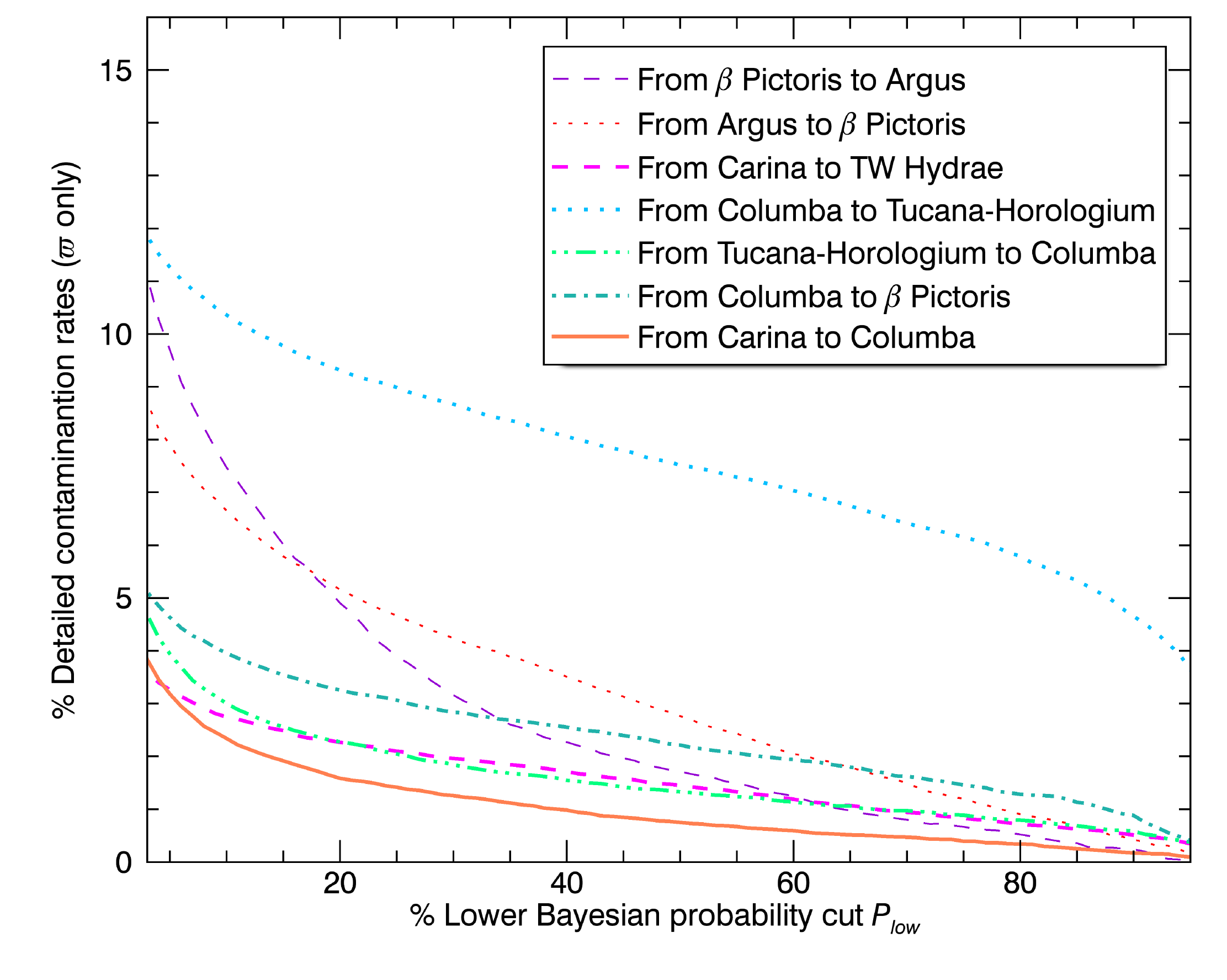}
		}
	\end{center}
	\caption{Cross contamination rates for NYAs considered in this work. Each curve represents a combination of contaminant to contaminated NYA. We only show the detailed contamination rates which have at least 3\% for a Bayesian probability $P_{H_k}$ =  5\%.}\label{fig:cont_beta}
\end{figure*}

As mentioned earlier, the fact that we use dependent observables in a naive Bayesian algorithm means that the Bayesian probabilities derived in this work are subject to be biased. To verify this, we have performed a Monte Carlo simulation, in which we draw 50~000 random synthetic objects from every SKM model described in Table~\ref{tab:NYA_coord} and use their synthetic characteristics to compute Bayesian probabilities in the same way than described earlier. Since we know from which SKM these synthetic objects were drawn in the first place, we can use this to assess the performance of our Bayesian analysis. We have assumed an IMF described in Section \ref{sec:prior} to assign masses to these synthetic objects, and in turn converted them to $M_{W1}$ magnitudes using the AMES-Cond isochrones \citep{2003A&A...402..701B} in combination with CIFIST2011 BT-SETTL atmosphere models (\citealp{2013MSAIS..24..128A}, \citealp{2013A&A...556A..15R}). In doing so, we have assumed a uniform age distribution spanning the age range of the hypothesis from which the synthetic object was drawn. Using $M_{W1}$, we have then assigned synthetic spectral types and NIR colors by using the photometric models described in Section~\ref{sec:photometry}. We have only included the young field hypothesis (not the old one) in this Monte Carlo analysis. Hence, the contamination rates that we derive in this section (and that are shown in Figures \ref{fig:contam}--\ref{fig:cont_beta}) are to be compared only to objects that display signs of youth. We discuss the contamination rates of objects with no evidence of youth at the end of this section. We have completed this Monte Carlo analysis four times: (1) without using neither radial velocity nor distance in the Bayesian analysis, (2) by using radial velocity only, (3) by using distance only and (4) by using both radial velocity and distance. The contamination rates are obtained by choosing a lower limit $P_{\mathrm{low}}$ to the Bayesian probability, then counting the number of times $N_{H_k \rightarrow H_l}$ where a synthetic object originating from the SKM of hypothesis $H_k$ has $P_{H_l} > P_{\mathrm{low}}$. We then define the correspondent fraction of contaminants as :

\begin{figure*}
	\begin{center}
		\subfigure[No radial velocity ($\nu$) or distance ($\varpi$)]{\label{fig:contama}
		 \includegraphics[width=0.45\textwidth]{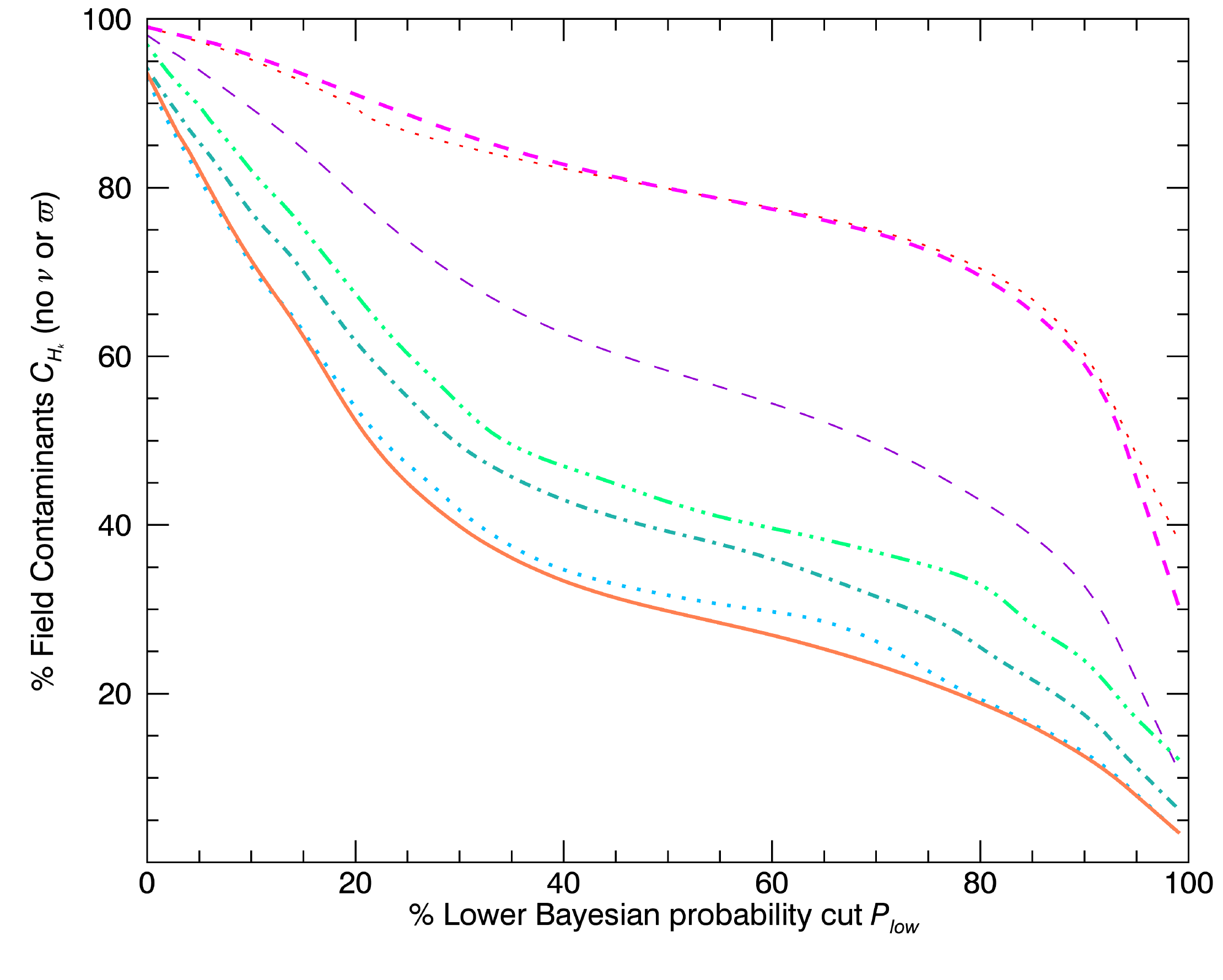}
		}
		\subfigure[Radial velocity ($\nu$) only]{\label{fig:contamb}
		 \includegraphics[width=0.45\textwidth]{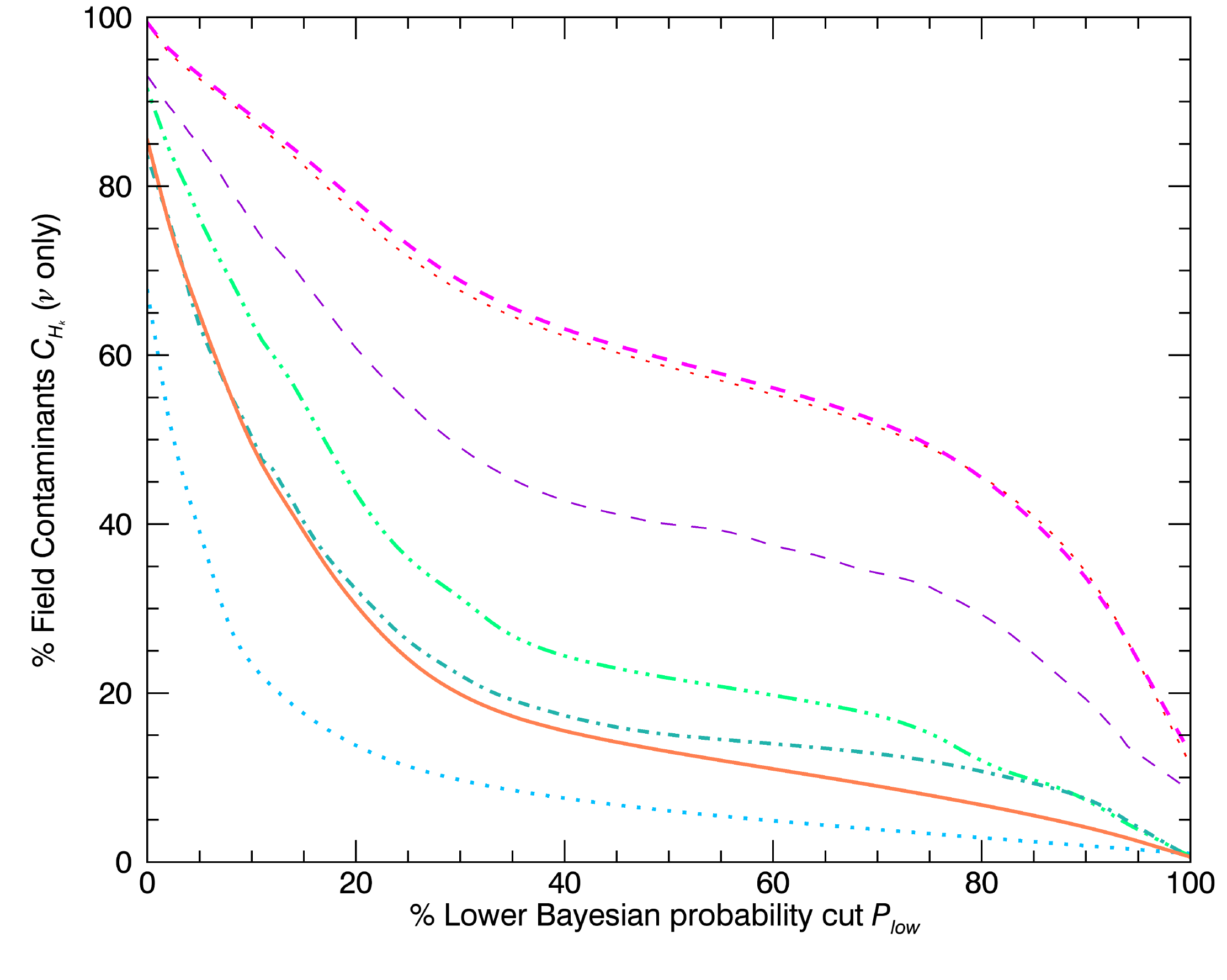}
		}
		\subfigure[Distance ($\varpi$) only]{\label{fig:contamc}
		 \includegraphics[width=0.45\textwidth]{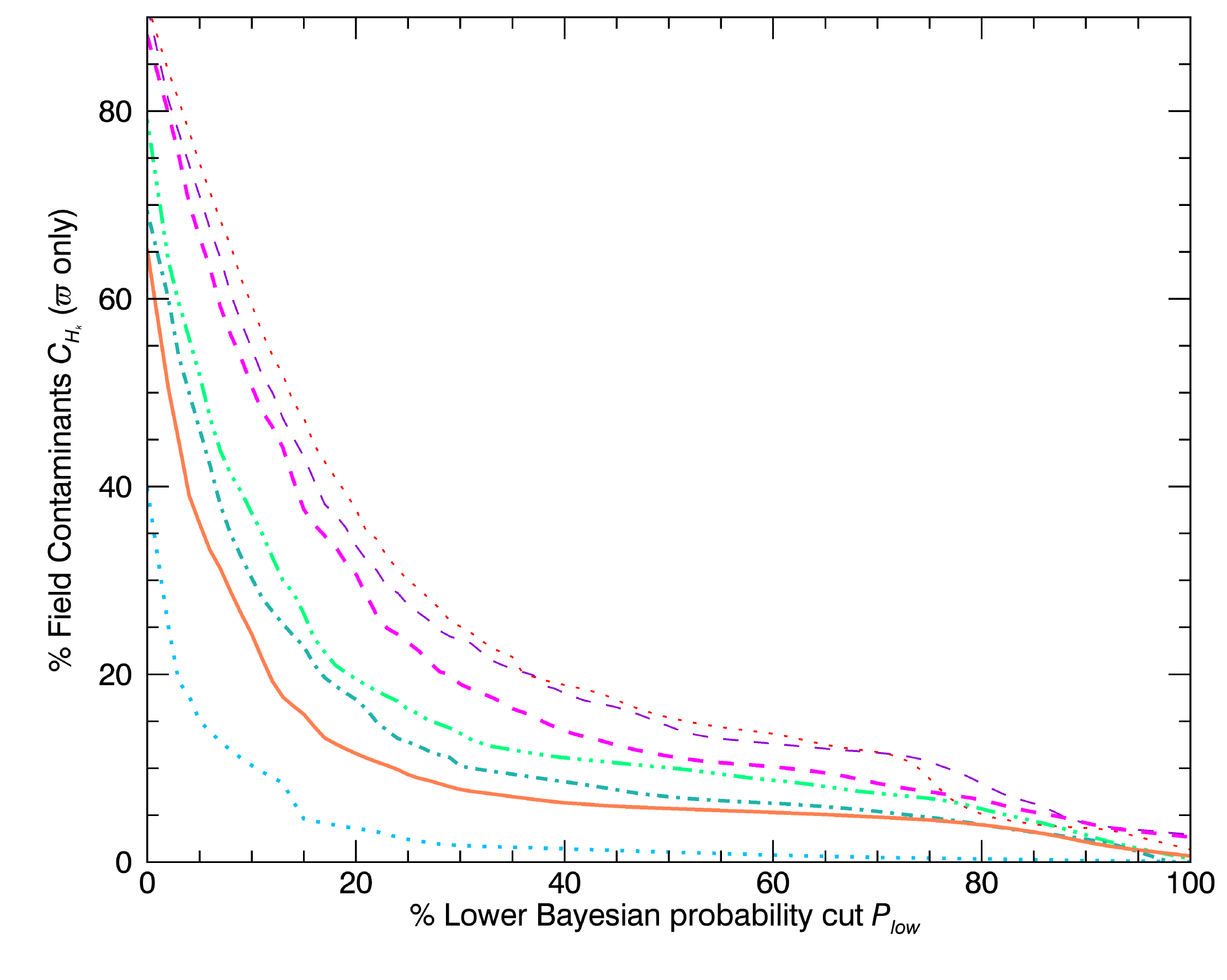}
		}
		\subfigure[Radial velocity ($\nu$) and distance ($\varpi$)]{\label{fig:contamd}
		 \includegraphics[width=0.45\textwidth]{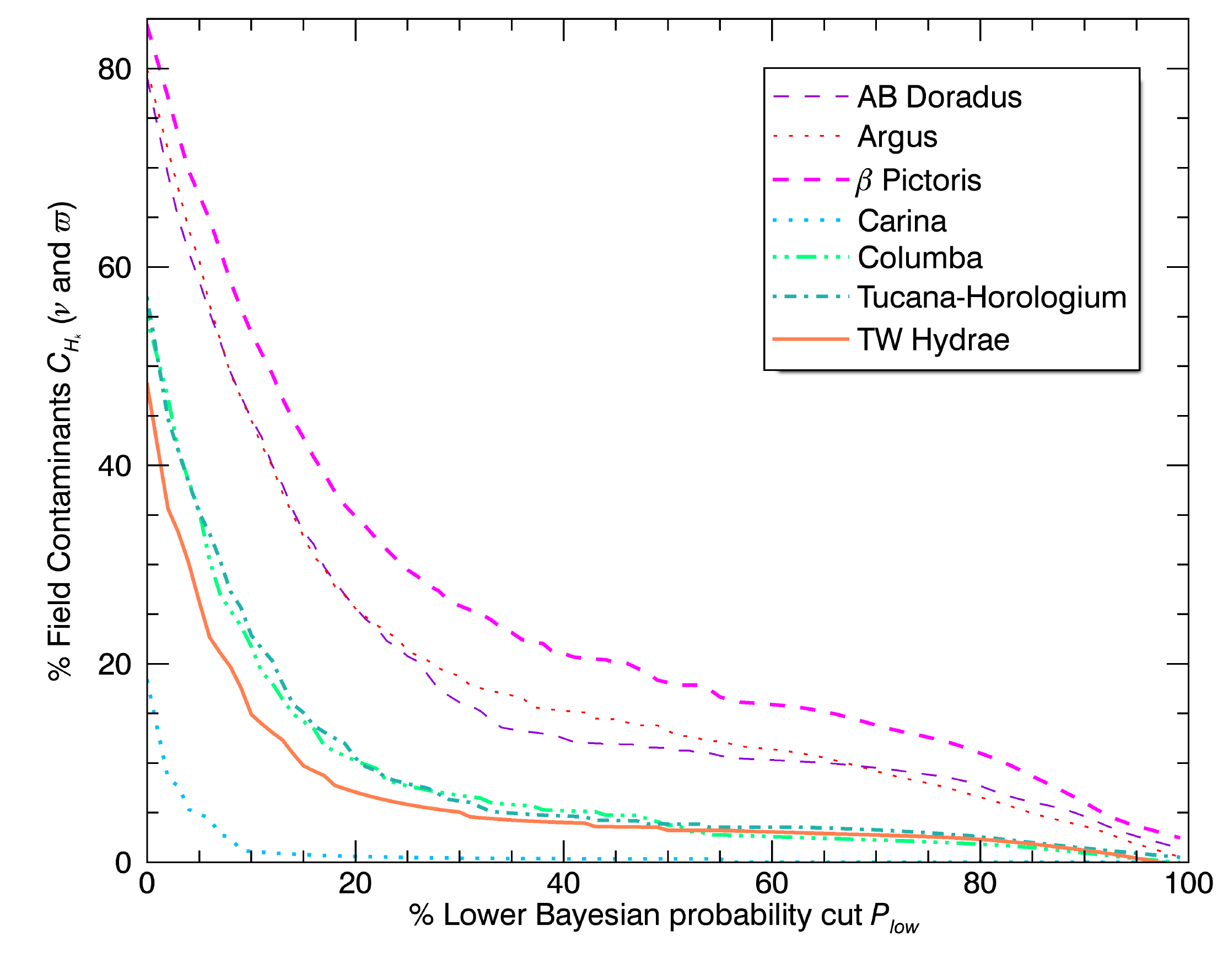}
		}
	\end{center}
	\caption{Field contamination rates in different NYAs, as a function of the chosen lower limit on Bayesian probability $P_{H_k}$. A fraction of $C_{H_k}\left(P_{\mathrm{low}}\right)$ of objects ending up in $H$ with $P_{H_k} > P_{\mathrm{low}}$ will be field contaminants. From upper-left to lower right, we show results by taking into account (1) no radial velocity and no parallax, (2) radial velocity only, (3) parallax only and (4) both radial velocity and parallax. In most cases, the field makes up for all contaminants. Exceptions where some NYAs contaminate other NYAs are shown in Figure~\ref{fig:cont_beta}.}
	\label{fig:contam}
\end{figure*}

\begin{equation}
	f_{H_l \rightarrow H_k}\left(P_{\mathrm{low}}\right) = \frac{N_{H_l \rightarrow H_k}\left(P_{\mathrm{low}}\right)}{N_{\mathrm{synth}}}.
\end{equation}

where $N_{\mathrm{synth}} = 50~000$ is the number of synthetic objects considered. We then rescale these synthetic populations according to the prior probabilities $P(H_l)$ described in Section~\ref{sec:prior}. In the cases where we do not have a distance or radial velocity measurement for a given object $\mathcal{O}$, we use statistical predictions yielded by our Bayesian analysis to adjust the population numbers $P(H_l)$ considered in this section. By doing this, we are counting how many synthetic objects drawn from every SKM could have properties alike those of a given object $\mathcal{O}$, for which we want to estimate the contamination probability. We thus expect that a total number $f_{H_l \rightarrow H_k}\left(P_{\mathrm{low}}\right)\cdot P(H_l)$ of objects drawn from the SKM of hypothesis $H_l$ will end up as contaminant candidates to hypothesis $H_k$ with $P_{H_k} > P_{\mathrm{low}}$. Consequently, there will be a fraction of contaminants $\mathcal{C}_{H_k}$ with similar properties to $\mathcal{O}$, which is a function of the low-cut Bayesian probability $P_{\mathrm{low}}$ :

\begin{equation}
	\mathcal{C}_{H_k}\left(P_{\mathrm{low}}\right) = \frac{\sum_l f_{H_l \rightarrow H_k} \cdot P(H_l)\ -\ f_{H_k \rightarrow H_k}\cdot P(H_k)}{\sum_l f_{H_l \rightarrow H_k}\cdot P(H_l)},
\end{equation}

The denominator corresponds to the total number of objects that end up as candidates to $H_k$ with $P_{H_k} > P_{\mathrm{low}}$, coming from all possible SKMs. The numerator is the same quantity, from which we subtract the number of objects that really originated from the SKM of $H_k$ in the first place. Hence, the numerator is equal to the number of objects from all associations \emph{other than} $H_k$ that ended up as contaminant candidates to $H_k$, i.e. the number of contaminants. \\

In Figure~\ref{fig:contam}, we present the fraction of young field contaminants without taking account of cross contamination between NYAs : 

\begin{equation}
	C_{H_k}\left(P_{\mathrm{low}}\right) = \frac{f_{yf \rightarrow H_k}\cdot P(H_{yf})}{f_{yf \rightarrow H_k}\cdot P(H_{yf}) + f_{H_k \rightarrow H_k} \cdot P(H_{k})},
\end{equation}

where index $l = yf$ refers to the \emph{young field}. Since the value for $P(H_k)$ is dependent on the object $\mathcal{O}$ for which we want to estimate the contamination rate (see Section~\ref{sec:prior}), we cannot capture all the information in only one such figure~; we would rather need such a figure for each object. We have thus chosen to display here a typical case by using values for $P(H_k)$ that vary smoothly and monotonically as a function of Bayesian probability in the same way that was observed in our sample, since object with a higher Bayesian probability of verifying a given $H_k$ generally have a higher prior $P(H_k)$. We can see that (1) the Bayesian probabilities derived in this work are generally biased, but comparable to the probability $(1-C_{H_k})$ that an object is not a field contaminant, (2) close-by NYAs such as ABDMG, $\beta$PMG and ARG, for which members are the most spread out in the whole sky, have a greater young field contamination rate, (3) adding a measurement of distance and radial velocity produces Bayesian probabilities that are even more biased towards the field and thus more conservative. This is particularly true whenever a distance measurement is used : then, even objects with very low (e.g. 30\%) Bayesian probabilities are unlikely ($<$~30\%) to be young field contaminants. It is interesting to note that the general shape of contamination rates indicate that Bayesian probabilities in the cases where $P_{H_k} > 50$\% tend to be overestimated whereas those with $P_{H_k} < 50$\% tend to be underestimated, with an apparent lack of objects having Bayesian probabilities around 50\%. This is precisely the expected behavior of a naive Bayesian classifier receiving dependent input variables (\citealp{Hand:2001tr}, \citealp{Russek:1983ed}). For a given hypothesis, there is always a maximum value for the Bayesian probability, which is close to but not exactly $P_{H_k} = 100\%$. The reason for this is that even if we consider an object whose \emph{XYZUVW} would lie exactly at the center of the SKM of a given NYA, there would be associated small, but non-zero Bayesian probabilities for every other hypothesis. Since the sum of all probabilities must be normalized to unity, no object will ever have exactly $P_{H_k} = 100\%$ for a particular hypothesis $H_k$, with the effect that the curves show large random excursions at $P_{H_k} > 95\%$. We have found that this generally happens around $P_{H_k} = 95\%$ for most NYAs. For this reason, even if we have used a very large number of synthetic objects in our Monte Carlo simulation, small number statistics inevitably occur at these very high Bayesian probabilities. We have thus corrected the contamination curves in this regime with polynomial fitting to avoid effects of the small number statistics. We remind that the results in Figure~\ref{fig:recov} rely on the assumption that objects under study display signs of youth. We expect to overestimate the contamination rates for objects with ages significantly lower than this, because there will be less field contaminants at lower ages. We chose not to include this consideration in the prior probabilities because one cannot efficiently constrain the age of a low-mass object based only on signs of low-gravity. \\

In Figure~\ref{fig:recov}, we present the recovery rate $R_{H_k} = f_{H_k \rightarrow H_k}\left(C_{\mathrm{low}}\right)$, the fraction of synthetic objects drawn from $H_k$ ending up as candidates to $H_k$. Hence, $R_{H_k}$ represents the expected fraction of true NYA members that will be recovered with the Bayesian method described here, depending on how many contaminants we allow in our output candidates sample. It can be seen that adding radial velocity or parallax measurements significantly increase the recovery rate. Furthermore, we can see that in absence of radial velocity and parallax measurements, our method will yield relatively small recovery rates for COL, ABDMG, $\beta$PMG and ARG unless we consider candidates with relatively high field contamination rates (by considering objects with low Bayesian probabilities). It should also be considered that lower-mass members to NYAs could be spread further than the bona fide members considered in building our SKM models. If this is the case, then the recovery rates presented here will be underestimated, since our SKMs will not be a fair representation of reality.\\

In Figure~\ref{fig:cont_beta}, we show the cross-contamination rates $\mathcal{C}_{H_l \rightarrow H_k}\left(P_{\mathrm{low}}\right)$ between NYAs :

\begin{equation}
	\mathcal{C}_{H_l \rightarrow H_k}\left(P_{\mathrm{low}}\right) = \frac{f_{H_l \rightarrow H_k}\cdot P(H_l)}{\sum_l f_{H_l \rightarrow H_k} \cdot P(H_l)},
\end{equation}

where $l$ does not include the field, for every combination yielding a contamination fraction higher than 3\% when considering Bayesian probabilities $P_{H_k} > 5\%$. These contamination rates apply to objects which are not field contaminants, and hence are applicable regardless of their age. In the case where neither radial velocity nor parallax is known, there are 3 combinations where we expect the cross-contamination rates to be relatively high (larger than 15\% for small Bayesian probabilities) : from ABDMG to $\beta$PMG, from ARG to $\beta$PMG and from COL to $\beta$PMG. When only radial velocity is known, this only happens from COL to $\beta$PMG, whereas when only parallax is known, the cross-contamination rates drop below 20\% for every NYA combination at any Bayesian probability. If both radial velocity and parallax are known, the cross-contamination rates drop even more, to rates always lower than 3\%. \\
 
There is a subclass of red objects considered in this work for which we do not have any other signs of youth. For those objects, we have used a similar contamination analysis than described here, but consider both (young and old) field hypotheses. We have found that the contamination rates do not significantly differ from those given in Figure~\ref{fig:contam} for a given Bayesian probability, which means that our Bayesian probabilities are biased in the same way whether or not we include the old field hypothesis.
 
 \subsection{Statistical Predictions for Distance and Radial Velocity}\label{sec:stat_pred}

\begin{figure*}
	\begin{center}
		\subfigure{
		 \includegraphics[width=0.45\textwidth]{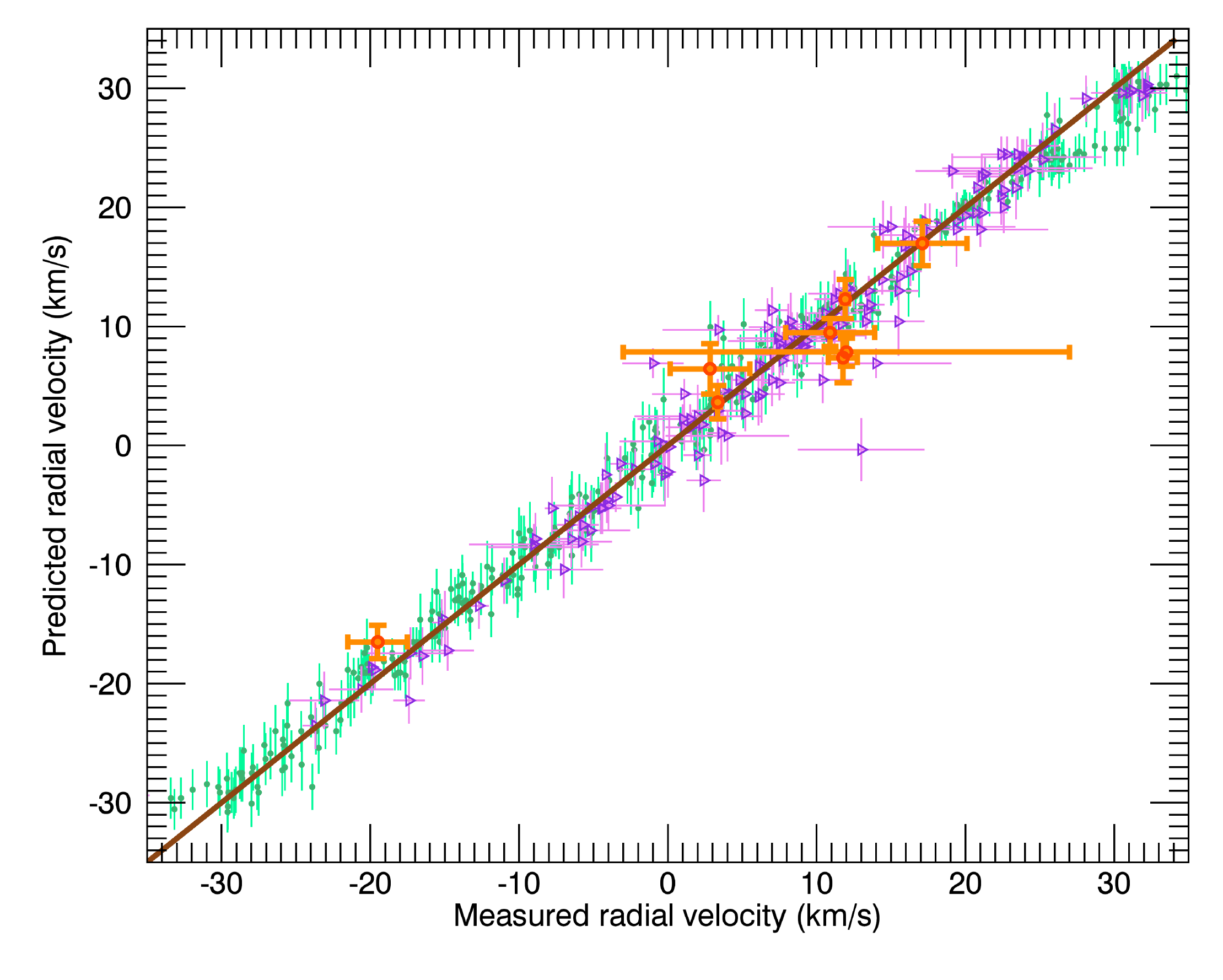}
		}
		\subfigure{
		 \includegraphics[width=0.45\textwidth]{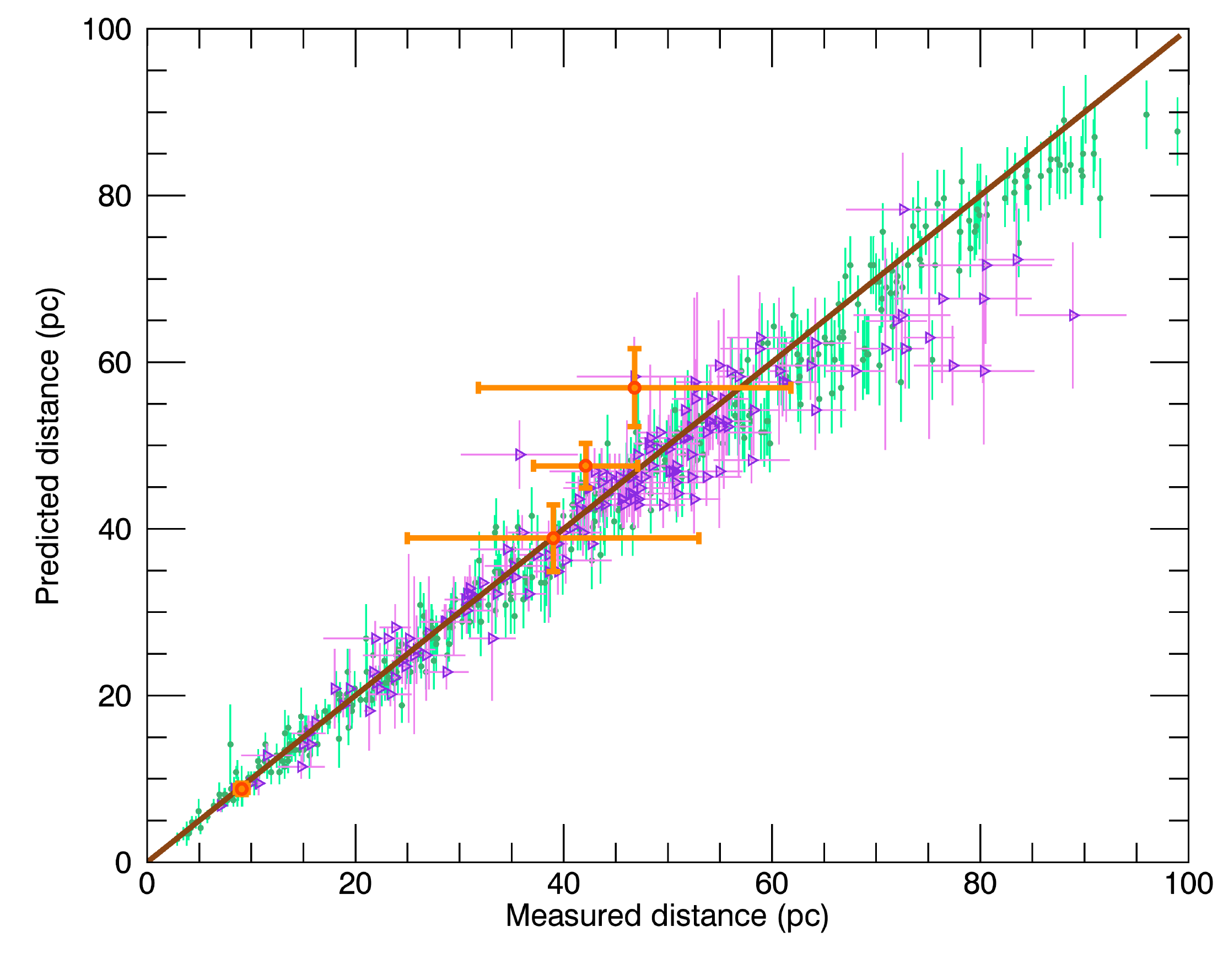}
		}
	\end{center}
	\caption{Performance of the statistical radial velocities and distances predictions for NYA candidates. Results from the Monte Carlo contamination analysis (small green dots), for existing bona fide members (purple open triangles) and candidates in our sample (orange, thick open circles) are displayed. For a better clarity, we only show 30 synthetic (green) data points per bins of 5 \kms\ or 5 pc. The reduced $\chi^2$ values of the blue dots (including those not displayed) are 1.1 and 1.6, respectively.}
	\label{fig:dstat}
\end{figure*}

We have used the Monte Carlo analysis described in the previous section to assess the performance of our Bayesian method in predicting the distance and radial velocity of a given object. To do this, we compare statistical distances and radial velocities to the actual values of input synthetic objects, in the case where we do not use radial or distance as input parameters in our Bayesian analysis. We have only included objects ending up as NYA candidates in this figure, since the predictions for field hypotheses are less precise, due to the intrinsic larger scatter in the likelihood PDFs of field objects. We show the results in Figure~\ref{fig:dstat}, as well as a similar analysis applied to known bona fide members of NYAs. In the latter case. We find that the agreement is generally very good between predictions and true values, with reduced $\chi^2$ values of 1.1 and 1.6 for the radial velocity and distance predictions, respectively. Our analysis can thus predict distances to precisions of 8.0\% and radial velocities to $1.6~\kms$. The higher $\chi^2$ value corresponding to distance predictions can be assigned to the fact that distance estimates tend to be slightly underestimated at large distances. A small fraction of bona fide members have outlier \emph{XYZUVW} parameters compared to the locus of their NYA, which is reflected in a larger scatter in their radial velocity and distance predictions, compared to synthetic objects. We also show that statistical predictions agree well with actual measurements for young objects in our sample.

\section{ANALYSIS OF PRESENT FAINT, BONA FIDE MEMBERS}\label{sec:bfide_ph}
 
\begin{figure*}
	\begin{center}
		\subfigure{\label{fig:bfidea}
		 \includegraphics[width=0.45\textwidth]{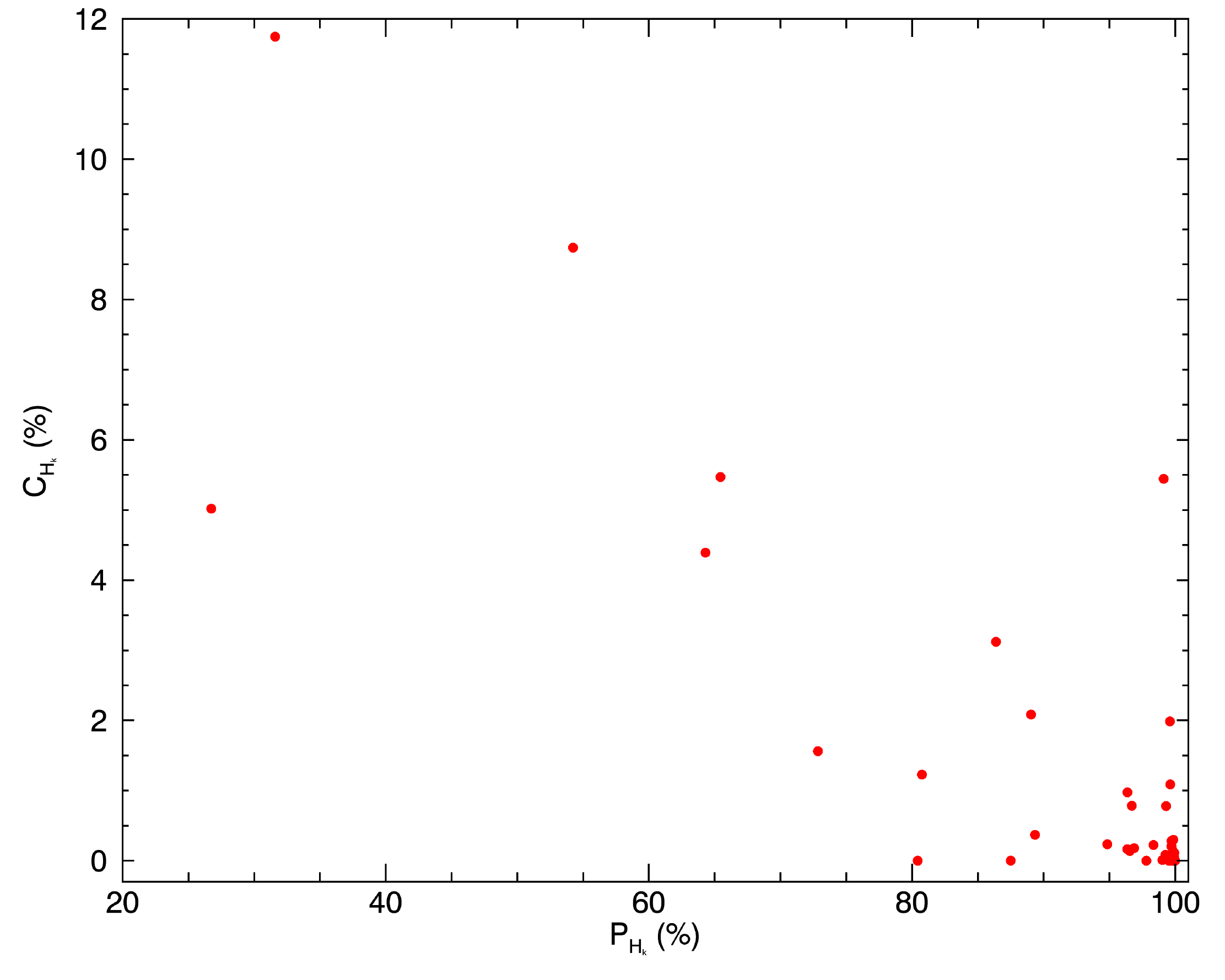}
		}
		\subfigure{\label{fig:bfideb}
		 \includegraphics[width=0.45\textwidth]{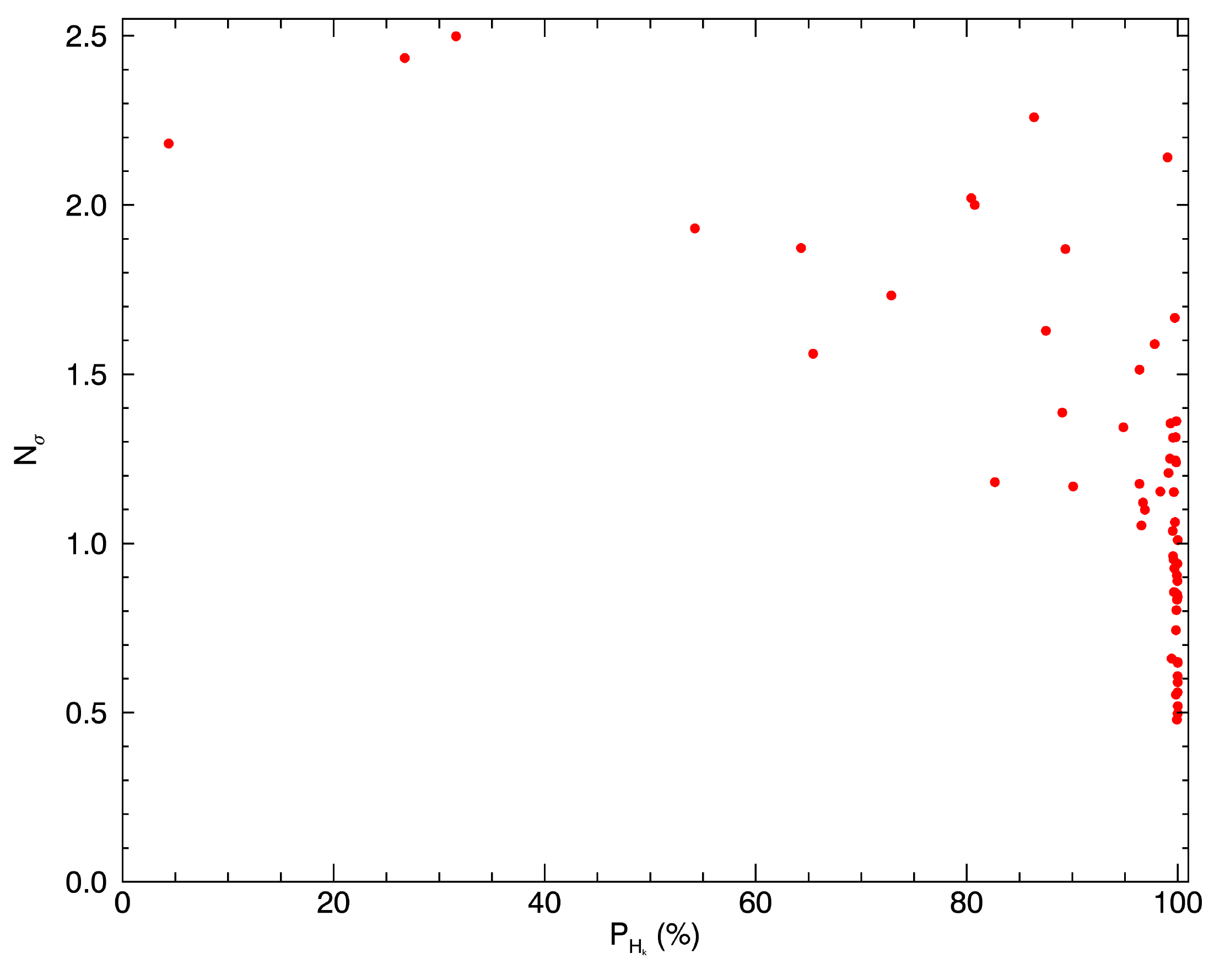}
		}
	\end{center}
	\caption{Resulting Bayesian probability $P_{H_k}$ and young field contamination rates $C_{H_k}$ for bona fide members with $M_{W1} > 3.0$ in the literature, analyzed with our modified Bayesian method (left). $N_\sigma$ distance from the center of the respective SKM of each object in this bona fide sample, as a function of the resulting $P_{H_k}$ (right). One can see that objects further from the center ($N_\sigma > 1.0$) generally have lower $P_{H_k}$, and that $C_{H_k}$ is anti-correlated with $P_{H_k}$, as expected. Bona fide members can have Bayesian probabilities as low as $P_{H_k} = 25\%$, but they generally have $C_{H_k} \lesssim 12\%$.}
	\label{fig:bfide}
\end{figure*}

 	We have applied our modified Bayesian analysis to all currently known bona fide members (see Section~\ref{sec:bonafide}) that have absolute $W1$ magnitudes higher than 3, so that we can use the photometric models described in Section~\ref{sec:photometry}. The young field contamination rates as a function of Bayesian probability is displayed in Figure~\ref{fig:bfidea} for each object in this sample. We can see that some outlier members presently considered as bona fide have Bayesian probabilities down to $P_{H_k} \sim 25\%$, but that they generally have low contamination rates $C_{H_k} \lesssim 12\%$, with the exception of 3 objects that we did not display~: \emph{2MASS~J17383964+6114160}, \emph{2MASS~J05365509--4757481} and \emph{2MASS~J05365685--4757528} have contamination rates of 78\%, 41\% and 37\%, respectively. All of them are 1.2 to 2.2 $\sigma$ away from the locus of their NYA. In Figure~\ref{fig:bfideb}, we display this $N_\sigma$ distance as a function of the Bayesian probability. We obtain $N_\sigma$ by propagating the error of the 6-dimensional distance of each object in the \emph{XYZUVW} parameter space, where we treat the width of each axis in the SKM as a measurement error over the central position of the SKM. We can see that objects with lower Bayesian probabilities are generally further from the center of the SKM. In particular, objects within 1$\sigma$ of the SKM center always have $P_{H_k} > 99\%$. Both $P_{H_k}$ and $C_{H_k}$ provide a quantitative framework for qualifying the membership of bona fide objects. Core members generally have high Bayesian probabilities $P_{H_k} \gtrsim 50\%$ and $C_{H_k}$ less than a few~\%, while peripheral ones are those characterized by lower $P_{H_k}$ (25~\textendash~50\%), yet with a modest contamination rate i.e., $C_{H_k} \lesssim 12\%$.

 
\section{RESULTS AND DISCUSSION}\label{sec:results}
	
	In Table~\ref{tab:mass}, we list all candidate members to NYAs from the input sample of young or red dwarfs described in Section~\ref{sec:youngl} that were recovered by our modified Bayesian analysis with a Bayesian probability $P_{H_k}$ corresponding to a field contamination rate $C_{H_k}$ lower than 90\%. We remind that Bayesian probabilities reported here cannot be directly compared to the values in \cite{2013ApJ...762...88M}, because of the different prior probabilities we have used. If we had set them to unity so that a comparison was possible, every object in the three sections of Table~\ref{tab:mass} would have $P_{H_k} \gtrsim 90\%$. We report even candidates with contamination rates as high as $C_{H_k} \sim$ 90\% to ensure high recovery rates (see Figure~\ref{fig:recov}). Only in the cases where objects display signs of youth, we have not included the \emph{old field} hypothesis in our Bayesian analysis. For all objects, we have used sky position, proper motion, NIR photometry, spectral types, radial velocity and trigonometric distance whenever they were available. There are a few objects for which a very low precision radial velocity is available \citep{2010ApJS..190..100K}, which we did not use because we have to assume that measurement errors are small in order to propagate them to errors on spatial velocities. We find a few core and peripheral bona fide members, 35 very strong candidate members for which $C_{H_k}$ is less than 15\%, 15 modest candidate members with $C_{H_k}$ between 15 and 70\%, and 6 low-probability candidate NYA members with $C_{H_k}$ between 70 and 90\%. For each of them, we give their NIR or optical spectral type, as well as the Bayesian probability, predicted radial velocity and distance associated with the NYA they most probably belong to. We use the $J$, $H$, $K_s$, $W1$ and $W2$ apparent magnitudes and statistical distances (or parallax measurements) for each object, along with the age of their most probable association, to determine their most probable mass using AMES-COND isochrones \citep{2003A&A...402..701B} in combination with CIFIST2011 BT-SETTL atmosphere models (\citealp{2013MSAIS..24..128A}, \citealp{2013A&A...556A..15R}) in a likelihood analysis. We thus report several \emph{planemo} candidates whose mass estimates lie entirely inside the planetary-mass regime, 9 of them being new, very strong candidates. In Figure~\ref{fig:density}, we show an example of the $P(\{O_i\},\nu,\varpi|H_k)i$ PDF for the ABDMG bona fide member \emph{2MASS~J03552337+1133437}. The very good agreement between measurements and predicted values for distance and radial velocity associated with the most probable hypothesis (ABDMG) illustrates the robustness of our analysis. Radial velocity and distance measurements were \emph{not} used as input parameters to generate this PDF. Similar figures for all objects in our sample are available at our group's website \url{www.astro.umontreal.ca/\textasciitilde gagne}. We give all the details on the output of our Bayesian analysis for each object in our sample in Tables~\ref{tab:results}~and~\ref{tab:dstat}.

\subsection{Comments on Individual Objects}\label{sec:indiv}

	In this section, we comment on the properties and previous knowledge of individual objects displayed in the first two sections of Table~\ref{tab:mass}. Those are objects that we identify as candidate members to NYAs, with a probability lower than 70\% of being field or young field contaminants. We also comment on objects for which our conclusions are different from those of other authors.\\

\subsubsection{Bona Fide Members}

\emph{2MASS~J01231125--6921379} (\emph{2MUCD~13056}) is a young M7.5 BD with Li absorption \cite{2009ApJ...705.1416R}. We find that it is a strong candidate to the THA with a predicted radial velocity of $9.9~\pm~2.5$~\kms\ and distance of $47.4~\pm~3.2$~pc. \cite{2009ApJ...705.1416R} measure a radial velocity $\nu~=~10.9~\pm~3$\kms\ and Riedel et al. (submitted to the ApJ) measures a trigonometric distance of $42.1~\pm~5$~pc, both agreeing well with our predictions, which means this object has $P_{H_k} > 99.9\%$ and $C_{H_k} < 0.1\%$ and an estimated mass of 56~\textendash~74~\Mjup. We have performed a likelihood analysis to constrain the age of this object by comparing its absolute NIR broadband photometry to BT-SETTL models. We find that the presence of Li absorption implies an age of $<$~80~Myr, which is consistent with the age of THA. We note that a mass of $<~65~$\Mjup,which would imply that this object does not burn Li at all, is only consistent with an age of $<$~50~Myr, and hence our present age constraint based on Li absorption remains valid. Since this object has everything needed to be considered as such, we propose it as a new 56~\textendash~74~\Mjup\ bona fide BD member to the THA, making it the latest-type current bona fide member to this association. \\

\begin{figure}
	\includegraphics[width=0.5\textwidth]{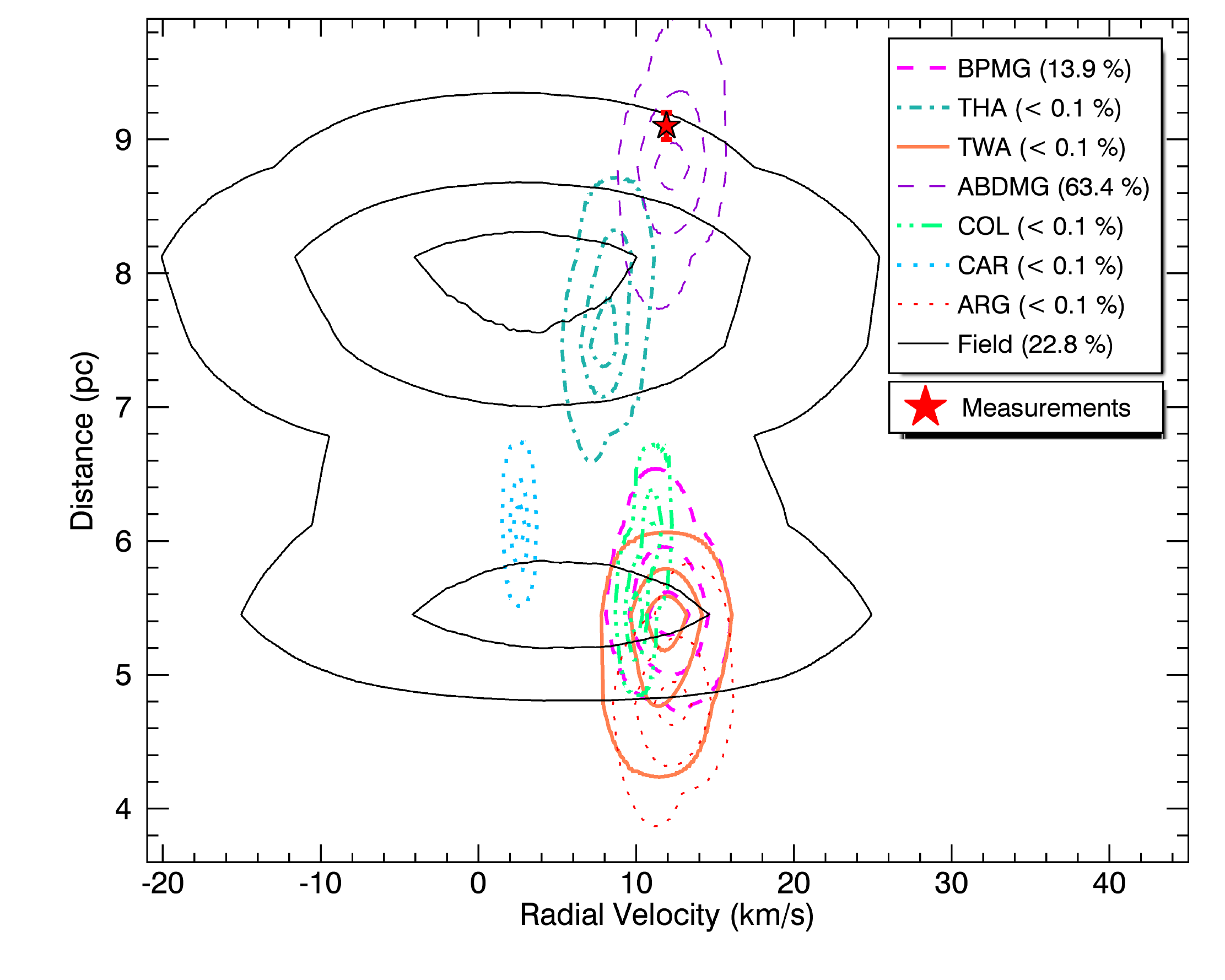}
	\caption{Probability density distributions $P(\{O_i\},v,\varpi|H_k)$ for \emph{2MASS~J03552337+1133437} obtained from a Bayesian analysis that did not use radial velocity or distance as input data, compared to the actual radial velocity and trigonometric distance measurements (red star). The three contour lines of each distribution encompass 10\%, 50\% and 90\% of their total Bayesian probability, the latter being indicated in parenthesis in the legend. We can see that the measurements agree very well with the predictions for the ABDMG hypothesis even if it did not use radial velocity or distance as input parameters. We have displayed the sum of the "single" and "binary" hypotheses PDFs for every hypothesis, which explains the bimodal shape of the field distribution. Similar figures for all candidates in Table~\ref{tab:mass} are available at our group's website \emph{www.astro.umontreal.ca/\textasciitilde gagne}.}
	\label{fig:density}
\end{figure}

\emph{2MASS~J03552337+1133437} (\emph{2MUCD~20171}) is an L5$\gamma$ BD, thus one of the latest known young dwarfs up to date. \cite{2010ApJ...723..684B} measured a radial velocity of $11.9 \pm 0.2$~\kms\ for this object. \cite{2013AJ....145....2F} reported this object as a young field BD with various signs of low gravity in its NIR spectrum as well as the presence of Li absorption, proposing an age of 50~\textendash~150~Myr, which is similar to the age range of the ABDMG, along with distance measurement of $8.2 \pm 0.9$~pc. \cite{2013AN....334...85L} then presented a more precise measurement of its parallax of $9.1 \pm 0.1$~pc, that, along with its radial velocity, allowed them to propose it as a new ABDMG bona fide BD. Here we combined both parallax measurements in an error-weighted average to find a value of $9.1 \pm 0.1$~pc, and confirm that this object should be considered as a 13\textendash 14~\Mjup\ BD bona fide member to the ABDMG, with $P_{H_k} = 99.7\%$ and $C_{H_k} = 0.1\%$. The predicted distance and radial velocity associated with the ABDMG are $8.5~\pm~0.4$~pc and $12.6 \pm 1.7$~\kms, respectively at 1.5$\sigma$ and 0.4$\sigma$ of the measured values (see Figure~\ref{fig:density}). Our analysis suggests that this object could be an unresolved binary. \\

\emph{2MASS~J11395113--3159214} (\emph{TWA~26}) is an over-luminous M9$\gamma$ dwarf  with signs of low-gravity in both its optical and NIR spectra. It has a triangular-shaped $H$-band continuum and \cite{2011A&A...529A..44W} derives a low surface gravity of log g = 3.5 by fitting atmosphere models to the whole NIR spectrum. \cite{2013ApJ...772...79A} classify this object as VL-G. \cite{2012ApJ...752...56F} measure a distance of $28.5 \pm 3.5$~pc for this object, and \cite{2013ApJ...762..118W} measure $42.0 \pm 4.5$~pc. \cite{2005ApJ...634.1385M} measure a radial velocity of $11.6 \pm 2$~\kms\ and propose it as a TWA member. Here we combine both distance measurements to get $33.5 \pm 15.3$~pc and find that it is a 16~\textendash~27~\Mjup\ bona fide member to TWA, with $P_{H_k} = 99.3\%$ and $C_{H_k} < 0.1\%$. It would be useful to clarify the reason why both distance measurements for this object disagree so much. \\

\subsubsection{Peripheral Candidates}

\emph{2MASS~J06085283--2753583} is an M9$\gamma$ dwarf with unusually red colors for its spectral type, Li absorption and signs of low-gravity in both its optical and NIR spectra. It displays a typical triangular-shaped $H$-band continuum and \cite{2013ApJ...772...79A} classify it as VL-G. \cite{2010ApJ...715L.165R} measure a radial velocity of $24.0~\pm~1.0~\kms$, report it as a strong candidate member to $\beta$PMG and estimate its age to be around 10~Myr based on atmospheric models fitting. \cite{2012ApJ...752...56F} report a trigonometric distance of $31.3~\pm~3.5$~pc, and \cite{2008ApJ...689.1295K} estimate its age to be younger than 100~Myr based on the strength of its Li feature. Here, we find that this object is a 15\textendash~23 \Mjup\ BD candidate member to COL, with $P_{H_k}$ = 3.7\% and $C_{H_k}$ = 4.0\%. We would thus classify this object as a peripheral COL bona fide member, rather than a member to $\beta$PMG. The reason why we find this is \emph{solely} due to the radial velocity measurement. If we did not use radial velocity as an input parameter, our Bayesian method would predict $\nu_s~=~20.1~\pm~1.5~\kms$ for $\beta$PMG and $\nu_s~=~22.7~\pm~1.3~\kms$ for COL. The latter is closer to the actual measurement, but even then it can seem surprising that the Bayesian probability for the $\beta$PMG hypothesis drops that much when including it, since it is at only 2.1~$\sigma$ of the predicted value for $\beta$PMG. To understand this, one must look closely at the radial velocity distribution for $\beta$PMG (see Figure~\ref{fig:priors})~; the distribution falls quite steeply after $\nu~=~20~\kms$. In other words, the radial velocity that was measured for \emph{2MASS~J06085283--2753583} is not allowed for in our SKM model for $\beta$PMG. This large sensitivity on radial velocity is due to the fact that this object is close to the anti-apex of both $\beta$PMG and COL. The \emph{XYZUVW} parameters of this object are $13.1~\pm~1.6$~pc, $-28.0~\pm~2.4$~pc, $-34.3~\pm~1.9$~pc, $-7.6~\pm~0.7$~\kms, $-18.6~\pm~0.8$~\kms\ and $-7.9~\pm~0.8$~\kms, respectively. Those are closer to the SKMs of $\beta$PMG than COL, which is consistent with the fact that we would classify it as a $\beta$PMG member without using the $\xi_\nu$ parameter. We conclude that the membership of this object is still ambiguous and that a better radial velocity measurement would be useful in investigating this further. COL membership could be ruled out by additional radial velocity measurements bringing it closer to $20~\kms$. \\

\emph{2MASS~J10220489+0200477} is an over-luminous M9$\beta$ dwarf with colors unusually red for its spectral type and signs of youth in its optical spectrum. \cite{2012ApJ...752...56F} measure its distance to be $38 \pm 16$~pc, and we combine the radial velocity measurements of \cite{2010AJ....139.1808S} and \cite{2008AJ....135..785W} into $-7.9 \pm 4.8$~\kms. We find that this object is a 34~\textendash~53~\Mjup\ candidate to ABDMG, albeit with a very low $P_{H_k} = 2.6\%$. This very low Bayesian probability is due to the mismatch of this object's Galactic motion compared to current bona fide members of ABDMG. The \emph{XYZUVW} parameters for this object are $-12.5~\pm~5.3$~pc, $-23.1~\pm~9.7$~pc, $27.5~\pm~11.6$~pc, $16.1~\pm~6.0$~\kms, $-60.3~\pm~27.6$~\kms\ and $-54.2~\pm~20.7$~\kms, respectively. This is 51~pc and 57~\kms\ away from the SKM of ABDMG. The first is not problematic since it is comparable to the scatter of bona fide members, however the kinematic mismatch is highly significant. However, our Monte Carlo analysis indicates that this is associated with a low $C_{H_k} = 6.0\%$ probability of being a young field contaminant. It is thus possible that this object could be a contaminant from a source that was not considered in this work. As an alternate interpretation, it would be tempting to see this case as a tentative indication of mass segregation, however this is at odds with current evidence \citep{2009AJ....137....1F} and a larger low-mass population would clearly be needed to assess this possibility. We also point out that a better distance and radial velocity measurements are crucial for better constraining the position of this object in the \emph{XYZUVW} parameter space. \\

\subsubsection{Contaminants From Other Associations}

\emph{2MASS~J03393521--3525440} (\emph{LP~944--20}) is an L0 dwarf with a triangular-shaped $H$-band continuum, Li absorption and signs of low gravity from atmospheric models fitting. \cite{1998MNRAS.296L..42T} estimates its age to be 475~\textendash~650~Myr. \cite{2003A&A...400..297R} proposed it as a candidate member to the Castor moving group (CAS) (~320~Myr) through a kinematic comparison with Castor members. \cite{2002AJ....124..519R} measure a radial velocity of $10~\pm~2~\kms$ where \cite{2009ApJ...705.1416R} measure $7.6~\pm~2.6~\kms$, and \cite{1996MNRAS.281..644T} measure a trigonometric distance of $5.0~\pm~0.1$~pc. We combine both radial velocity measurements to obtain $9.3~\pm~1.7~\kms$. Our Bayesian analysis indicates that this object is a candidate member to ARG with $P_{H_k} = 17.5\%$, however we did not include CAS in our set of hypotheses. By performing a simpler Bayesian analysis similar to that presented in \cite{2013ApJ...762...88M} but including the CAS hypothesis, we find that the CAS hypothesis has $P_{H_k} = 99.7\%$ whereas ARG has a negligible probability (remember those probabilities are strongly biased). This means that \emph{2MASS~J03393521--3525440} is indeed a better fit to CAS than ARG. We have used \emph{XYZUVW} values of $-5.3~\pm~12.5$~pc, $4.7~\pm~15.7$~pc, $0.0~\pm~16.3$~pc, $-13.3~\pm~5.7$~\kms, $-8.5~\pm~2.8$~\kms\ and , $-8.8~\pm~4.5$~\kms\ respectively for the CAS hypothesis, which were obtained from members presented in Table~1 of \cite{1998A&A...339..831B}. \\

\emph{2MASS~J23134727+2117294} (\emph{NLTT~56194}) is an M7.5 dwarf with X-ray emission and signs of low-gravity in its optical spectrum. Based on its X-ray emission and various spectroscopic features, \cite{2009ApJ...699..649S} estimate its age to be between 100 and 300 Myr. Based on this age estimate and the kinematics of \emph{2MASS~J23134727+2117294}, \cite{2012ApJ...758...56S} propose that it is a candidate member to the Castor moving group, and measure a radial velocity of $-1.6 \pm 0.3$. Here we find it is a $\beta$PMG candidate with $P_{H_k} = 22.3\%$. However, if we include the Castor hypothesis in a simpler analysis similar to that of \cite{2013ApJ...762...88M} without using photometry, we find that the kinematics of this object clearly better match the Castor hypothesis, with a Bayesian probability $P_H >$ 99.9\%, at a predicted distance of $16.8 \pm 2.7$~pc. We thus propose that this object is a candidate member to the Castor moving group, which would imply its mass to be between 81 and 94 \Mjup. The predicted radial velocity associated with the Castor hypothesis is $-0.6 \pm 2.8$~\kms, at only 0.4$\sigma$ of the measurement. \\

\subsubsection{Candidates with High Probability}

\emph{2MASS~J00040288--6410358} is an object with signs of low gravity in its optical spectrum and NIR colors unusually red for its L1$\gamma$ spectral type. It has already been proposed as a THA candidate member by \cite{2010ApJS..190..100K}, in agreement with our results : we find $P_{H_k} = 99.7\%$ and $C_{H_k} = 0.5\%$. If it is actually a member to the THA with, would have a mass between 13 and 14 \Mjup, which would place it in the planetary-mass regime.\\

\emph{2MASS~J00065794--6436542} is an L0 object displaying H$\alpha$ emission and signs of low gravity in its optical spectrum. Here we propose it as a 21~\textendash~41~\Mjup\ strong BD candidate member to the THA, with $P_{H_k} > 99.9\%$ and $C_{H_k} = 0.2\%$. Our analysis suggests that this object could be an unresolved binary. \\

\emph{2MASS~J00192626+4614078} (\emph{2MUCD~10013}) is an M8 dwarf with high rotational velocity, Li absorption and signs of low-gravity in its NIR spectrum. \cite{2009ApJ...705.1416R} estimated its age to be less than several hundred Myr based on its Li absorption, and \cite{2013ApJ...772...79A} characterized it as an Intermediate-Gravity (INT-G) dwarf. \cite{2009ApJ...705.1416R} measure a radial velocity of $-19.5 \pm 2.0$~\kms\ for this object. Here, we find that it is a 78~\textendash~94~\Mjup\ LMS candidate to ABDMG, with $P_{H_k} = 88.0\%$ and $C_{H_k} = 3.9\%$. The predicted radial velocity associated with the ABDMG hypothesis is of $-17.0 \pm 1.4$~\kms, at 1$\sigma$ of the measured value. \\

\emph{2MASS~J00325584--4405058} is an L0$\gamma$ dwarf with colors too red for its spectral type and signs of low-gravity in both its optical and NIR spectra. \cite{2013ApJ...772...79A} characterize it as a Very-Low Gravity (VL-G) dwarf. \cite{2012ApJ...752...56F} report and a trigonometric distance of 26.4~$\pm$~3.3~pc for this object. Taking these measurements into account, we find that this object is a 10~\textendash~12~\Mjup\ \emph{planemo} candidate member to $\beta$PMG with $P_{H_k}$ = 91.8\% and $C_{H_k} = 0.2\%$. \\

\emph{2MASS~J00374306--5846229} is another red L0$\gamma$ object with signs of low gravity in its optical spectrum. It was not previously recognized as a NYA candidate member, but here we propose it as a strong 13~\textendash~15 \Mjup\ candidate to the THA, with $P_{H_k} = 97.3\%$ and $C_{H_k} = 0.7\%$. Our analysis suggests that this object could be an unresolved binary. \\

\emph{2MASS~J00413538--5621127} (\emph{2MUCD~20035}) is reported in \cite{2010A&A...513L...9R} as a nearby, young M8 BD with Li absorption, signs of accretion and a most probable age of 10 Myr. The authors note that its sky position and proper motion indicate that this object is a probable member of the Tucana-Horologium association. \cite{2010ApJ...722..311L} indicate that this object is an unresolved M6.5 + M9 binary. Here we also find that \emph{2MASS~J00413538--5621127} is a strong candidate member to THA. Furthermore its proposed age of 10~Myr agrees well with the 10\textendash~40~Myr age range for the THA. Its predicted radial velocity $\nu$~=~6.4~$\pm$~2.4~\kms\ agrees relatively well with the combined measurement $\nu$~=~2.8~$\pm$~1.9~\kms\ from \cite{2010ApJ...723..684B} and \cite{2009ApJ...705.1416R}, which yields $P_{H_k} > 99.9\%$ and $C_{H_k} = 0.2\%$. We estimate the masses of each component to be 14~\textendash~41~\Mjup\ and 18~\textendash~41~\Mjup. \\

\emph{2MASS~J00452143+1634446} (\emph{2MUCD~20037}) is a BD with signs of low gravity in its optical spectrum, H$\alpha$ emission and NIR colors unusually red for its L3.5 spectral type. We propose it as a new 13\textendash~14 \Mjup\ strong candidate member to the ARG. Its predicted radial velocity of $\nu$~=~3.4~$\pm$~1.3~\kms\ agrees very well with the actual measurement $\nu$~=~3.4~$\pm$~0.2~\kms, which yields $P_{H_k} > 99.9\%$ and $C_{H_k} = 1.8\%$. \\

\emph{2MASS~J00470038+6803543} is a peculiar L7 dwarf with extremely red colors for its spectral type. \cite{2012AJ....144...94G} and \cite{2013ApJS..205....6M} identify this object as possibly very dusty, over-metallic or young, which could explain its odd nature. After obtaining a NIR spectrum at a better resolution, \cite{2013PASP..125..809T} identify that this object has signs of low-gravity such as weaker-than-normal atomic lines. Here, we identify that this object is a strong candidate member to ABDMG, with $P_{H_k} = 98.2\%$ and $C_{H_k} = 2.4\%$. This object would have a very low-mass of 11~\textendash~15~\Mjup\ if membership is confirmed. \\

\emph{2MASS~J01033203+1935361} is an L6$\beta$ dwarf with signs of low-gravity in both its optical and NIR spectra. It has unusually red NIR colors for its spectral type and a typical triangular-shaped $H$-band continuum. \cite{2012ApJ...752...56F} measure a trigonometric distance of $21.3~\pm~3.4$~pc for this object. Here, we find that it is a strong 10~\textendash~11~\Mjup\ \emph{planemo} candidate member to ARG, with $P_{H_k} = 76.0\%$ and $C_{H_k} = 0.1\%$. \\

\emph{2MASS~J01174748--3403258} is an L1 dwarf whose NIR spectrum was reported by \cite{2011A&A...529A..44W} as fitting best with theoretical atmosphere models at a relatively low gravity of 4.5~dex. More recently, \cite{2013ApJ...772...79A} report that this object has a typical triangular-shaped $H$-band continuum as well as weak alkali lines, classifying it as an intermediate-gravity dwarf. Here we propose that this object is a high probability 13~\textendash~14~\Mjup\ candidate member to the THA, with $P_{H_k} = 99.3\%$ and $C_{H_k} = 1.0\%$. \\

\emph{2MASS~J01225093--2439505} is an M3.5 + L5 binary system in which the primary displays X-ray emission and the secondary has unusually red NIR colors for its spectral type, as well as a triangular-shaped $H$-band continuum. \cite{2013ApJ...774...55B} report a radial velocity measurement of $9.6~\pm~0.7$~\kms\ and propose that this object could be a young candidate member to ABDMG, however we find here that it is rather a candidate member to $\beta$PMG, with $P_{H_k} = 98.2\%$ and $C_{H_k} = 3.4\%$. If we do not include the radial velocity measurement, it is a better match to ABDMG. However, the radial velocity measurement being at 2.7$\sigma$ from the $15.6~\pm~2.1$~\kms\ prediction for ABDMG, but only at 0.5$\sigma$ from the $10.6~\pm~1.7$~\kms\ prediction for $\beta$PMG, we conclude that it is a candidate member to $\beta$PMG rather than ABDMG. We note that our proper motion measurement arising from a cross-correlation of 2MASS and \emph{WISE} ($\mu_\alpha = 89.7 \pm 7.9$ \masyr, $\mu_\delta = -108.9 \pm 8.6$ \masyr) is discrepant from that previously reported in UCAC4 \citep{2012yCat.1322....0Z} and PPMXL (\citealp{2010AJ....139.2440R}; $\mu_\alpha = 89.7 \pm 7.9$ \masyr, $\mu_\delta = -108.9 \pm 8.6$ \masyr), resulting in a large error of $24.2$ \masyr in our adopted value for $mu_\alpha$, which also favors the $\beta$PMG hypothesis over ABDMG. It would thus be useful to get a better measurement of the proper motion of this object to address the possibility that it is a member to ABDMG. We have used NIR photometry reported in \cite{2013ApJ...774...55B} to estimate a mass of 5~\textendash~6~\Mjup\ for the secondary and 67~\textendash~89~\Mjup\ for the primary. \\

\emph{2MASS~J01415823--4633574} is an L0$\gamma$ dwarf with several indicators of youth. Its optical and NIR spectra both display signs of low-gravity, including a triangular-shaped $H$-band continuum, its NIR colors are unusually red for its spectral type, it displays H$\alpha$ emission and \cite{2011A&A...529A..44W} report that its NIR spectrum is best fitted by models with log~g~=~4. \cite{2006ApJ...639.1120K} report that this object should have an age comprised between 1 and 50~Myr, and that it could be a member either of the THA or $\beta$PMG. Here we find that this object is a very strong 14~\textendash~20~\Mjup\ candidate member to the THA with a Bayesian probability of 99.7\%, associated to a field contamination probability of $C_{H_k} = 0.1\%$. Its predicted radial velocity and distance are $\nu$~=~7.6~$\pm$~2.4~\kms\ and $\varpi$~=~41.4~$\pm$~2.8~pc if it is a member of the THA, or $\nu$~=~14.1~$\pm$~1.7~\kms\ and $\varpi$~=~28.9~$\pm$~2.4~pc if it is a member of the $\beta$PMG. The radial velocity measurement $\nu$~=~12~$\pm$~15 from \cite{2006ApJ...639.1120K} is not precise enough to verify either of these two hypotheses. However, we find that this object has a significantly higher probability of being a member to the THA even if we do not take this measurement into account. Our analysis also suggests that this object could be an unresolved binary. \\

\emph{2MASS~J02215494--5412054} and \emph{2MASS~J02251947--5837295} have both been reported as low-gravity M9 dwarfs (\citealp{2008AJ....135..580R}, \citealp{2009AJ....137....1F}), but we found no mention of them as being a candidates to any NYA. Here we propose that both objects are very strong 16~\textendash~26~\Mjup\ and 20~\textendash~32~\Mjup\ BD candidates to the THA with $P_{H_k} = > 99.9\%$ and $C_{H_k} = 0.2\%$. \\

\emph{2MASS~J02235464--5815067}, \emph{2MASS~J02340093--6442068} and \emph{2MASS~J03231002--4631237} (\emph{2MUCD~20157}) are three L0$\gamma$ dwarfs unusually red for their spectral types, with signs of low gravity in their optical spectra. Furthermore, \emph{2MASS~J03231002--4631237} shows Li absorption. Here we report that all of them are very strong 13~\textendash~15 \Mjup\ BD candidate member to THA, with $P_{H_k} > 99.9\%$ ($C_{H_k} = 0.1\%$), $P_{H_k} = 99.9\%$ ($C_{H_k} = 0.2\%$) and $P_{H_k} = 98.4\%$ ($C_{H_k} = 1.2\%$), respectively. Our analysis suggests that both \emph{2MASS~J02235464--5815067} and \emph{2MASS~J03231002--4631237} could be unresolved binaries. \\

\emph{2MASS~J02411151--0326587} is an L0$\gamma$ dwarf with colors too red for its spectral type, signs of low-gravity in both its optical and NIR spectra and a triangular-shaped $H$-band continuum. \cite{2013ApJ...772...79A} categorize this as a VL-G object. Here we propose this object as a THA BD candidate, with $P_{H_k} = 79.1\%$ and $C_{H_k} = 1.1\%$, and that it would have a mass comprised between 13~\textendash~14 \Mjup\ if it is actually a member. \\

\emph{2MASS~J03264225--2102057} (\emph{2MUCD~10184}) is an L4 dwarf with colors too red for its spectral type and Li absorption. \cite{2007AJ....133..439C} suggests that this object should be younger than 500~Myr based on the strength of its Li absorption. We find that this object is a 13~\textendash~15 \Mjup\ BD candidate member to ABDMG, with $P_{H_k} = 98.9\%$ and $C_{H_k} = 1.3\%$. Our analysis suggests that this object could be an unresolved binary.\\

\emph{2MASS~J03421621--6817321} (\emph{2MUCD~10204}) is an L2 dwarf that was reported by \cite{2009AJ....137....1F} as having colors too red for its spectral type. We find that even if we do not have strong indicators of youth for this object, it is still a very strong 11~\textendash~13 \Mjup\ \emph{planemo} candidate member to THA, with $P_{H_k} = 98.8\%$ and $C_{H_k} = 5.6\%$. Our analysis also suggests that this object could be an unresolved binary. \\

\emph{2MASS~J03572695--4417305} is an L0$\beta$ binary system unusually red for its spectral type with subtle signs of low gravity in its unresolved optical spectrum. \cite{2003AJ....126.1526B} report this object as a binary system with an angular separation of 0\textquotedbl.098 and a position angle of 174\textdegree. \cite{2010ApJ...722..311L} obtained resolved spectral types of M9 and L1.5 for the two components, and estimate their age to be around 100~Myr because of their low surface gravity. Here we report this unresolved system as a very strong 14~\textendash~15 \Mjup\ candidate member to THA, with $P_{H_k} = 99.6\%$ and $C_{H_k} = 1.2\%$. \\

\emph{2MASS~J04210718--6306022} (\emph{2MUCD~10268}) is an L5$\gamma$ dwarf with unusually red colors for its spectral type and signs of low-gravity in both its optical and NIR spectra. This object also displays Li absorption, and here we report that it is a \emph{planemo} candidate member to ARG with $P_{H_k} = 98.1\%$ and $C_{H_k} = 8.0\%$, with a mass of 10~\textendash~11 \Mjup. \\

\emph{2MASS~J04362788--4114465} is a peculiar M8 dwarf with signs of low-gravity in both its optical and NIR spectra, which \cite{2013ApJ...772...79A} classify as VL-G. Here we find that this object is a very strong 32~\textendash~49 \Mjup\ BD candidate member to COL, with $P_{H_k} = 96.0\%$ and $C_{H_k} = 9.1\%$. \\

\emph{2MASS~J04433761+0002051} (\emph{2MUCD~10320}) is an M9$\gamma$ dwarf with signs of low gravity in its optical spectrum, a high rotational velocity, NIR colors unusually red for its spectral type, and displaying H$\alpha$ emission and Li absorption. \cite{2008ApJ...689.1295K} report that the strength of its Li absorption is compatible with an age of $<~100$~Myr, and \cite{2012AJ....143...80S} proposes it as a candidate member to the ABDMG, and \cite{2009ApJ...705.1416R} measure a radial velocity of $17.1 \pm 3.0$~\kms. This measurement agrees within 0.06$\sigma$ of the predicted $17.3 \pm 1.8$~\kms\ value for the $\beta$PMG hypothesis. Here, we find that this object is probably not a member of the ABDMG, but rather a strong candidate 15~\textendash~16~\Mjup\ BD member to the $\beta$PMG, with $P_{H_k} = 99.8\%$ and $C_{H_k} = 3.4\%$. Schlieder (priv. comm.) agrees with our result that this object should rather be a $\beta$PMG candidate. The reason for their claim that this object is a candidate to ABDMG arises from their use of optical data in deriving a proper motion measurement of $\mu_\alpha = 48$ \masyr, $\mu_\delta = -122$ \masyr, which is at 3.3$\sigma$ of the one presented here ($\mu_\alpha = 35.9 \pm 7.7$ \masyr, $\mu_\delta = -98.0 \pm 8.2$ \masyr). Our analysis suggests that this object could be an unresolved binary. \\

\emph{2MASS~J05184616--2756457} (\emph{2MUCD~10381}) is an unusually bright L1$\gamma$ dwarf with very red colors for its spectral type and signs of low gravity in both its optical and NIR spectra. It also shows a typical triangular-shaped $H$-band continuum, and \cite{2013ApJ...772...79A} classify it as VL-G. \cite{2012ApJ...752...56F} measure a trigonometric distance of $46.8 \pm 15.0$~pc. Here we report this object as a 13\textendash~22~\Mjup\ candidate member to COL, with $P_{H_k} = 96.2\%$ and $C_{H_k} = 0.7\%$. The predicted distance for the COL hypothesis is of $51.8~\pm~5.6$~pc, which is at 0.3$\sigma$ from the measured value. However, it would be desirable to increase the precision of the current distance measurement, which still only has a 3$\sigma$ significance. Our analysis suggests that this object could be an unresolved binary. \\

\emph{2MASS~J05361998--1920396} (\emph{2MUCD~10397}) is an L2$\gamma$ dwarf with unusually red colors for its spectral type and signs of low-gravity in its optical spectrum. This object displays a triangular-shaped $H$-band continuum and \cite{2013ApJ...772...79A} classify it as VL-G. \cite{2012ApJ...752...56F} measure a trigonometric distance of $39.0 \pm 14.0$~pc for this object. Here we report that it is a 11\textendash~14~\Mjup\ candidate member to COL, with $P_{H_k} = 95.2\%$ and $C_{H_k} = 0.7\%$. The predicted distance associated with the COL hypothesis is of $40.2 \pm 3.2$~pc, which is at 0.1$\sigma$ from the measured value. However, it would be desirable to increase the precision of the current distance measurement, which only has a 2.8$\sigma$ significance. \\

\emph{2MASS~J12451416--4429077} (\emph{TWA~29}) is an over-luminous M9.5p dwarf with H$\alpha$ emission and signs of low-gravity in both its optical and NIR spectra. It has a typical triangular-shaped $H$-band continuum and \cite{2011A&A...529A..44W} derives a marginally low surface gravity of log g = 4.5 by fitting atmosphere models to its NIR spectrum. It has been identified by \cite{2007ApJ...669L..97L} as a candidate member to the TWA, and \cite{2013AAS...22113703W} measure a trigonometric distance of $79.0 \pm 12.9$~pc. Here we also find that this object is a 17~\textendash~19~\Mjup\ BD candidate to TWA, with $P_{H_k} = 93.3\%$ and $C_{H_k} = 0.4\%$. The predicted distance associated with the TWA hypothesis is of $74.6~\pm~6.8$~pc, at only 0.3$\sigma$ of the measured value. \\

\emph{2MASS~J16471580+5632057} is a peculiar L9 dwarf with colors unusually red for its spectral type. \cite{2012ApJS..201...19D} measure a distance of $8.6 \pm 2.2$~pc for this object. Without making any assumption on its age, we find that it is a 4~\textendash~6~\Mjup\ candidate to ARG, with $P_{H_k} = 26.3\%$ and $C_{H_k} = 3.3\%$. If we do not include the distance measurement, the Bayesian probability is $P_{H_k} < 0.1\%$. \\

\emph{2MASS~J20004841--7523070} (\emph{2MUCD~20845}) is an M9 dwarf with signs of low gravity in its optical spectrum and NIR colors unusually red for its spectral type. \cite{2010MNRAS.409..552G} indicate that this object could be a member of the Castor moving group, but that further spectroscopic study is needed to assess its membership. They also measure a radial velocity of $11.8 \pm 1.0$~\kms\ for this object. The Castor moving group is not considered in the results presented here because of its age older than 100~Myr, however we have performed a simpler Bayesian analysis without using photometry (see \cite{2013ApJ...762...88M}) but including the Castor hypothesis, and found that it only had a 3.1\% Bayesian probability (versus 72.3\% for the $\beta$PMG hypothesis), associated to a predicted distance of $18.9~\pm~4.4$~pc. Here we rather propose it as a 18\textendash~27~\Mjup\ BD candidate member to the $\beta$PMG, with $P_{H_k} = 96.6\%$ and $C_{H_k} = 4.0\%$, and a predicted distance of $33.3^{+3.2}_{-2.8}$~pc. We suggest that the best way to completely rule out the Castor membership would be a measurement of its parallax. Our analysis suggests that this object could be an unresolved binary. \\

\emph{2MASS~J21011544+1756586} (\emph{**~BOY~11}) is an L7.5 dwarf with unusually red colors for its spectral type and a typical triangular-shaped $H$-band continuum. \cite{2011A&A...529A..44W} estimate a marginally low surface gravity of log g = 4.5 by fitting atmosphere models to its NIR spectrum. However, we consider that none of these signs of youth are strong enough to assume an age of $<$~1~Gyr for this object. \cite{2010ApJ...711.1087K} report that this is an unresolved binary and \cite{2004AJ....127.2948V} measure a distance of $33.2~\pm~3.8$~pc. Without making any assumption about the age of this object, we find that it is a 11~\textendash~12~\Mjup\ \emph{planemo} candidate member to ABDMG, with $P_{H_k} = 26.8\%$ and $C_{H_k} = 4.2\%$. Our analysis suggests that this object could be an unresolved binary. \\

\emph{2MASS~J21140802--2251358} is a very red L7 object identified by \cite{2013ApJ...777L..20L} to be a \emph{planemo} candidate member to $\beta$PMG. They report a trigonometric distance of $24.6~\pm~1.4$~pc for this object. Here, we find that this object is indeed a strong 8\textendash~9~\Mjup\ \emph{planemo} candidate member to the $\beta$PMG, with $P_{H_k} = 99.7\%$ and $C_{H_k} = 0.1\%$. \\

\emph{2MASS~J21265040--8140293} is an L3$\gamma$ dwarf with unusually red colors for its spectral type and signs of low-gravity in its optical spectrum. We find that this object is a 13~\textendash~14~\Mjup\ candidate to THA, with $P_{H_k} = 94.5\%$ and $C_{H_k} = 0.5\%$. Our analysis indicates that this object could be an unresolved binary system. \\

\emph{2MASS~J22064498--4217208} is an L2 dwarf with Li absorption displaying unusually red colors for its spectral type. Here we find that without making any assumption on its age, it is a 18~\textendash~21~\Mjup\ BD candidate member to ABDMG with $P_{H_k} = 95.3\%$ and $C_{H_k} = 14.1\%$. \\

\emph{2MASS~J22443167+2043433} (\emph{2MUCD~20968}) is an L6.5 lithium dwarf with signs of low gravity in its NIR spectrum, and NIR colors unusually red for its spectral type. \cite{2011A&A...529A..44W} suggest a value for log~g~=~3.5 based on atmospheric models fitting to its NIR spectrum. We found that this object is a strong candidate member to the ABDMG, with $P_{H_k} = 99.6\%$ and $C_{H_k} = 0.5\%$. We estimate a mass of 11~\textendash~12~\Mjup\ if membership is confirmed. Our analysis suggests that this object could be an unresolved binary. \\

\emph{2MASS~J23225299--6151275} is an L2$\gamma$ BD with signs of low gravity in its optical spectrum and NIR colors unusually red for its spectral type (\citealp{2008AJ....135..580R}, \citealp{2009AJ....137.3345C}, \citealp{2013AJ....145....2F}). We propose it as a new strong 12~\textendash~13 \Mjup\ candidate to the THA, with $P_{H_k} > 99.9\%$ and $C_{H_k} = 0.3\%$. We also report that we have identified a common proper-motion primary LMS at an angular separation of 16\textquotedbl.6: \emph{2MASS~J23225240--6151114}, an M5 which has a proper motion of $\mu_\alpha = 80.2 \pm 3.7$ \masyr, $\mu_\delta = -69.5 \pm 9.3$ \masyr, as inferred from its 2MASS and \emph{WISE} positions. This measurement is within 0.27$\sigma$ and 0.37$\sigma$ of the $\mu_\alpha$ and $\mu_\delta$ proper motion of the companion, respectively. The UCAC4 \citep{2012yCat.1322....0Z} proper-motion is consistent with it. If the system is at the statistical distance of 43.0~$\pm$~2.4~pc predicted for the THA hypothesis, then the physical separation would be 714~$\pm$~40~AU. The predicted statistical distance for the young field hypothesis is of 57.0$^{+7.6}_{-9.6}$~pc, which would bring the physical separation of the system to 946$^{+126}_{-159}$~AU. If the THA hypothesis is verified, the M5 primary would have a mass comprised between 34 and 37 \Mjup, and thus the system would have a mass ratio of $q$ = 0.35$^{+.03}_{-.05}$. \\

\LongTables
\begin{deluxetable*}{l|lllllllc}
\tablecolumns{9}
\tablecaption{Age and Mass Estimates of Candidates \label{tab:mass}}
\tablehead{\colhead{Name} & \colhead{SpT\tablenotemark{d}} & \colhead{$C_{H_k}$} & \colhead{$P_{H_k}$} & \colhead{NYA} & \colhead{Reported} & \colhead{Mass} & \colhead{$v_{rs}$\tablenotemark{f}} & \colhead{$d_{s}$\tablenotemark{f}}\\
\colhead{} & \colhead{} & \colhead{\%} & \colhead{\%}  & \colhead{} & \colhead{Candidate\tablenotemark{e}} & \colhead{($M_{\mathrm{Jup}}$)} & \colhead{($\kms$)} & \colhead{(pc)}}
\startdata
\cutinhead{Bona fide members}
J0123--6921 & M7.5 & $< 0.1$\tablenotemark{a}\tablenotemark{b} & $> 99.9$ & THA & $\cdots$ & $56\-- 74$ & $9.9 \pm 2.5$ & $47.4 \pm 3.2$ \\
J0355+1133 & L5$\gamma$ & $0.1$\tablenotemark{a}\tablenotemark{b} & $99.7$ & ABDMG\tablenotemark{c} & ABDMG (49) & $13\-- 14$ & $12.6 \pm 1.7$ & $8.5 \pm 0.4$ \\
J1139--3159 & M9$\gamma$ & $< 0.1$\tablenotemark{a}\tablenotemark{b} & $99.3$ & TWA\tablenotemark{c} & $\cdots$ & $16\-- 27$ & $11.3 \pm 2.2$ & $46.6 \pm 4.4$ \\
\cutinhead{Peripheral candidates}
J0608--2753 & M9$\gamma$ & $1.5$\tablenotemark{a}\tablenotemark{b} & $3.7$ & COL & $\beta$PMG (66) & $16\-- 24$ & $22.7 \pm 1.3$ & $42.6 \pm 7.6$ \\
J1022+0200 & M9$\beta$ & $6.0$\tablenotemark{a}\tablenotemark{b} & $2.6$ & ABDMG & $\cdots$ & $34\-- 53$ & $9.6 \pm 15.0$ & $16.5 \pm 1.2$ \\
\cutinhead{Contaminants from other associations}
J0339--3525 & M9 & $\cdots$ & $99.7$\tablenotemark{g} & CAS & CAS (65) & $44\-- 45$ & $14.4 \pm 3.3$ & $6.8 \pm 2.6$ \\
J2313+2117 & M7.5 & $\cdots$ & $95.8$\tablenotemark{g} & CAS & CAS (74) & $81\-- 94$ & $-0.6 \pm 2.8$ & $16.8 \pm 2.7$ \\
\cutinhead{Candidates with high probability}
J0004--6410 & L1$\gamma$ & $0.5$ & $99.7$ & THA\tablenotemark{c} & THA (42) & $13\-- 14$ & $6.8 \pm 2.9$ & $47.4 \pm 3.2$ \\
J0006--6436 & L0 & $0.2$ & $> 99.9$ & THA\tablenotemark{c} & $\cdots$ & $21\-- 41$ & $6.5 \pm 2.5$ & $43.4 \pm 2.8$ \\
J0019+4614 & M8 & $3.9$\tablenotemark{a} & $88.0$ & {\scriptsize ABDMG} & {\scriptsize ABDMG (75)} & $78\-- 94$ & $-17.0 \pm 1.4$ & $37.4 \pm 2.8$ \\
J0032--4405 & L0$\gamma$ & $0.2$\tablenotemark{b} & $91.8$ & {\scriptsize $\beta$PMG} & $\cdots$ & $10\-- 11$ & $11.6 \pm 1.7$ & $26.1 \pm 2.0$ \\
J0037--5846 & L0$\gamma$ & $0.7$ & $97.3$ & THA\tablenotemark{c} & $\cdots$ & $13\-- 15$ & $6.8 \pm 2.5$ & $47.8 \pm 3.2$ \\
J0041-5621 & M6.5+M9 & $0.2$\tablenotemark{a} & $> 99.9$ & {\scriptsize THA}\tablenotemark{c} & {\scriptsize THA (63)} & $14\-- 41$ & $6.5 \pm 2.4$ & $41.8 \pm 2.4$ \\
J0045+1634 & L2$\beta$ & $1.8$\tablenotemark{a} & $99.9$ & {\scriptsize ARG} & $\cdots$ & $13\-- 14$ & $3.4 \pm 1.3$ & $13.3 \pm 0.8$ \\
J0047+6803 & L7p\tablenotemark{c} & $2.4$ & $98.2$ & ABDMG & $\cdots$ & $11\-- 15$ & $-20.4 \pm 1.1$ & $10.5 \pm 0.8$ \\
J0103+1935 & L6$\beta$ & $0.1$\tablenotemark{b} & $76.0$ & {\scriptsize ARG} & $\cdots$ & $10\-- 11$ & $8.6 \pm 2.1$ & $15.3 \pm 1.2$ \\
J0117--3403 & L1\tablenotemark{c} & $1.0$ & $99.3$ & THA & $\cdots$ & $13\-- 14$ & $3.4 \pm 2.1$ & $40.6 \pm 2.0$ \\
J0122--2439 & M3.5+L5\tablenotemark{d} & $3.4$\tablenotemark{a} & $92.8$ & {\scriptsize $\beta$PMG} & {\scriptsize ABDMG (BW13)} & $5\-- 89$ & $10.6 \pm 1.7$ & $22.5 \pm 2.8$ \\
J0141--4633 & L0$\gamma$ & $0.1$\tablenotemark{a} & $99.7$ & {\scriptsize THA}\tablenotemark{c} & {\scriptsize THA/$\beta$PMG (42)} & $14\-- 20$ & $7.6 \pm 2.4$ & $41.4 \pm 2.8$ \\
J0221--5412 & M9 & $0.2$ & $> 99.9$ & THA & $\cdots$ & $16\-- 26$ & $10.2 \pm 2.2$ & $41.0 \pm 2.4$ \\
J0223--5815 & L0$\gamma$ & $0.1$ & $> 99.9$ & THA\tablenotemark{c} & $\cdots$ & $14\-- 15$ & $10.6 \pm 2.4$ & $43.4 \pm 2.8$ \\
J0225--5837 & M9 & $0.2$ & $> 99.9$ & THA & $\cdots$ & $20\-- 32$ & $10.7 \pm 2.4$ & $43.8 \pm 2.8$ \\
J0234--6442 & L0$\gamma$ & $0.2$ & $99.9$ & THA & THA (42) & $13\-- 14$ & $10.9 \pm 2.5$ & $45.8 \pm 2.8$ \\
J0241--0326 & L0$\gamma$ & $1.1$ & $79.1$ & THA & $\cdots$ & $13\-- 14$ & $5.1 \pm 2.5$ & $49.8 \pm 3.2$ \\
J0323--4631 & L0$\gamma$ & $1.2$ & $98.4$ & THA\tablenotemark{c} & $\cdots$ & $14\-- 15$ & $12.6 \pm 2.4$ & $49.4 \pm 3.2$ \\
J0326--2102 & L4 & $1.3$ & $98.9$ & ABDMG\tablenotemark{c} & $\cdots$ & $13\-- 15$ & $23.1 \pm 2.1$ & $26.1 \pm 2.0$ \\
J0342--6817 & L2 & $5.6$ & $98.8$ & THA\tablenotemark{c} & $\cdots$ & $11\-- 13$ & $13.1 \pm 2.2$ & $50.2 \pm 3.6$ \\
J0357--4417 & L0$\beta$ & $1.2$ & $99.6$ & THA\tablenotemark{c} & $\cdots$ & $14\-- 15$ & $14.2 \pm 2.2$ & $48.6 \pm 3.2$ \\
J0421--6306 & L5$\gamma$ & $8.0$ & $98.1$ & ARG & $\cdots$ & $10\-- 11$ & $9.7 \pm 2.2$ & $16.5 \pm 1.2$ \\
J0436--4114 & M8p & $9.1$ & $96.0$ & COL & $\cdots$ & $32\-- 49$ & $22.0 \pm 2.0$ & $44.2 \pm 6.4$ \\
J0443+0002 & M9$\gamma$ & $3.4$\tablenotemark{a} & $99.8$ & {\scriptsize $\beta$PMG}\tablenotemark{c} & {\scriptsize ABDMG (75)} & $17\-- 19$ & $16.9 \pm 2.0$ & $25.7 \pm 3.2$ \\
J0518--2756 & L1$\gamma$ & $0.7$\tablenotemark{b} & $96.2$ & {\scriptsize COL}\tablenotemark{c} & $\cdots$ & $14\-- 22$ & $22.9 \pm 1.7$ & $51.8 \pm 5.6$ \\
J0536--1920 & L2$\gamma$ & $0.7$\tablenotemark{b} & $95.2$ & {\scriptsize COL} & $\cdots$ & $12\-- 13$ & $22.7 \pm 1.7$ & $40.2 \pm 3.2$ \\
J1245--4429 & M9.5p & $0.4$\tablenotemark{b} & $93.3$ & {\scriptsize TWA}\tablenotemark{c} & {\scriptsize TWA (51)} & $17\-- 19$ & $9.9 \pm 2.1$ & $81.8 \pm 8.4$ \\
J1647+5632 & L9p\tablenotemark{d} & $3.3$\tablenotemark{b} & $26.3$ & {\scriptsize ARG} & $\cdots$ & $4\-- 5$ & $-10.9 \pm 3.1$ & $14.5 \pm 1.2$ \\
J2000--7523 & M9 & $4.0$\tablenotemark{a} & $96.6$ & {\scriptsize $\beta$PMG}\tablenotemark{c} & {\scriptsize CAS (26)} & $19\-- 27$ & $6.4 \pm 2.4$ & $32.9 \pm 3.2$ \\
J2101+1756 & L7.5 & $4.2$\tablenotemark{b} & $26.8$ & {\scriptsize ABDMG}\tablenotemark{c} & $\cdots$ & $11\-- 12$ & $-19.8 \pm 2.0$ & $24.9 \pm 1.6$ \\
J2114--2251 & L7\tablenotemark{d} & $0.1$\tablenotemark{b} & $99.7$ & {\scriptsize $\beta$PMG}\tablenotemark{c} & {\scriptsize $\beta$PMG (LI13)} & $8\-- 9$ & $-6.4 \pm 1.7$ & $22.1 \pm 1.6$ \\
J2126--8140 & L3$\gamma$ & $0.5$ & $94.5$ & THA\tablenotemark{c} & $\cdots$ & $13\-- 14$ & $8.2 \pm 2.4$ & $45.0 \pm 2.8$ \\
J2206--4217 & L2 & $14.1$ & $95.3$ & ABDMG & $\cdots$ & $18\-- 21$ & $7.6 \pm 2.0$ & $28.5 \pm 1.6$ \\
J2244+2043 & L6.5 & $0.5$ & $99.6$ & ABDMG\tablenotemark{c} & $\cdots$ & $11\-- 12$ & $-15.5 \pm 1.7$ & $18.5 \pm 1.2$ \\
J2322--6151 & L2$\gamma$ & $0.3$ & $> 99.9$ & THA & $\cdots$ & $12\-- 13$ & $4.8 \pm 2.5$ & $43.0 \pm 2.4$ \\
\cutinhead{Candidates with modest probability}
J0033--1521 & L4$\beta$ & $21.8$ & $31.9$ & ARG & $\cdots$ & $9\-- 11$ & $2.3 \pm 1.3$ & $17.3 \pm 1.6$ \\
J0129+3517 & L4\tablenotemark{c} & $18.4$ & $43.5$ & ARG & $\cdots$ & $9\-- 11$ & $6.4 \pm 2.0$ & $28.5 \pm 3.2$ \\
J0253+3206 & M7p & $29.7$ & $25.5$ & $\beta$PMG & $\cdots$ & $13\-- 15$ & $5.7 \pm 2.4$ & $35.8 \pm 2.8$ \\
J0303--7312 & L2$\gamma$ & $66.1$ & $4.4$ & THA & THA (42) & $12\-- 14$ & $12.1 \pm 2.7$ & $53.0 \pm 3.6$ \\
J0406--3812 & L0$\gamma$ & $60.7$ & $2.1$ & COL & COL (42) & $12\-- 14$ & $21.3 \pm 3.4$ & $69.4 \pm 9.2$ \\
J0619--2903 & M6 & $22.0$ & $80.7$ & COL\tablenotemark{c} & $\cdots$ & $15\-- 23$ & $24.2 \pm 2.0$ & $55.8 \pm 6.0$ \\
J0632--5010 & L3 & $61.1$ & $1.3$ & ABDMG & $\cdots$ & $10\-- 14$ & $30.8 \pm 1.4$ & $10.5 \pm 4.8$ \\
J0642+4101 & L/Tp\tablenotemark{c} & $52.0$ & $49.5$ & ABDMG & $\cdots$ & $11\-- 12$ & $0.6 \pm 1.5$ & $17.3 \pm 0.8$ \\
J0652--5741 & M8$\beta$ & $49.7$\tablenotemark{b} & $3.3$ & {\scriptsize ABDMG}\tablenotemark{c} & $\cdots$ & $29\-- 34$ & $29.2 \pm 1.3$ & $45.8 \pm 5.2$ \\
J1004+5022 & L3$\beta$ & $29.6$ & $32.2$ & ABDMG & $\cdots$ & $22\-- 28$ & $-10.7 \pm 3.5$ & $26.1 \pm 3.6$ \\
J1600--2456 & M7.5p\tablenotemark{c} & $59.0$ & $0.1$ & ABDMG\tablenotemark{c} & $\cdots$ & $11\-- 13$ & $-6.9 \pm 2.0$ & $20.5 \pm 1.2$ \\
J1956--7542 & L0$\gamma$ & $55.3$ & $16.6$ & THA\tablenotemark{c} & $\cdots$ & $13\-- 14$ & $6.4 \pm 2.7$ & $59.8 \pm 4.4$ \\
J2148+4003 & L6 & $36.6$ & $48.1$ & ARG & $\cdots$ & $6\-- 7$ & $-9.2 \pm 1.3$ & $4.9 \pm 0.4$ \\
J2208+2921 & L3$\gamma$ & $53.8$ & $10.1$ & $\beta$PMG & $\cdots$ & $9\-- 11$ & $-10.6 \pm 2.0$ & $35.4 \pm 3.6$ \\
J2351+3010 & L5.5 & $62.7$ & $47.0$ & ARG & $\cdots$ & $9\-- 11$ & $-1.5 \pm 1.3$ & $20.9 \pm 2.0$ \\
\cutinhead{Candidates with low probability}
J0126+1428 & L4$\gamma$ & $76.7$ & $3.4$ & $\beta$PMG & $\cdots$ & $7\-- 9$ & $6.0 \pm 4.5$ & $38.6 \pm 6.0$ \\
J0512--2949 & L4.5 & $77.9$ & $15.4$ & $\beta$PMG & $\cdots$ & $5\-- 7$ & $19.7 \pm 1.5$ & $12.9 \pm 2.0$ \\
J0712--6155 & L1$\beta$ & $62.8$\tablenotemark{b} & $2.6$ & {\scriptsize ABDMG}\tablenotemark{c} & $\cdots$ & $27\-- 40$ & $29.0 \pm 2.0$ & $43.0 \pm 6.4$ \\
J1547--2423 & M9 & $88.0$ & $0.1$ & ARG & $\cdots$ & $13\-- 14$ & $-19.7 \pm 2.2$ & $23.3 \pm 2.8$ \\
J2013--2806 & M9 & $70.7$ & $44.0$ & $\beta$PMG & $\cdots$ & $14\-- 16$ & $-7.4 \pm 2.4$ & $44.2 \pm 4.8$ \\
J2213--2136 & L0$\gamma$ & $80.5$ & $3.1$ & $\beta$PMG & $\cdots$ & $12\-- 13$ & $-1.9 \pm 1.8$ & $45.0 \pm 3.6$
\enddata
\tablenotetext{a}{This result takes into account a radial velocity measurement.}
\tablenotetext{b}{This result takes into account a parallax measurement.}
\tablenotetext{c}{The binary hypothesis has a higher probability.}
\tablenotetext{d}{Spectral types with this mention are near-infrared. Other ones are optical.}
\tablenotetext{e}{Objects for which membership was already suspected. See Table \ref{tab:input} for references and abbreviations.}
\tablenotetext{f}{Statistical predictions associated with the most probable NYA. For the actual measurements when available, see Table~\ref{tab:input}.}
\tablenotetext{g}{This probability was obtained from a simpler Bayesian analysis (see \citealp{2013ApJ...762...88M}) that makes the assumption of uniform prior probabilities.}
\end{deluxetable*}

\subsubsection{Candidates with Modest Probability}

\emph{2MASS~J00332386--1521309} is an L4$\beta$ dwarf with colors too red for its spectral type and subtle signs of low-gravity in its optical spectrum. \cite{2013ApJ...772...79A} characterize its NIR spectrum as a normal Field-Gravity (FLD-G) dwarf. The only NIR gravity indicator that is not clearly consistent with FLD-G is the shape of the $H$-band continuum that could be triangular, however the quality of the available data is not sufficient to say more about this. We propose this object as a weak candidate to ARG, with $P_{H_k} = 31.9\%$ and $C_{H_k} = 21.8\%$. If it is actually a member of ARG, it would have a mass between 9 and 11 \Mjup. \\

\emph{2MASS~J01291221+3517580} is an unusually red L4 dwarf with Li absorption, with no clear evidence of youth. We find that, without making any assumption on its age, this object is a 9~\textendash~11~\Mjup\ candidate member to ARG with $P_{H_k} = 7.2\%$ and $C_{H_k} = 67.1\%$. \\

\emph{2MASS~J02530084+1652532} is an M7 dwarf for which models fitting suggest a marginally low log~g~$\sim$~4.5 \citep{2011A&A...529A..44W}. Without making any assumption on its age, we find that this object is a 13~\textendash~15~\Mjup\ BD candidate member to ARG with $P_{H_k} = 25.5\%$ and $C_{H_k} = 29.7\%$. A measurement of its radial velocity and distance, as well as a thorough analysis of its spectral properties would be needed to confirm this. \\

\emph{2MASS~J03032042--7312300} is an L2$\gamma$ dwarf with colors too red for its spectral type and signs of low-gravity in its optical spectrum. Here, as also reported in \cite{2010ApJS..190..100K}, we find that this is a candidate member to THA albeit a weak one, with $P_{H_k} = 4.4\%$ and $C_{H_k} = 66.1\%$, which would make it a 12~\textendash~14 \Mjup\ object. \\

\emph{2MASS~J04062677--3812102} is an L0$\gamma$ dwarf with unusually red colors  for its spectral type and signs of low gravity in both its optical and NIR spectra. It also displays the typical triangular-shaped $H$-band continuum characteristic of low-gravity. \cite{2013ApJ...772...79A} classified this object as VL-G. \cite{2010ApJS..190..100K} reported that the good match of this object's optical spectrum to that of \emph{2MASS~J0141--4633} suggests an age of $\sim$ 30~Myr, and that its sky location furthermore strengthens the hypothesis of this object being a member of COL. Here we find that this object effectively has a good match to the properties of COL, but we find it is quite a weak candidate member with $P_{H_k} = 2.1\%$ and $C_{H_k} = 60.7\%$.. However, if we consider that this object effectively has an age of 30~Myr, the probability that it is a field contaminant would drop below $C_{H_k} < 5\%$. If it is actually a member of COL, we estimate its mass to be between 12 and 14 \Mjup. \\

\emph{2MASS~J06195260--2903592} is an M6 dwarf unusually red for its spectral type and reported as having signs of low gravity in its optical spectrum by \cite{2003AJ....126.2421C}. \cite{2013ApJ...772...79A} estimate the age of this object to be $\sim$ 10~Myr because it displays a circumstellar disk (which could also explain its reddening). We find that this object is a good 15\textendash~23~\Mjup\ candidate member to COL, with $P_{H_k} = 80.7\%$ and $C_{H_k} = 22.0\%$. The lower-end mass estimate is more probable because of the circumstellar disk, and for the same reason $C_{H_k}$ is probably pessimistic. Our analysis suggests that this object could be an unresolved binary. \\

\emph{2MASS~J06322402--5010349} is an L3 dwarf with strong Li absorption. Without making any assumption on its age, we find that it is a modest 10~\textendash~14~\Mjup\ candidate member to ABDMG with $P_{H_k} = 1.3\%$ and $C_{H_k} = 61.1\%$. A measurement of its radial velocity and distance, as well as a thorough analysis of its spectral properties would be needed to confirm this. \\

\emph{2MASS~J06420559+4101599} is a very peculiar object identified by \cite{2013ApJS..205....6M} as an URL dwarf. It has a NIR spectrum that is badly fit by any known L or T dwarfs. It has an extremely red continuum and a classification using solely the $J$-band would result in a T spectral type, however this object shows no sign of CH$_4$, which is inconsistent with it being a T dwarf. These peculiar properties could result from a very dusty photosphere at the L/T transition, and \cite{2013ApJS..205....6M} report that low-gravity or metallicity could not provide the whole explanation. They have thus classified this object as L/Tp Here we identify that without making any assertion about this object's age, it comes out as a weak candidate member to ABDMG, with $P_{H_k} = 49.5\%$ and $C_{H_k} = 52.0$. If this object turns out to be a member of ABDMG, it would have a mass of approximately 11~\textendash~12~\Mjup, which means that this could be a \emph{planemo} at the L/T transition. If we could find evidence that this system is young, the probability that it is a field contaminant would also be lower. A measurement of its distance could significantly strengthen the proposition that this is a member of ABDMG. \cite{2013ApJS..205....6M} report on two more systems that resemble this one~: \emph{J1738+6142} and \emph{J0754+7909}. We find that none of them have kinematics coherent with any of the NYAs considered here. Being able to restrict the age of \emph{J0642+4101} to that of ABDMG would be of great interest in understanding the physical nature of this odd object, we thus urge that measuring its distance and radial velocity should be a priority. \\

\emph{2MASS~J06524851--5741376} (\emph{2MUCD~10601}) is an M8$\beta$ dwarf with unusually red colors for its spectral type and subtle signs of low-gravity in its optical spectrum. \cite{2012A&A...548A..33C} identifies this system as a tight binary with an angular separation of 0\textquotedbl.23, a mass ratio of $q$ $\sim$~0.7\textendash~0.8 and a semi-major axis of 5~\textendash~6 AU. \cite{2012ApJ...752...56F} measure a trigonometric distance of $32.0 \pm 3.3$~pc. Here we report this system as a BD binary candidate to ABDMG, with $P_{H_k} = 3.3\%$ and $C_{H_k} = 49.7\%$. The low Bayesian probability is due to the fact that the predicted distance value associated with the ABDMG hypothesis is of $45.8^{+5.2}_{-4.8}$~pc, at 2.4$\sigma$ of the measured value. If this system is confirmed as a member of ABDMG, the mass of each component would be approximately 21 to 33~\Mjup. \\

\emph{2MASS~J10042066+5022596} is an L3$\beta$ dwarf with unusually red colors for its spectral type, Li absorption and signs of low-gravity in both its optical and NIR spectra. It has a typical triangular-shaped $H$-band continuum, and \cite{2013ApJ...772...79A} report it as VL-G. This object is a companion to \emph{G~196--3}, a bright co-moving M3 LMS at 17\textquotedbl.7 with a radial velocity of -0.7~$\pm$~1.2 \kms\ \citep{2012ApJ...758...56S}. \cite{2008ApJ...676.1281M} report an age estimate of 60 to 300~Myr for \emph{2MASS~J10042066+5022596}, however \cite{2004ApJ...600.1020M} state that it could be younger. Here we find that it comes out as a weak 22~\textendash~28~\Mjup\ BD candidate member to ABDMG, with $P_{H_k} = 32.2\%$ and $C_{H_k} = 29.6\%$. At the predicted distance of 26.1~$\pm$~3.6~pc, this would mean that this object is at a physical separation of 462~$\pm$~64~AU. Since the companion is masked by its bright primary in \emph{WISE} data, we did not use \emph{WISE} photometry and did not measure a proper motion from the 2MASS and \emph{WISE} data for this object. As a result, we did not consider photometry at all in the Bayesian analysis, which means that the true contamination rate for this object could be somewhat higher, since our Monte Carlo contamination analysis made use of the 2MASS and \emph{WISE} photometry. The radial velocity of the parent star is within 0.4$\sigma$ of the CAR hypothesis, which is associated with a statistical radial velocity prediction of -1.8~$\pm$~2.8~\kms. For a system of approximately 0.4~\Msol at this separation, the expected variation in radial velocity is of the order of 1~\kms, hence the binary nature of this object should not affect our conclusions. If we thus include this radial velocity measurement in it, the Bayesian probability associated to the CAR hypothesis increases to $P_H$ = 97.1\%, but still yields a high $C_{H_k} \sim 85\%$. The reason for this is that such a low radial velocity and high proper motion are unlikely to come from CAR in our SKM models (see Figure~\ref{fig:priors}). We thus conclude that this object's membership is quite ambiguous, and that a measurement of its distance is needed to decide whether it is a candidate member to ABDMG or CAR. It is also possible that the SKM model for CAR is still not a fair representation of reality, since we know only 7 bona fide members in this NYA. Finding more members to CAR will allow to investigate this further. \\

\emph{2MASS~J16002647--2456424} is a peculiar M7.5 dwarf with signs of low-gravity in its NIR spectrum. We find that it is a weak 11~\textendash~13~\Mjup\ \emph{planemo} candidate member to ABDMG with a $P_{H_k} = 0.1\%$ and $C_{H_k} = 59.0\%$. Even if the field contamination probability seems weak for such a low Bayesian probability, we stress that this result should be interpreted with caution since \emph{2MASS~J16002647--2456424} has a sky position close to the Upper Scorpius association. It is thus likely that this object is a member to Upper Scorpius, which was not considered in our analysis. \\

\emph{2MASS~J19564700--7542270} is an L0$\gamma$ dwarf with unusually red colors for its spectral type and signs of low-gravity in its optical spectrum. We find that this object is a 13~\textendash~14~\Mjup\ BD candidate to THA, with $P_{H_k} = 16.6\%$, $C_{H_k} = 55.3\%$, and signs that it could be an unresolved binary system. \\

\emph{2MASS~J21481633+4003594} is an L6.5 dwarf with NIR colors unusually red for its spectral type, a triangular-shaped $H$-and continuum and weaker-than-normal alkali lines. Atmosphere models fitting also suggests that this is a young object with log~g~$\sim$~4 \citep{2011A&A...529A..44W}. Here, we find that this object is a moderate 6~\textendash~7~\Mjup\ \emph{planemo} candidate to ARG, with $P_{H_k} = 48.1\%$ and $C_{H_k} = 36.6\%$. \\

\emph{2MASS~J22081363+2921215} is an L3$\gamma$ dwarf with a triangular-shaped $H$-band continuum that display signs of youth in its optical spectrum. It shows Li absorption and has NIR colors unusually red for its spectral type. Here, we find that it is a moderate 9~\textendash~11~\Mjup\ \emph{planemo} candidate member to $\beta$PMG, with $P_{H_k} = 10.1\%$ and $C_{H_k} = 53.8\%$. \\

\emph{2MASS~J23512200+3010540} is a peculiar L5 dwarf with unusually red NIR colors for its spectral type, as reported by \cite{2010ApJS..190..100K}. We find that it is a moderate 9~\textendash~11~\Mjup\ \emph{planemo} candidate to ARG, with $P_{H_k} = 47.0\%$ and $C_{H_k} = 62.7\%$. A measurement of its radial velocity and distance would be needed to confirm this. \\

\subsubsection{Candidates not Uncovered with our Method}

\emph{2MASS~J09510459+3558098} (\emph{NLTT~22741}) is an M4.5 dwarf displaying X-ray emission. \cite{2009ApJ...699..649S} estimated its age to be comprised between 40 and 300~Myr, and then \cite{2009ApJ...699..649S} proposed it as a candidate member to THA. Here, we find that without considering the radial velocity measurement of 10.2 $\pm$ 0.2 \kms\ from \cite{2012ApJ...758...56S}, it only has a Bayesian probability $P_{H_k}$ = 16.1\% for ABDMG, with a predicted radial velocity of --3.9~$\pm$~1.8~\kms, as well as small Bayesian probabilities of $P_{H_k}$ = 0.2\% for TWA and $P_{H_k}$ = 0.3\% for CAR. When the radial velocity measurement is added, Bayesian probabilities fall below 0.01\% for every NYA hypothesis, which is associated to a $>$~99.9\% probability that this object is a young field contaminant. This object has an L6 co-moving companion displaying signs of youth for which \cite{2012ApJS..201...19D} measured a distance of  62~$\pm$~27~pc, which further weakens the hypothesis that this object is a candidate member to any NYA considered here. \\

\emph{2MASS~J13142039+1320011} (\emph{**~Law~2}) is an over-luminous M7 dwarf with H$\alpha$ and X-ray emission. \cite{2012AJ....143...80S} report that this object is a likely member of ABDMG, based on its sky position, proper motion from the LSPM catalog and parallax \citep{2009AJ....137.3632L}. However, even if our proper motion measurement agrees within 1$\sigma$ to that in LSPM, we find a Bayesian probability of less than $P_{H_k}$ = 0.1\% for the ABDMG hypothesis when we do not include the distance measurement. A distance of 21.3~$\pm$~1.2~pc is predicted for the ABDMG hypothesis, which is similar to that predicted by \citeauthor{2012AJ....143...80S} (\citeyear{2012AJ....143...80S} ; 20.1 $\pm$ 1.0 pc). However, when we add the measured distance 16.4 $\pm$ 0.8 pc, the Bayesian probability for all NYA hypotheses are less than 0.01\%. \\

\subsubsection{Discussion}

Results presented here and in \cite{2013ApJ...762...88M} show that Bayesian analysis is a powerful tool for searching for new candidate members to NYAs that are significantly spread on the sky, even without having access to radial velocity and parallax measurements. With the modified version presented here which is adapted to later-than-M5 objects, it can be now conceivable to build a credible sample of BD and \emph{planemo} candidates to NYAs. This fraction might be even lower if there are still missing bona fide members in the A0~\textendash~M0 spectral-type range. However, there are some limitations to the present method that could potentially be complemented by other methods such as traceback analysis : (1) We expect to miss a fraction of true members, which would be hard to differentiate with field contaminants unless we have measurements of their radial velocity and parallax. This is especially true for ARG, ABDMG and $\beta$PMG. (2) Potential outlier members with \emph{XYZUVW} values significantly different from the locus values of their NYA, might not be uncovered by our method unless we slowly build up our SKM model by iteratively adding bona fide members with relatively low Bayesian probabilities such as \emph{2MASS~J06085283--2753583}. (3) Our analysis is model-dependent and thus results are vulnerable to change if the SKM or photometric models described earlier are not a good representation of reality. Several improvements could still be brought to our method, including the addition of older NYAs such as Castor and Carina-Near, and yet a better treatment of photometric sequences when we know more about broad-band photometry of young BDs (e.g., see J. Filippazzo et al., in preparation). If the IMF of NYAs is not significantly different than that from the field, one can expect that currently known members are only the tip of the iceberg, accounting for only 10\% of their total population. This consideration has motivated us to initiate a systematic all-sky survey for more later-than-M5 members to NYAs in the 2MASS and \emph{WISE} catalogs, which will be the subject of an upcoming paper. The very first results of this survey can be found in \cite{2013arXiv1307.1127G}.

\section{SUMMARY AND CONCLUSIONS}\label{sec:conclusions}

We have presented several modifications to the Bayesian inference method introduced by \cite{2013ApJ...762...88M} in order to assess the probabilities that late-type objects are members to several NYAs. In particular, we introduced the use of NIR colors and spectral types in order to calibrate the distance hypotheses for later-than-M5 objects, as well as improved our spatial and kinematic modeling of NYAs by representing their \emph{XYZ} and \emph{UVW} distributions as rotated ellipsoids. We have also presented a thorough contamination analysis to assess the significance of the results yielded by this method. We have then identified several LMS, BD and \emph{planemo} candidate members to NYAs, which were already recognized for displaying various signs of youth, or for having redder-than-normal NIR colors. We also provide statistical predictions of their radial velocities and distances if they are actual members, so that these hypotheses might be tested against observation in the coming years (see, e.g., J. K. Faherty et al., in preparation). We report on 35 very strong $> M5$ candidate members to NYAs, from which 25 are assigned a membership to a NYA for the first time. We also propose \emph{2MASS~J01231125--6921379} as a new M7.5 bona fide members to THA. We independently confirm that \emph{2MASS~J03552337+1133437} should be considered as a bona fide members to ABDMG and question the possibility that \emph{2MASS~J06085283--2753583} could be a member of COL instead of $\beta$PMG. We also report \emph{2MASS~J23225240--6151114} as an M5 common proper-motion primary to the L2$\gamma$ BD \emph{2MASS~J23225299--6151275}, this system being a strong candidate member to THA. We note that \emph{2MASS~J00470038+6803543} and \emph{2MASS~J22244381--0158521}, which are extremely red L dwarfs with no clear evidence of youth, are strong candidate members to ABDMG. Finally, we show that a dozen candidates unveiled here could be free-floating planetary-mass objects if their membership is confirmed. Radial velocity and parallax measurements are needed to confirm their membership. An online we tool as well as additional figures and information on NYAs can be found at our group's website \url{www.astro.umontreal.ca/\textasciitilde gagne}.

\nocite{2011ApJ...727...62R}
\nocite{2012A&A...548A..26D}
\nocite{2013A&A...553L...5D}
\nocite{2013AN....334...85L}
\nocite{2013prpl.conf2G024F}
\nocite{2013AJ....145....2F}

\nocite{2003A&A...409..523R}\nocite{1997A&A...320..440H}\nocite{1997A&A...320..428H}\nocite{1996A&A...305..125R}\nocite{1987PAICz..69..323R}\nocite{1987A&A...180...94B}\nocite{1986A&A...157...71R}\nocite{2012A&A...538A.106R}
\nocite{2013ApJ...762...88M}
\nocite{2000ApJ...535..959Z}\nocite{2004ApJ...613L..65Z}\nocite{2001ApJ...562L..87Z}\nocite{2006ApJ...649L.115Z}\nocite{2001ASPC..244..122Z}\nocite{2011ApJ...732...61Z}\nocite{2001ApJ...549L.233Z}\nocite{2001ApJ...559..388Z}\nocite{2004ARA&A..42..685Z}\nocite{2011ApJ...727...62R}\nocite{2003ApJ...593.1074G}\nocite{2000ApJ...542..464C}\nocite{2008ApJ...687.1264M}
\nocite{2008A&A...491..829K}\nocite{2009AJ....137....1F}\nocite{2008ASPC..384..119C}\nocite{2007ApJ...669L..97L}\nocite{2010ApJ...714...45L}\nocite{2011ApJ...732...56G}\nocite{2008ApJ...676.1281M}\nocite{2003AJ....126.2421C}\nocite{2009AJ....137.3345C}\nocite{2009ApJ...699..649S}\nocite{2012ApJ...758...56S}\nocite{2011ApJ...727....6S}\nocite{2011ASPC..448..481R}\nocite{2009IAUS..258.....M}\nocite{2010ARA&A..48..581S}\nocite{2010MNRAS.409..552G}\nocite{2004ApJ...600.1020M}\nocite{2008ApJ...689.1295K}\nocite{2000AJ....120..447K}\nocite{2010ApJS..190..100K}\nocite{2006ApJ...639.1120K}\nocite{2006ApJ...643.1160L}\nocite{2010A&A...519A..93B}\nocite{2011A&A...527A..24L}\nocite{2011MNRAS.418.1231D}\nocite{2007MNRAS.374..372L}\nocite{2010AJ....140..119S}\nocite{2011MNRAS.411..117K}\nocite{2012AJ....143..114S}
\nocite{2010ApJ...711.1087K}\nocite{2003AJ....126.1526B}\nocite{2010ApJ...722..311L}\nocite{2005A&A...435L...5F}\nocite{2006AJ....132..891R}\nocite{2006MNRAS.368.1917L}\nocite{2008MNRAS.384..150L}\nocite{2010ApJ...722..311L}\nocite{2010ApJ...715..561A}
\nocite{2008AJ....136.2483A}\nocite{2011ApJ...732...61Z}\nocite{2004ApJ...613L..65Z}\nocite{2004ARA&A..42..685Z}\nocite{2009A&A...508..833D}\nocite{2008hsf2.book..757T}\nocite{2005ApJ...634.1385M}\nocite{2001ASPC..244..104M}\nocite{2000ApJ...544..356M}
\nocite{2007ApJ...669.1167B}\nocite{2008ApJ...689.1127M}\nocite{2002A&A...382..563B}\nocite{2006A&A...458..805B}\nocite{2010A&A...519A..93B}\nocite{2005yCat.1297....0Z}\nocite{2009yCat.1315....0Z}\nocite{2003A&A...402..701B}\nocite{2010AJ....139.2440R}\nocite{2011ASPC..448..481R}\nocite{2010ApJ...715L.165R}\nocite{1996AJ....112.2799H}\nocite{2011AJ....142..104R}\nocite{2010AJ....140..897R}\nocite{2009NewA...14..615F}\nocite{2008hsf2.book..757T}\nocite{2011A&A...527A..24L}\nocite{2008ApJ...689.1295K}\nocite{2008AJ....135..580R}\nocite{2004MNRAS.355..363L}
\nocite{2008AJ....136.1290R}

\acknowledgments

The authors would like to thank Jacqueline Faherty, Emily Rice, Adric Riedel, Philippe Delorme, Ben Oppenheimer, C\'eline Reyl\'e, Sandie Bouchard, Am\'elie Simon and Brendan Bowler for useful comments and discussions. Thanks to Annie Robin for help with the Besan\c con Galactic model. We would like to address special thanks to Adric Riedel for generously sharing valuable parallax data with our team. This work was supported in part through grants from the Fond de Recherche Qu\'eb\'ecois \textendash Nature et Technologie and the Natural Science and Engineering Research Council of Canada. This research has made use of the SIMBAD database and VizieR catalogue access tool, operated at Centre de Donn\'ees astronomiques de Strasbourg (CDS), France \citep{2000A&AS..143...23O}. This research has benefitted from the M, L, and T dwarf compendium housed at \url{http://DwarfArchives.org} and maintained by Chris Gelino, Davy Kirkpatrick, and Adam Burgasser. This publication makes use of data products from the Two Micron All Sky Survey, which is a joint project of the University of Massachusetts and the Infrared Processing and Analysis Center/California Institute of Technology, funded by the National Aeronautics and Space Administration and the National Science Foundation (\citealp{2006AJ....131.1163S}, \citealp{2003yCat.2246....0C}). This publication makes use of data products from the Wide-field Infrared Survey Explorer, which is a joint project of the University of California, Los Angeles, and the Jet Propulsion Laboratory/California Institute of Technology, funded by the National Aeronautics and Space Administration \citep{2012yCat.2311....0C}. This research has benefitted from the SpeX Prism Spectral Libraries, maintained by Adam Burgasser at \url{http://www.browndwarfs.org/spexprism}. This research has made use of the NASA/ IPAC Infrared Science Archive, which is operated by the Jet Propulsion Laboratory, California Institute of Technology, under contract with the National Aeronautics and Space Administration. We thank our anonymous referee for our initial manuscript and for several insightful comments that greatly improved the overall quality and clarity of this work.

\bibliographystyle{apj}
\bibliography{Papers_Library}
\mbox{~}

\newpage
\clearpage
\onecolumngrid
\LongTables
\begin{landscape}
\tabletypesize{\scriptsize}
\renewcommand{\tabcolsep}{0.5mm}
\begin{deluxetable}{lccccccccccccccccccccl}
\tablecolumns{22}
\tablecaption{Input Sample\label{tab:input}}
\tablehead{
\colhead{2MASS} & \colhead{Common} & \colhead{Opt.} & \colhead{NIR} & \colhead{} & \multicolumn{3}{c}{2MASS} & \colhead{} & \multicolumn{2}{c}{\emph{WISE}} & \colhead{} & \colhead{$\mu_\alpha$} & \colhead{$\mu_\delta$} & \colhead{$\mu$\tablenotemark{e}} & \colhead{$v_{\mathrm{rad}}$} & \colhead{$v_{\mathrm{rad}}$} & \colhead{$d_\pi$} & \colhead{$d_\pi$} & \colhead{Signs}  & \colhead{Reported\tablenotemark{g}} & \colhead{Other\tablenotemark{d}}\\
\cline{6-8}
\cline{10-11}
\colhead{Designation} & \colhead{Name} & \colhead{SpT} & \colhead{SpT} & \colhead{} & \colhead{$J$} & \colhead{$H$} & \colhead{$K_s$} & \colhead{} & \colhead{W1} & \colhead{W2} & \colhead{} & \colhead{(mas yr$^{-1}$)} & \colhead{(mas yr$^{-1}$)} & \colhead{Ref.} & \colhead{(km s$^{-1}$)} & \colhead{Ref.} & \colhead{(pc)} & \colhead{Ref.} & \colhead{of Youth\tablenotemark{a}} & \colhead{Cand.} & \colhead{References}
}
\startdata
\cutinhead{Young Objects}
J00034227-2822410 & 2MUCD 10002 & M7.5 & M7 & & $13.07$ & $12.38$ & $11.97$ & & $11.67$ & $11.50$ & & $280.3 \pm 1.5$ & $-123.3 \pm 1.7$ & 22 & $11.6 \pm 1.1$ & 74 & $38.9 \pm 1.2$ & 22,74,82 & O?ITXU & $\cdots$ & 2,23\\
J00040288-6410358 & $\cdots$ & L1$\gamma$ & $\cdots$ & & $15.79$ & $14.83$ & $14.01$ & & $13.37$ & $12.94$ & & $64.6 \pm 8.0$ & $-49.0 \pm 17.0$ & 42 & $\cdots$ & $\cdots$ & $\cdots$ & $\cdots$ & OR & THA (42) & 25\\
J00065794-6436542 & $\cdots$ & L0 & $\cdots$ & & $13.39$ & $12.66$ & $12.17$ & & $11.72$ & $11.39$ & & $83.3 \pm 3.6$ & $-65.1 \pm 10.2$ & 60 & $\cdots$ & $\cdots$ & $\cdots$ & $\cdots$ & OH & $\cdots$ & 54\\
J00115060-1523450 & $\cdots$ & $\cdots$ & M7.5 & & $15.93$ & $15.64$ & $15.26$ & & $14.99$ & $14.79$ & & $100.7 \pm 19.0$ & $-159.4 \pm 19.4$ & $\cdots$ & $\cdots$ & $\cdots$ & $\cdots$ & $\cdots$ & T?M & $\cdots$ & 6,87\\
J00192626+4614078 & 2MUCD 10013 & M8 & M8 & & $12.60$ & $11.94$ & $11.50$ & & $11.26$ & $11.00$ & & $121.3 \pm 4.3$ & $-77.7 \pm 6.3$ & 69 & $-19.5 \pm 2.0$ & 63 & $\cdots$ & $\cdots$ & ITVL & ABDMG (75) & 2,23,75\\
J00250365+4759191 & 2MUCD 13016 & L4 & $\cdots$ & & $14.84$ & $13.67$ & $12.90$ & & $11.74$ & $11.57$ & & $275.0 \pm 0.7$ & $11.7 \pm 0.8$ & 22 & $\cdots$ & $\cdots$ & $43.8 \pm 1.7$ & 82 & RL & $\cdots$ & 16,23\\
J00274197+0503417 & PC 0025+0447 & M9.5 & L0 & & $16.19$ & $15.29$ & $14.96$ & & $14.62$ & $14.14$ & & $10.5 \pm 0.4$ & $-0.8 \pm 0.3$ & 22 & $\cdots$ & $\cdots$ & $72.5 \pm 8.4$ & 19 & OT & $\cdots$ & 2,23\\
J00325584-4405058 & $\cdots$ & L0$\gamma$ & L0 & & $14.78$ & $13.86$ & $13.27$ & & $12.82$ & $12.49$ & & $117.8 \pm 4.3$ & $-91.6 \pm 4.3$ & 24 & $\cdots$ & $\cdots$ & $26.4 \pm 3.3$ & 24 & OTR & $\cdots$ & 17,2,23,62\\
\enddata
\end{deluxetable}
\clearpage
\newpage
\end{landscape}
\twocolumngrid

\clearpage
\onecolumngrid
\LongTables
\begin{landscape}
\tabletypesize{\scriptsize}
\renewcommand{\tabcolsep}{0.6mm}
\begin{deluxetable}{ll|ccccl|ccccl|ccccl|ccccl|ccccl|ccccl|ccccl|cccc|}
\tablecolumns{41}
\tablecaption{Bayesian Probabilities for Young and Red Candidates\tablenotemark{a} \label{tab:results}}
\tablehead{
\colhead{} & \colhead{} & \multicolumn{4}{c}{TWA} & \colhead{} & \multicolumn{4}{c}{$\beta$PMG} & \colhead{} & \multicolumn{4}{c}{THA} & \colhead{} & \multicolumn{4}{c}{COL} & \colhead{} & \multicolumn{4}{c}{CAR} & \colhead{} & \multicolumn{4}{c}{ARG} & \colhead{} & \multicolumn{4}{c}{ABDMG}  & \colhead{} & \multicolumn{4}{c}{Young Field}\\
\cline{3-6}
\cline{8-11}
\cline{13-16}
\cline{18-21}
\cline{23-26}
\cline{28-31}
\cline{33-36}
\cline{38-41}
\colhead{Name} & \colhead{} & \colhead{$P$} & \colhead{$P_{v}$} & \colhead{$P_{\pi}$} & \colhead{$P_{v+\pi}$} & \colhead{} & \colhead{$P$} & \colhead{$P_{v}$} & \colhead{$P_{\pi}$} & \colhead{$P_{v+\pi}$} & \colhead{} & \colhead{$P$} & \colhead{$P_{v}$} & \colhead{$P_{\pi}$} & \colhead{$P_{v+\pi}$} & \colhead{} & \colhead{$P$} & \colhead{$P_{v}$} & \colhead{$P_{\pi}$} & \colhead{$P_{v+\pi}$} & \colhead{} & \colhead{$P$} & \colhead{$P_{v}$} & \colhead{$P_{\pi}$} & \colhead{$P_{v+\pi}$} & \colhead{} & \colhead{$P$} & \colhead{$P_{v}$} & \colhead{$P_{\pi}$} & \colhead{$P_{v+\pi}$} & \colhead{} & \colhead{$P$} & \colhead{$P_{v}$} & \colhead{$P_{\pi}$} & \colhead{$P_{v+\pi}$} & \colhead{} & \colhead{$P$} & \colhead{$P_{v}$} & \colhead{$P_{\pi}$} & \colhead{$P_{v+\pi}$}
}
\startdata
J0001+1010 & $ $ & $0.0$ & \ldots & \ldots & \ldots & $ $ & $0.0$ & \ldots & \ldots & \ldots & $ $ & $0.0$ & \ldots & \ldots & \ldots & $ $ & $0.0$ & \ldots & \ldots & \ldots & $ $ & $0.0$ & \ldots & \ldots & \ldots & $ $ & $0.0$ & \ldots & \ldots & \ldots & $ $ & $0.0$ & \ldots & \ldots & \ldots & $ $ & $\textcolor{red}{\mathbf{0.5}}$ & \ldots & \ldots & \ldots \\
J0003-2822 & $ $ & $0.0$ & $0.0$ & $0.0$ & $0.0$ & $ $ & $0.1$ & $0.0$ & $0.0$ & $0.0$ & $ $ & $0.0$ & $0.0$ & $0.0$ & $0.0$ & $ $ & $0.0$ & $0.0$ & $0.0$ & $0.0$ & $ $ & $0.0$ & $0.0$ & $0.0$ & $0.0$ & $ $ & $0.0$ & $0.0$ & $0.0$ & $0.0$ & $ $ & $0.0$ & $0.0$ & $0.0$ & $0.0$ & $ $ & $99.9$ & $100.0$ & $100.0$ & $\textcolor{red}{\mathbf{100.0}}$ \\
J0004-6410 & $ $ & $0.0$ & \ldots & \ldots & \ldots & $ $ & $0.0$ & \ldots & \ldots & \ldots & $ $ & $\textcolor{red}{\mathbf{99.6}}$\tablenotemark{b} & \ldots & \ldots & \ldots & $ $ & $0.0$ & \ldots & \ldots & \ldots & $ $ & $0.0$ & \ldots & \ldots & \ldots & $ $ & $0.0$ & \ldots & \ldots & \ldots & $ $ & $0.0$ & \ldots & \ldots & \ldots & $ $ & $0.3$\tablenotemark{b} & \ldots & \ldots & \ldots \\
J0006-6436 & $ $ & $0.0$ & \ldots & \ldots & \ldots & $ $ & $0.0$ & \ldots & \ldots & \ldots & $ $ & $\textcolor{red}{\mathbf{100.0}}$\tablenotemark{b} & \ldots & \ldots & \ldots & $ $ & $0.0$ & \ldots & \ldots & \ldots & $ $ & $0.0$ & \ldots & \ldots & \ldots & $ $ & $0.0$ & \ldots & \ldots & \ldots & $ $ & $0.0$ & \ldots & \ldots & \ldots & $ $ & $0.0$ & \ldots & \ldots & \ldots \\
J0011-1523 & $ $ & $0.0$ & \ldots & \ldots & \ldots & $ $ & $0.0$ & \ldots & \ldots & \ldots & $ $ & $0.0$ & \ldots & \ldots & \ldots & $ $ & $0.0$ & \ldots & \ldots & \ldots & $ $ & $0.0$ & \ldots & \ldots & \ldots & $ $ & $0.0$ & \ldots & \ldots & \ldots & $ $ & $0.0$ & \ldots & \ldots & \ldots & $ $ & $0.0$ & \ldots & \ldots & \ldots \\
\enddata
\tablenotetext{a}{Bayesian Parameter without using radial velocity or parallax ($P$), including radial velocity information only ($P_{v}$), including parallax information only ($P_{\pi}$) or including radial velocity and parallax information ($P_{v+\pi}$). Parameters in red are the highest ones (not including the old field hypothesis) using all available observables. We stress the fact that one cannot directly interpret $P$ as an actual probability.}
\tablenotetext{b}{The Bayesian probability $P$ associated to a binary hypothesis was higher.}
\end{deluxetable}
\clearpage
\end{landscape}
\twocolumngrid

\clearpage
\onecolumngrid
\LongTables
\begin{landscape}
\tabletypesize{\tiny}
\renewcommand{\tabcolsep}{0.35mm}
\renewcommand{\arraystretch}{1.5}
\begin{deluxetable}{ll|lll|lll|lll|lll|lll|lll|lll|lll|ll|}
\tablecolumns{28}
\tablecaption{Statistical Distances and Radial Velocities\tablenotemark{a} \label{tab:dstat}}
\tablehead{
\colhead{} & \colhead{} & \multicolumn{2}{c}{TWA} & \colhead{} & \multicolumn{2}{c}{$\beta$PMG} & \colhead{} & \multicolumn{2}{c}{THA} & \colhead{} & \multicolumn{2}{c}{COL} & \colhead{} & \multicolumn{2}{c}{CAR} & \colhead{} & \multicolumn{2}{c}{ARG} & \colhead{} & \multicolumn{2}{c}{ABDMG} & \colhead{}  & \multicolumn{2}{c}{Young Field} & \colhead{}  & \multicolumn{2}{c}{Old Field}\\
\cline{3-4}
\cline{6-7}
\cline{9-10}
\cline{12-13}
\cline{15-16}
\cline{18-19}
\cline{21-22}
\cline{24-25}
\cline{27-28}
\colhead{Name} & \colhead{} & \colhead{$d_s$} & \colhead{$v_{\mathrm{rad}}$} & \colhead{} & \colhead{$d_s$} & \colhead{$v_{\mathrm{rad}}$} & \colhead{} & \colhead{$d_s$} & \colhead{$v_{\mathrm{rad}}$} & \colhead{} & \colhead{$d_s$} & \colhead{$v_{\mathrm{rad}}$} & \colhead{} & \colhead{$d_s$} & \colhead{$v_{\mathrm{rad}}$} & \colhead{} & \colhead{$d_s$} & \colhead{$v_{\mathrm{rad}}$} & \colhead{} & \colhead{$d_s$} & \colhead{$v_{\mathrm{rad}}$} & \colhead{} & \colhead{$d_s$} & \colhead{$v_{\mathrm{rad}}$} & \colhead{} & \colhead{$d_s$} & \colhead{$v_{\mathrm{rad}}$}
}
\startdata
J0001+1010 & & $15.3 \pm 1.6$ & $-12.0 \pm 2.4$ & & $63.8^{+11.6}_{-6.4}$ & $-5.5 \pm 3.9$ & & $55.0^{+4.0}_{-3.6}$ & $-2.6 \pm 3.6$ & & $84.2^{+8.0}_{-8.8}$ & $-4.0 \pm 7.2$ & & $60.2^{+6.0}_{-4.8}$ & $-5.0 \pm 3.9$ & & $83.4^{+8.0}_{-9.2}$ & $-6.7 \pm 5.6$ & & $87.8^{+8.0}_{-8.4}$ & $1.2 \pm 6.5$ & & $124.7^{+21.2}_{-15.2}$ & $-11.0 \pm 10.7$ & & $\textcolor{red}{\mathbf{147.1^{+25.6}_{-22.8}}}$ & $\textcolor{red}{\mathbf{-13.0 \pm 17.5}}$\\
J0003-2822 & & $16.1 \pm 1.2$ & $3.9 \pm 1.3$ & & $15.3^{+1.2}_{-0.8}$ & $6.1 \pm 1.4$ & & $18.1^{+1.2}_{-1.6}$ & $-5.0 \pm 1.7$ & & $18.1^{+1.2}_{-1.6}$ & $-2.3 \pm 1.3$ & & $19.3^{+0.4}_{-0.8}$ & $0.1 \pm 0.4$ & & $16.1^{+2.0}_{-1.2}$ & $2.2 \pm 1.3$ & & $22.1^{+0.8}_{-1.2}$ & $13.1 \pm 2.0$ & & $\textcolor{red}{\mathbf{34.2^{+6.4}_{-4.8}}}$ & $\textcolor{red}{\mathbf{-0.2 \pm 7.4}}$ & & $\ldots$ & $\ldots$\\
J0004-6410 & & $27.7^{+2.8}_{-2.4}$ & $6.7 \pm 1.8$ & & $36.6 \pm 4.0$ & $11.6 \pm 2.5$ & & $\textcolor{red}{\mathbf{47.8^{+3.6}_{-2.8}\tablenotemark{b}}}$ & $\textcolor{red}{\mathbf{6.8 \pm 3.1\tablenotemark{b}}}$ & & $54.6 \pm 5.6\tablenotemark{b}$ & $10.0 \pm 3.2\tablenotemark{b}$ & & $55.8^{+4.8}_{-5.2}\tablenotemark{b}$ & $15.6 \pm 5.1\tablenotemark{b}$ & & $53.0^{+6.4}_{-6.0}\tablenotemark{b}$ & $-0.1 \pm 2.9\tablenotemark{b}$ & & $53.0 \pm 5.2\tablenotemark{b}$ & $20.7 \pm 3.2\tablenotemark{b}$ & & $56.6^{+7.6}_{-9.2}\tablenotemark{b}$ & $10.3 \pm 8.6\tablenotemark{b}$ & & $\ldots$ & $\ldots$\\
\enddata
\tablenotetext{a}{Values in red are the ones for which the hypothesis had the highest Bayesian probability (including the field hypotheses) when using all available observables.}
\tablenotetext{b}{Object for which the Bayesian probability ($P_{H_k}$) for the binary hypothesis has a higher probability.}
\end{deluxetable}
\clearpage
\end{landscape}
\clearpage
\twocolumngrid

\end{document}